\documentclass[english]{extarticle}
\usepackage[T1]{fontenc}
\usepackage{geometry}
\usepackage{booktabs}
\usepackage{amsmath}
\usepackage{amsthm}
\usepackage{graphicx}

\makeatletter

\providecommand{\tabularnewline}{\\}

\newcommand{\laddress}[1]{
	\par {\raggedright #1
	\vspace{1.4em}
	\noindent\par}
}

\usepackage{color}

\@ifundefined{showcaptionsetup}{}{%
 \PassOptionsToPackage{caption=false}{subfig}}
\usepackage{subfig}
\makeatother

\usepackage{babel}
\begin{document}
\title{Dimits shift, avalanche-like bursts, and Solitary propagating structures in the two-field Flux-Balanced
Hasegawa-Wakatani model for plasma edge turbulence}
\author{Di Qi \textsuperscript{a,b}, Andrew J. Majda \textsuperscript{a,b},
and Antoine Cerfon \textsuperscript{a}}
\maketitle

\laddress{\textsuperscript{a }Department of Mathematics, Courant Institute
of Mathematical Sciences, New York University, New York, NY 10012}

\laddress{\textsuperscript{b }Center for Atmosphere and Ocean Science, Courant
Institute of Mathematical Sciences, New York University, New York,
NY 10012}
\begin{abstract}
We show that the recently introduced two-field flux-balanced Hasegawa-Wakatani (BHW) model captures the key features of drift-wave turbulent transport mediated by zonal flows observed in more complete and accurate gyrokinetic simulations, such as the existence of a nonlinear upshift of the treshold for drift wave turbulence driven transport, often called the Dimits shift, as well as non-local transport with avalanche bursts and solitary propagating structures. Because of the approximations made in the BHW model, these observations are made for the particle flux instead of the heat flux more commonly studied in ion temperature gradient (ITG) driven turbulence in fluid or gyrokinetic codes. Many of these features are not seen in other Hasegawa-Wakatani models, which confirms the critical role of the electron dynamics parallel to the magnetic field lines. To address questions regarding the role of boundary conditions on the drift-wave zonal flow dynamics, we apply our model to both a channel domain geometry and the more typical doubly periodic geometry. We only observe strong soliton-like solutions in the particle flux for the channel geometry, in the vicinity of the boundaries, where strong velocity shear and density gradients are generated which are absent in the doubly periodic simulations. Changing the aspect ratio of the simulation domain also has a significant effect. In domains which are elongated in the radial direction, more complex multiscale dynamics takes place, with multiple zonal jets interacting with each other, and large scale avalanches.
\end{abstract}

\section{Introduction}

The zonal flow -- drift wave interaction is known to play a fundamental role in setting the observed level of radial particle and heat transport in magnetically confined plasmas 
\cite{hammett1993,waltz1994,Beer1994,hortonreview1999,rogers2000,diamond2005zonal,itohzonalphysics2006}. The self-organization of the plasma from a turbulent state to a more laminar state with zonal flows can lead to a sharp reduction of turbulent transport, sometimes even a full suppression \cite{dimits2000simulation,dimits2000comparisons}. Furthermore, regimes with a high level of turbulent transport are not accurately modeled by theories relying on a local diffusion paradigm \cite{difpradalier2010}, and are instead better described by approaches allowing for nonlocal, nondiffusive transport, characterized by avalanche bursts and solitary traveling structures \cite{candy2003anomalous,mcmillan2009avalanchelike,difpradalier2010,gorler2011,vanwyk2016,vanwyk2016,mcmillan2018,ivanovArxiv}, in which zonal flows also play a key role \cite{beyer2000nondiffusive,benkadda2001bursty,gorler2011,ivanovArxiv}. While these observations rely on a large amount of numerical evidence, and the importance of these physical processes in modeling and predicting transport in magnetic confinement fusion devices is not disputed, theoretical understanding of the fundamental mechanisms at play and reduced models with accurate yet inexpensive predictive capabilities remain scarce.

Although reduced fluid models can rarely be derived self-consistently from first principles for the values of the plasma parameters measured in magnetic confinement fusion experiments, they have been useful to identify the elementary building blocks required for the emergence of zonal flows from drift wave turbulence \cite{dewar2007zonal,numata2007bifurcation,majda2018flux,qi2019flux}, the associated reduction of turbulence driven transport \cite{stonge2017,zhu2019theory,ivanovArxiv}, as well as the nonlocal transport characterized by ballistic bursts called avalanches \cite{sarazin2000,beyer2000nondiffusive} and long-time-coherent, soliton-like traveling structures, for which the term ``ferdinons" was recently coined \cite{zhou2019solitary,ivanovArxiv}. In this article, we demonstrate that our recently proposed flux-balanced Hasegawa-Wakatani (BHW) model \cite{majda2018flux,qi2019flux} is one of the simplest models which captures all of these characteristic features  of turbulent transport in magnetic confinement fusion devices. It is therefore a powerful model to study the complex interplay between the fundamental physics mechanisms responsible for the observed level of transport, but also to test and benchmark mathematical methods for model reduction as well as uncertainty quantification, which, if successful, may then be tested on the more accurate gyrokinetic model and the significantly more computationally intensive gyrokinetic simulations.
 
There have been three different Hasegawa-Wakatani (HW) models introduced in the plasma physics community: the original HW (OHW) model  \cite{hasegawa1983plasma}, the modified HW (MHW) model \cite{numata2007bifurcation}, and our flux-balanced HW model \cite{majda2018flux}. The BHW model differs from the other two by its treatment of the electron response parallel to the magnetic lines, which is constructed so as to ensure, in contrast to the OHW and MHW models, that there is not net radial flux of electrons in the adiabatic electron limit. In this article, we will show that certain characteristics of magnetized turbulent transport discussed in the previous paragraph are only captured by the BHW model, and not the other HW models. This is in particular true for what is known as the Dimits shift, namely the nonlinear upshift
in the critical gradient required for the onset of significant turbulent transport \cite{dimits2000comparisons,zhu2019theory,ivanovArxiv}. Our results thus show that the electron dynamics is a critical component to the existence of a true Dimits shift, with a sharp transition in the observed level of transport from a turbulent flow regime to a regularized zonal flow regime, which is absent in the other HW models. This result is in agreement with the conclusion St-Onge reached with the Terry-Horton model \cite{stonge2017}. Observe that the Dimits shift usually refers to the nonlinear upshift for the onset of strong heat flux, and is usually considered in the context of the ion temperature gradient (ITG) instability \cite{dimits2000comparisons}. Since our BHW model does neither have a temperature gradient nor a heat flux, we will study the Dimits shift for the onset of strong particle transport, driven by a density gradient drift instability, as was done in other models \cite{stonge2017,zhu2019theory}.

In addition, this article will investigate, within the framework of the BHW model, the role that boundary conditions play on the dynamics. The dominant point of view in the magnetic confinement fusion community has long been that since turbulent structures are very much smaller than the system size in the directions perpendicular to the magnetic field, the local turbulence properties can be expected to mostly depend on local gradients, and are not affected by boundaries in a meaningful way. As a result, a large number of simulations are done with periodic boundary conditions in the directions perpendicular to the magnetic field, both for reduced fluid models and for detailed kinetic models \cite{beer1995,numata2007bifurcation,xanthopoulos2011,stonge2017,faber2015,martin2018}. However, the nonlocal nature of the transport characteristics, and the large scale structures which emerge as a result of the plasma self-organization raise questions regarding the validity of this approach. Recent results with non-periodic boundary conditions \cite{mcmillan2018}, and with domains with different aspect ratios \cite{qi2019zonal} suggest that boundary conditions indeed have an impact on the resulting turbulence and transport. In this article, we will therefore study the emergence of zonal flows and of large scale structures, and the existence of a Dimits shift as well as of nonlocal transport for two different boundary conditions: a channel
domain geometry, where the radial component of the velocity and the density perturbation are set to zero at the inner edge and outer edge of the computational domain \cite{qi2019channel}, and the periodic boundary conditions more commonly used for such study. Furthermore, we also investigate the dependence of our results on the aspect ratio of the computational domain. A computational
domain which is elongated in the radial direction includes many more intermediate
scales, and their interaction creates richer dynamics with the generation
of multiple jets interaction and stronger intermittent bursts in the
transient regime.

In the remainder of this article, the flux-balanced Hasegawa-Wakatani
model and its fundamental properties are described
in Section \ref{sec:The-flux-balanced-HW}. The existence of a true Dimits shift and a sharp transition from a turbulence dominated, high flux regime to a regime with strong zonal structures and very small particle flux are studied for the channel domain geometry in Section \ref{sec:Dimits-shift}. The avalanche-like structures in the turbulent transport of particles analyzed in Section \ref{sec:Avalanche-like-structures}, with a comparison between different HW models and different aspect ratios of the computational domain.
In Section \ref{sec:Dimits-shift-Periodic}, we revisit the simulations of the previous sections with doubly periodic boundary conditions imposed instead of the channel geometry, and again compare the MHW and BHW models. Finally, our results
are summarized in Section \ref{sec:Summary}.

\section{The flux-balanced Hasegawa-Wakatani Model in plasma edge turbulence\label{sec:The-flux-balanced-HW}}

Below, and for the remainder of the article, we will identify the distinct contributions from the zonal states and from the non-zonal
fluctuations by decomposing the electrostatic potential $\varphi$
and particle density $n$ into the zonal mean
states $\overline{\varphi},\overline{n}$ and their fluctuations about
the zonal mean $\tilde{\varphi},\tilde{n}$ as follows:
\[
\varphi=\overline{\varphi}+\tilde{\varphi},\;n=\overline{n}+\tilde{n},\quad\overline{f}\left(x\right)=L_{y}^{-1}\int_{0}^{L_{y}}f\left(x,y\right)dy.
\]

Generically, the Hasegawa-Wakatani models \cite{hasegawa1983plasma,numata2007bifurcation,majda2018flux}
describe the drift wave -- zonal flow interactions in a two-field
coupled system in a shearless slab geometry. The system supports unstable drift waves driven by a background ion density gradient and a non-adiabatic electron response parallel to the magnetic field. 
The\emph{ flux-balanced Hasegawa-Wakatani} (BHW) model was recently introduced in \cite{majda2018flux}, and is conveniently expressed in terms of the flux-balanced potential vorticity $q=\nabla^{2}\varphi-\tilde{n}$
and the density fluctuation $n$, as follows:\addtocounter{equation}{0}\begin{subequations}\label{plasma_balance}
\begin{eqnarray}
\frac{\partial q}{\partial t}+\mathbf{u}\cdot\nabla q-\kappa\frac{\partial\varphi}{\partial y} & = & \mu\Delta q,\label{eq:plasma_balance1}\\
\frac{\partial n}{\partial t}+\mathbf{u}\cdot\nabla n+\kappa\frac{\partial\varphi}{\partial y} & = & \alpha\left(\tilde{\varphi}-\tilde{n}\right)+\mu\Delta n.\label{eq:plasma_balance2}
\end{eqnarray}
\end{subequations}Here $\varphi$ is the electrostatic potential,
$n$ is the density fluctuation from the background density profile
$n_{0}\left(x\right)$, and $\mathbf{u}\equiv\nabla^{\bot}\varphi=\left(-\partial_{y}\varphi,\partial_{x}\varphi\right)$
is the $E\times B$ velocity field, where $x$ can be interpreted as the coordinate for the radial direction, and $y$ the coordinate for the poloidal direction, if our shearless slab geometry is interpreted as an approximation of the edge of a magnetic confinement device. The constant background density
gradient $\kappa=-\nabla\ln n_{0}$ is a consequence of the assumption of an exponential
background density profile in the plasma region of interest: $n_{0}\left(x\right)=\exp\left(-\kappa x\right)$. $\alpha$ is a constant which is often called the
``adiabaticity parameter" \cite{numata2007bifurcation}, and which is proportional to the inverse of the parallel resistivity. The terms $\mu\Delta q$ and $\mu\Delta n$ can be viewed as dissipation due to the collisional diffusion of the electrons perpendicular to the magnetic field \cite{majda2018flux}, and have the same dissipation coefficient $\mu$ for the vorticity and for the density \cite{majda2015intermittency}. 

We stress that in the BHW model (\ref{plasma_balance}), the poloidally averaged
density $\overline{n}$ is removed from the potential
vorticity $q$. By doing so, we were able to prove rigorously \cite{majda2018flux}, as well as confirm numerically \cite{majda2018flux,qi2019flux}, that in the limit of adiabatic electrons, $\alpha\rightarrow+\infty$, the BHW model converges to the following one-field model:
\begin{equation}
\frac{\partial q_{\mathrm{MHM}}}{\partial t}+\nabla^{\bot}\varphi\cdot\nabla q_{\mathrm{MHM}}-\kappa\frac{\partial\varphi}{\partial y}=\mu\Delta q_{\mathrm{MHM}},\quad q_{\mathrm{MHM}}=\nabla^{2}\varphi-\tilde{\varphi},\label{eq:plasma_onelayer}
\end{equation}
which is called the \emph{modified Hasegawa-Mima} (MHM) model \cite{dewar2007zonal}, and where we draw the attention of the reader to the fact that the zonally averaged potential $\overline{\varphi}$ is absent in the definition of the MHM potential vorticity: $q_{\mathrm{MHM}}=\nabla^{2}\varphi-\tilde{\varphi}$. In the BHW model, there is therefore no net radial transport of electrons in the limit of adiabatic electrons, which is what one would expect physically \cite{Dorland1993,dorland1993gyrofluid,dewar2007zonal,pushkarev2013}.
In contrast, when the modified Hasegawa-Wakatani (MHW) model is expressed in terms of the potential vorticity and the density, the ``unbalanced''
potential vorticity $q_{MHW}=\nabla^{2}\varphi-n$ appears, which contains the mean state contribution $\overline{n}$. In the formal limit ``$\alpha=+\infty$'', the MHW model also agrees with the MHM model, but the convergence of the MHW model to this limit is problematic, in the sense that for $\alpha$ large but finite, the dynamics of the MHW model can be significantly different from that of the MHM model, as we have shown previously \cite{majda2018flux,qi2019flux}, with a considerable amount of kinetic energy maintained in the small scale turbulence in the MHW model, which is not fully quenched by the zonal structures. This subtle difference will play a major role in the remainder of this article, as it is the reason why a true Dimits shift will be observed in the BHW model, with a sharp transition between a high transport regime and a low transport regime, and not seen in the same way for the MHW model.

\subsection{Boundary conditions for the computational domain geometry}\label{sec:bdry_cond}

As discussed in the introduction, the two-dimensional flow is usually defined on a rectangular domain, i.e. $\mathbf{x}=\left(x,y\right)\in\mathcal{D}=\left[-L_{x},L_{x}\right]\times\left[-L_{y},L_{y}\right]$, with periodic boundary condition in both $x$ and $y$ directions, which for our model (\ref{plasma_balance}) would imply:
\begin{equation}
\begin{aligned}q\left(x,-L_{y}\right)=q\left(x,L_{y}\right), & \quad q\left(-L_{x},y\right)=q\left(L_{x},y\right),\\
n\left(x,-L_{y}\right)=n\left(x,L_{y}\right), & \quad n\left(-L_{x},y\right)=n\left(L_{x},y\right).
\end{aligned}
\label{eq:boundary_peri}
\end{equation}
The above doubly periodic boundary conditions offer a convenient setup to apply the standard Fourier decomposition to all quantities of interest, which can be computed efficiently with the Fast Fourier Transform (\emph{FFT}).

However, the emergence of large-scale structures from the nonlinear drift wave dynamics, sometimes comparable to the size of the computational domain, raises questions regarding the role of boundary conditions and the applicability of periodic boundary conditions to this problem, particularly in the ``radial" $x$ direction. In this article, we therefore also consider a channel domain geometry \cite{qi2019channel}, with periodic boundary conditions in the ``poloidal" $y$ direction, and zero radial flow and density fluctuation imposed along the radial edges $x=0,L_{x}$ of the domain:

\begin{equation}
\begin{aligned}q\left(x,-L_{y}\right)=q\left(x,L_{y}\right), & \quad u\left(0,y\right)=u\left(L_{x},y\right)=0,\\
n\left(x,-L_{y}\right)=n\left(x,L_{y}\right), & \quad n\left(0,y\right)=n\left(L_{x},y\right)=0.
\end{aligned}
\label{eq:boundary_channel}
\end{equation}
We do not have a rigorous justification derived from first principles for these boundary boundary conditions, whose primary merit are to be an alternative to the usual periodic boundary conditions. Still, these boundary conditions are reasonable in the sense that they would be physically appropriate for a plasma in contact with a solid wall at $x=L_{x}$, and such that a strong internal transport barrier is present at $x=0$, preventing transport across this flux surface. We also note that these boundary conditions for the density perturbation have been imposed in other reduced fluid models for the study of bursty turbulent transport with avalanches \cite{sarazin2000}.

\subsection{Stability analysis for the growth rate in drift wave turbulence\label{subsec:Linear-instability}}

Linear stability analysis of the BHW model (\ref{plasma_balance})
highlights the driving mechanism for the generation of non-zonal fluctuations
in the starting transient state. We write the solutions $\left(\tilde{\varphi},\tilde{n}\right)$ to the linearized BHW equations as normal modes,
\[
\tilde{\varphi}=\hat{\varphi}\exp\left(i\left(\mathbf{k\cdot x}-\omega t\right)\right),\quad\tilde{n}=\hat{n}\exp\left(i\left(\mathbf{k\cdot x}-\omega t\right)\right),
\]
where $\omega\equiv\omega\left(\mathbf{k}\right)$ is the wave frequency
for the corresponding wave vector, and where the subscript $\mathbf{k}$ for $\hat{\varphi}_{\mathbf{k}}$ and $\hat{n}_{\mathbf{k}}$ have been omitted for the simplicity of the notation. Non-zonal solutions are such that $k_{y}\neq 0$. 

We first consider the non-dissipative case, with $\mu=0$. Plugging the normal mode solutions in the linearized BHW equations, we find the following solutions to the dispersion equation \cite{majda2018flux,qi2019linking}: 
\begin{equation}
\omega^{\pm}=\frac{\alpha}{2}\frac{1+k^{2}}{k^{2}}\left[\pm\Gamma^{\frac{1}{4}}\cos\frac{\theta}{2}-i\left(1\mp \Gamma^{\frac{1}{4}}\sin\frac{\theta}{2}\right)\right],\label{eq:eigenvalues}
\end{equation}
with
\[
\Gamma\left(\mathbf{k};\frac{\kappa}{\alpha}\right)=1+16\gamma^{2},\quad\gamma\left(\mathbf{k};\frac{\kappa}{\alpha}\right)=\frac{\kappa}{\alpha}\frac{k_{y}k^{2}}{\left(1+k^{2}\right)^{2}}=\frac{\omega_{*}k^2}{\alpha(1+k^2)},\qquad\omega_{*}=\frac{\kappa k_{y}}{1+k^2}.
\]
where $\omega_{*}$ is called the drift wave frequency. From this expression, we can see that the question of whether a mode is stable or not only depends on the physical parameters $\kappa$ and $\alpha$ through the ratio $\kappa/\alpha$. For $\alpha\neq 0,\kappa\neq 0$, all non-zonal modes have a solution with positive imaginary part, so they are all unstable to this drift-wave instability. In the limit $\alpha\rightarrow+\infty$, the imaginary part of the frequency goes to zero for all modes, which means that adiabatic electrons stabilize the instability. In other words, the instability is driven by the combination of the background density gradient and the electron resistivity parallel to the magnetic field lines. The BHW model is thus a model for the study of electrostatic resistive drift wave turbulence, just like the MHW model \cite{numata2007bifurcation}. 

Let us now turn to the more interesting and physically relevant dissipative case, with $\mu>0$. The homogeneous damping operator with strength $\mu$
acts as a stabilizing effect for the system, acting most strongly on the small-scale modes. The damping effect can be easily accounted for by considering solutions to the linearized equations with dissipation of the form:
\[
\begin{aligned}\tilde{\varphi}^{\pm}= & \hat{\varphi}e^{-\mu k^{2}t}e^{-i\omega^{\pm}t}e^{i\mathbf{k\cdot x}},\\
\tilde{n}^{\pm}= & \hat{n}e^{-\mu k^{2}t}e^{-i\omega^{\pm}t}e^{i\mathbf{k\cdot x}}.
\end{aligned}
\]
The cross-field collisional diffusion of the electrons acts as a counterbalance to the drift resistive instability. A mode is stable when this damping effect is larger than the linear growth rate. If we write the linear growth rate as $\sigma^{+}=\mathrm{Im}\omega^{+}$, the criterion can be written as
\begin{equation}
\mu \geq\frac{\sigma^{+}}{k^2}\:\Leftrightarrow\;\mu\geq\frac{\alpha}{2}\frac{1+k^2}{k^4}\left[-1+\frac{1}{\sqrt{2}}\left(1+\sqrt{1+16\frac{\kappa^2}{\alpha^2}\frac{k_{y}^2k^4}{(1+k^2)^4}}\right)^{1/2}\right].\label{eq:stability_lim}
\end{equation}
From the criterion (\ref{eq:stability_lim}), we can conclude that in the presence of dissipation in the BHW model, short wavelength modes will be stable, while large wavelength modes will be unstable to the drift resistive instability. This instability corresponds to the first excited drift wave state. In the subsequent, nonlinear phase, a secondary instability \cite{qi2019zonal} takes place through the nonlinear transfer of energy from the transient fluctuating
modes to the zonal directions, and dominant zonal jets are created, as we will see in our numerical results.

Note that in the limit $\frac{\kappa}{\alpha}\ll 1$, a criterion can be derived for the overall linear stability of the system, by Taylor expanding the right-hand side of (\ref{eq:stability_lim}) and retaining only the term which is first order in $\frac{\kappa^2}{\alpha^2}$:

\begin{equation}
\mu\geq\frac{\kappa^{2}}{\alpha}\max_{k_{x},k_{y}}\frac{k_{y}^{2}}{\left(k^{2}+1\right)^{3}}=\frac{4}{27}\frac{\kappa^{2}}{\alpha}
\end{equation}

\subsection{Conserved quantities and their dynamical equations}

The nonlinear interactions conserve two important quantities in the BHW model (\ref{plasma_balance}). They are the potential enstrophy $W$ and total energy $E$ defined
as
\begin{equation}
W=\frac{1}{2}\int_\mathcal{D} q^{2}dxdy=\frac{1}{2}\int_{\mathcal{D}}\left(\nabla^{2}\varphi-\tilde{n}\right)^{2}dxdy,\quad E=\frac{1}{2}\int_{\mathcal{D}}\left(\left|\nabla\varphi\right|^{2}+n^{2}\right)dxdy,\label{eq:energy_plasma}
\end{equation}
First, the dynamical equation for the total enstrophy $W$ is determined by the potential vorticity equation
\begin{equation}
\frac{dW}{dt}=\kappa\int_{\mathcal{D}}\tilde{u}\tilde{n}dxdy\;-\mu\int_{\mathcal{D}}\left|\nabla q\right|^{2}dxdy.\label{eq:eqn_ens}
\end{equation}
We thus see that the total particle flux
\begin{equation}
\Gamma=\int_\mathcal{D}\tilde{u}\tilde{n}dxdy,\label{eq:particle_flux}
\end{equation}
acts as a forcing effect exciting drift waves. Second, the dynamical
equation for the total energy $E$ can be derived in a similar way
based on the vorticity and density equations
\begin{equation}
\frac{dE}{dt}=\int_\mathcal{D}\left(\kappa+\overline{v}\right)\left(\tilde{u}\tilde{n}\right)dxdy-\alpha\int_\mathcal{D}\left(\tilde{n}-\tilde{\varphi}\right)^{2}dxdy\:-\mu\int_\mathcal{D}\left(\left|\Delta\varphi\right|^{2}+\left|\nabla n\right|^{2}\right)dxdy.\label{eq:eqn_ene}
\end{equation}
The additional term due to the mean velocity $\overline{v}=\partial_{x}\overline{\varphi}$
advection in the energy equation represents the zonal flow transport
of the particle flux, $\overline{\tilde{u}\tilde{n}}$. We observe that the negative-definite term $\alpha\int_\mathcal{D}\left(\tilde{n}-\tilde{\varphi}\right)^{2}dxdy$ appears in the equation for the evolution of the energy $E$ as an energy sink due to the electron collisional parallel resistivity, while this term is absent from the equation for the evolution of the enstrophy $W$. 

\subsubsection{Additional conserved quantities in the channel geometry}

For the channel geometry, conservation equations for two additional quantities can be derived, namely the \emph{impulses} for the vorticity $I_{q}$ and density $I_{n}$,
defined as the first moments of the zonal potential vorticity $\bar{q}=\partial_{x}^{2}\bar{\varphi}$
and the zonal density $\bar{n}$:
\begin{equation}
I_{q}=\int_{0}^{L_{x}}x\bar{q}dx,\quad I_{n}=\int_{0}^{L_{x}}x\bar{n}dx.\label{eq:impulse}
\end{equation}
The dynamical equations for the impulses are \cite{qi2019channel}
\begin{equation}
\frac{dI_{q}}{dt}=-\Gamma+\mu L\partial_{x}\bar{q}_{L},\quad\frac{dI_{n}}{dt}=\Gamma+\mu L\partial_{x}\bar{n}_{L}.\label{eq:dyn_impul}
\end{equation}
where $\partial_{x}\bar{q}_{L}\equiv\partial_{x}\bar{q}\left(t,L_{x}\right)$ and $\partial_{x}\bar{n}_{L}\equiv\partial_{x}\bar{n}\left(t,L_{x}\right)$
are quantities evaluated at the right, i.e. outer, boundary $x=L_{x}$. The impulse dynamics (\ref{eq:dyn_impul}) thus link the zonal profiles with the total particle flux $\Gamma=\int_{\mathcal{D}}\tilde{u}\tilde{n}dxdy$
appearing in both the enstrophy and energy dynamics, (\ref{eq:eqn_ens})
and (\ref{eq:eqn_ene}). The second terms on the right-hand side of the equations are due to electron collisional diffusion.

The impulse equations (\ref{eq:dyn_impul}) imply useful additional conserved quantities
for the competition between zonal states and fluctuation modes in
the channel domain geometry. For the diffusionless case with $\mu=0$, the combination of the two impulse equations 
first gives the conservation of the total impulse, $I=I_{q}+I_{n}$.
The equations for the impulses can also be combined with the enstrophy equation (\ref{eq:eqn_ens}) for $W$, which for $\mu=0$ is only driven by the total particle flux. For the channel BHW model without diffusion ($\mu=0$) we thus obtain the following conservation equation:
\begin{equation}
\frac{d}{dt}\left(W+\kappa I_{q}\right)=0,\quad W+\kappa I_{q}=W-\kappa I_{n},\label{eq:conserv2}
\end{equation}
where the identity between the impulses $I_{n}=-I_{q}$ is used, assuming that the initial state is such that $I_{n}=-I_{q}$. Otherwise, a similar result holds, only shifted by a constant associated with the initial conditions. The
above conservation law offers a useful relation for the balance between
the fluctuation modes (represented by the enstrophy fluctuation $\tilde{W}$)
and the zonal state (represented by the impulse $I$ and zonal component of the enstrophy 
$\overline{W}$). Especially, assuming a system starting from small
initial fluctuations, an upper bound for the total enstrophy in the non-zonal fluctuations averaged over the domain area $A=L_{x}L_{y}$ can
be obtained from the above conservation law:
\[
\frac{1}{A}\int_\mathcal{D}\left(\tilde{q}^{2}-\tilde{q}_{0}^{2}\right)dxdy\leq\kappa^{2}\sum_{k=k_{x}}k^{-2}\leq\frac{L_{x}^{2}}{3}\kappa^{2}.
\]
This provides a useful quantification for the saturated total fluctuation
in the flow field from the drift instability. A detailed discussion for
the derivation of the conserved quantities and numerical confirmation of the saturation bound in the channel domain can be found in \cite{qi2019channel}.

\section{Dimits shift and statistical transition in the channel geometry\label{sec:Dimits-shift}}

The Dimits shift usually refers to the nonlinear upshift of the
critical temperature gradient for the rapid transition from a low transport regime to a strong turbulent transport regime, which is observed in gyrokinetic simulations of ion-temperature-gradient turbulence
\cite{dimits2000simulation,dimits2000comparisons,weiland2019drift}.
In our BHW model, which does neither have a temperature gradient nor a heat flux, but does have a density gradient and particle transport, an analogous Dimits shift together with a sharp transport regime transition are observed for the total particle flux. In this section, we characterize this transport transition and the corresponding properties of the magnetized plasma turbulence in our channel geometry. These results will be compared with the Dimits shift we obtain for the 
doubly periodic boundary conditions in Section \ref{sec:Dimits-shift-Periodic}.

\subsection{Numerical treatment of the boundary conditions}

For our numerical simulations of the BHW model, we implemented an efficient pseudo-spectral scheme \cite{qi2019flux}. For the doubly periodic computational domain (\ref{eq:boundary_peri})
with $\mathbf{x}=\left(x,y\right)\in\left[-L_{x},L_{x}\right]\times\left[-L_{y},L_{y}\right]$,
the state variables are expanded in Fourier modes according to:
\begin{equation}
\varphi=\sum_{k}\hat{\varphi}_{k}e^{i\mathbf{k\cdot x}},\quad n=\sum_{k}\hat{n}_{k}e^{i\mathbf{k\cdot x}},\quad\zeta=\nabla^{2}\varphi=-\sum_{k}k^{2}\hat{\varphi}_{k}e^{i\mathbf{k\cdot x}},\label{eq:fluc_mode}
\end{equation}
with spectral wave vector $\mathbf{k}=\left(k_{x},k_{y}\right)=\left(\frac{\pi}{L_{x}}n_{x},\frac{\pi}{L_{y}}n_{y}\right)$
and $k^{2}=\left|\mathbf{k}\right|^{2}$, and integer indices $n_{x}=-\frac{N_{x}}{2}+1,\cdots,\frac{N_{x}}{2}$,
and $n_{y}=-\frac{N_{y}}{2}+1,\cdots,\frac{N_{y}}{2}$ truncated at
a large wavenumber. Zonal modes are represented by wavenumbers
$k_{x}$ and $k_{y}=0$, and non-zonal fluctuation modes require
$k_{y}\neq0$. Observe that larger domain sizes $L_{x},L_{y}$ will allow for more intermediate scales in the corresponding directions in the spectral domain.

For the channel domain geometry (\ref{eq:boundary_channel}), we define
the flow solutions on the computational domain with $\mathbf{x}=\left(x,y\right)\in\left[0,L_{x}\right]\times\left[-L_{y},L_{y}\right]$.
The boundary conditions are periodic in the $y$ direction, and such that there are no density fluctuations and no $x$-directed flow at the boundaries $x=0,L_{x}$ of the domain. The latter is satisfied by imposing the following conditions, for any $y\in[-L_{y},L_{y}]$: 
\begin{equation}
\begin{aligned}\varphi\left(0,y,t\right) & = \varphi\left(L_{x},y,t\right)=0,\\
n\left(0,y,t\right) & = n\left(L_{x},y,t\right)=0.
\end{aligned}
\label{eq:boundary}
\end{equation}
Note that strictly speaking, the left and right edges only need to be lines with constant potential (not necessarily zero) for the channel boundary conditions to be satisfied. However, we can set the potential to zero at both edges without loss of generality because the potential is always defined to within a free constant, which we fix in such a way that $\varphi=0$ at the left edge, and because the BHW model is Galilean invariant \cite{majda2018flux}, which means that our results will be insensitive to the addition of a large scale profile of the form $Vx$ to the potential profile. 

In order to have an efficient pseudo-spectral scheme for the numerical simulations for the channel domain, we use the numerical trick already implemented in \cite{qi2019channel}, namely we
extend the original state variables $f$ to an odd periodic extension $f^{\mathrm{odd}}$ in the $x$ direction:
\[
f^{\mathrm{odd}}\left(x,y\right)=\begin{cases}
f\left(x,y\right), & 0\leq x\leq L_{x},\\
-f\left(-x,y\right), & -L_{x}\leq x<0,
\end{cases}
\]
with $f^{\mathrm{odd}}\left(-L_{x},y\right)=f^{\mathrm{odd}}\left(0,y\right)=f^{\mathrm{odd}}\left(L_{x},y\right)\equiv0$
as required for an odd periodic function. This extended function $f^{\mathrm{odd}}$
is doubly periodic on the extended
domain $\left[-L_{x},L_{x}\right]\times\left[-L_{y},L_{y}\right]$ and naturally satisfies the channel boundary
condition (\ref{eq:boundary}) on the constrained half domain $\left[0,L_{x}\right]\times\left[-L_{y},L_{y}\right]$.
In practice, $f^{\mathrm{odd}}$ is constructed as a Fourier
expansion in the $y$ direction and sine expansion in the $x$ direction
\[
f^{\mathrm{odd}}\left(x,y\right)=\sum_{k_{x}}i\hat{a}_{k_{x}}\left(y\right)e^{ik_{x}x},\quad\hat{a}_{k_{x}}\left(y\right)=\sum_{k_{y}}\hat{f}_{k_{x},k_{y}}e^{ik_{y}y},
\]
with $\hat{a}_{k_{x}}$ real coefficients from the sine transform. With this approach, the simulations for the channel geometry can be computed using the same pseudo-spectral
code as the code for the doubly periodic simulations on the domain $\left[-L_{x},L_{x}\right]\times\left[-L_{y},L_{y}\right]$, with the only modification that one must take the odd decomposition of
the functions.

For our simulations, the initial state is taken to be Gaussian random fields for the vorticity and the density, with zero mean and very small variance. The basic domain size is $L_{x}=L_{y}=20$ along the two directions,
and an elongated zonal length case $L_{x} = 100$ will be compared, which shows richer dynamics. We change the values of the model parameters $\kappa$ and
$\alpha$ to investigate the transitions in the turbulent flow. The collisional diffusion parameter is kept relatively small, either $\mu=1\times10^{-3}$
or $\mu=1\times10^{-4}$ depending on the cases, and diffusionless cases with $\mu=0$ will also be considered. In addition, hyperviscosities, $-\nu\Delta^{s}q$
and $-\nu\Delta^{s}n$, are added to each equation respectively in our numerical implementations, with small coefficient $\nu=7\times10^{-21}$
and high order $s=8$, in order to dissipate the energy in the underresolved smallest
scales. All numerical simulations are based on a standard pseudo-spectral scheme with $N=256$ discretization points along each direction. A
fourth-order explicit-implicit Runge-Kutta scheme is used for the time integration with the implicit part only for the stiff hyperviscosity term.

\subsection{Sharp transport transition from zonal jet dominated regime to drift wave turbulence dominated regime}

In the framework of the BHW model, a Dimits shift is a nonlinear upshift of the threshold for strong particle flux. In other words, it is the presence of a region in parameter space in which one would expect moderate to large particle flux according to linear drift wave analysis, but near-zero flux is in fact observed, because zonal flows quench the drift instability and prevent the radial transport of particles. We now investigate this phenomenon in detail.

In the BHW model (\ref{plasma_balance}), the parameters which control the linear drift wave instability to excite small-scale non-zonal fluctuation modes are the adiabaticity parameter $\alpha$ and the density gradient $\kappa$. The collisional diffusion parameter $\mu$, on the other hand, acts on both the vorticity and
density as a homogeneous damping effect, which is stabilizing. From the
linear stability analysis of the BHW model in Section \ref{subsec:Linear-instability}
as well as in \cite{qi2019linking,qi2019flux}, we know that for given values of $\kappa$ and $\mu$, there is a critical $\alpha$ value, which we call $\alpha_{\mathrm{cr}}\left(\kappa,\mu\right)$, above which the system is linearly stable. Likewise, for given values of $\alpha$ and $\mu$, there is a critical $\kappa$ value, which we call $\kappa_{\mathrm{cr}}\left(\alpha,\mu\right)$, below which the system is linearly stable. 

\begin{figure}
\includegraphics[scale=0.36]{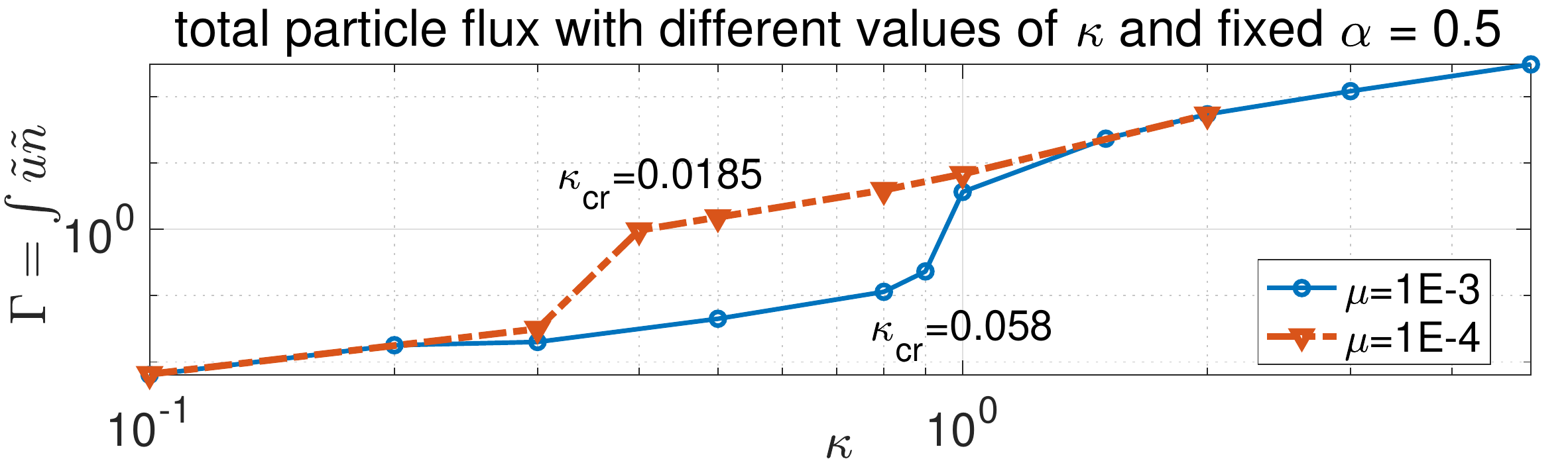}\includegraphics[scale=0.36]{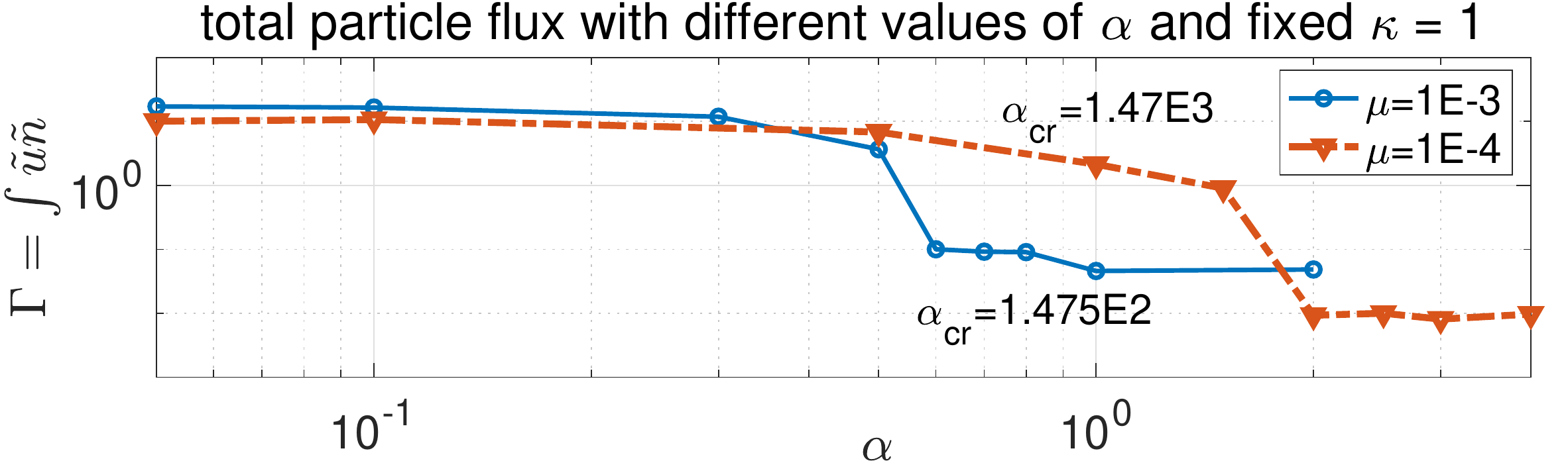}

\caption{Total particle flux $\Gamma=\int_{\mathcal{D}}\tilde{u}\tilde{n}dxdy$ as a function
of the parameters $\alpha$ and $\kappa$ from direct numerical solutions
of the BHW model in the channel domain geometry. The curves corresponding to two different values of the collisional diffusion parameter $\mu$, $\mu=1\times10^{-3}$ and $\mu=1\times10^{-4}$, are compared.
The critical values of $\kappa_{\mathrm{cr}}$ and $\alpha_{\mathrm{cr}}$
for the onset of the linear instability are listed near the corresponding
lines.\label{fig:Total-particle-flux}}
\end{figure}

In our numerical simulations, for the linearly stable regime, with all unstable modes damped by the dissipation, the flow is regular with no turbulent structures. When one enters the linearly unstable regime, obtained for $\alpha<\alpha_{\mathrm{cr}}$ or $\kappa>\kappa_{\mathrm{cr}}$,
we see richer dynamical structures induced by the drift wave instability. Still, in a large region above the threshold for the linear instability, the excited drift waves observed in the transient state are completely eliminated through the nonlinear coupling between zonal and fluctuation modes. This nonlinear interaction mechanism is described in detail in \cite{qi2019zonal} using the secondary instability analysis. In this intermediate regime, we therefore see the generation of dominant, long-lived zonal flows, and near-zero non-zonal turbulent fluctuations, as we will show explicitly in the next section. As a result, there is a near-zero total particle flux $\Gamma=\int_{\mathcal{D}}\tilde{u}\tilde{n}dxdy$, mostly determined by random numerical fluctuations. As one further decreases $\alpha$ or increases $\kappa$, to second critical parameter
values we call $\alpha_{\mathrm{turb}}\left(\kappa,\mu\right)$ or $\kappa_{\mathrm{turb}}\left(\alpha,\mu\right)$,
the secondary instability saturates, and we see the generation of strong
turbulent flows which are not quenched by zonal flows; this is the starting point for strong zonal transport. Before these critical values are reached, in the regime $\alpha_{\mathrm{turb}}<\alpha<\alpha_{\mathrm{cr}}$
or $\kappa_{\mathrm{cr}}<\kappa<\kappa_{\mathrm{turb}}$, the flow converges to regular zonal structures with little turbulence remaining after the starting transient state. The Dimits shift in the BHW model can be represented by the distance $\alpha_{\mathrm{cr}}-\alpha_{\mathrm{turb}}$ in the adiabaticity parameter, or $\kappa_{\mathrm{turb}}-\kappa_{\mathrm{cr}}$ in the density gradient parameter. The existence of the Dimits shift and a sharp transition in the level of particle transport for $\alpha=\alpha_{\mathrm{turb}}<\alpha_{\mathrm{cr}}$ or  $\kappa=\kappa_{\mathrm{turb}}>\kappa_{\mathrm{cr}}$ in the BHW model can be clearly seen in Figure \ref{fig:Total-particle-flux}, which shows the change in the total
particle flux as the parameter $\kappa$ changes in the range $\left[0.1,5\right]$
with fixed $\alpha=0.5$, and as the parameter $\alpha$ changes in
the range $\left[0.05,5\right]$ with fixed $\kappa=1$, for two
different values of the collisional diffusion parameter $\mu$: $\mu=10^{-4}$ and $\mu=10^{-3}$. We observe that, as expected, a larger value of
$\mu$ pushes the transition point to turbulence to a larger value of $\kappa$ or a smaller value of $\alpha$, requiring a stronger instability to induce strong
transport. Still the sudden phase transition is seen for both the smaller and the larger value of $\mu$. Finally note that for $\alpha$ and $\mu$ fixed, the particle flux continually increases as one increases $\kappa$ into the drift wave turbulence dominated regime. On the other hand, for $\kappa$ and $\mu$ fixed, the particle flux seems to asymptote as one decreases $\alpha$ towards zero, and would in fact decrease if we plotted the particle flux for even smaller values of $\alpha$ \cite{qi2019flux}. This is because the growth rate of the linear instability always increases with $\kappa$, whereas that growth rate decreases and converges to zero in the limit $\alpha\rightarrow 0$. The parameter values for $\alpha$, $\kappa$ and $\mu$ are picked according to the reference values in \cite{dewar2007zonal,hasegawa1983plasma} and the previous simulation tests \cite{qi2019channel,qi2019flux} where the flows display the most interesting structures and statistics. From our numerical simulations, we computed the equilibrium total particle flux in the final statistical steady state by taking the time average of 5000 time steps, which averages out the random, sometimes sharp fluctuations in the time series (see Figure \ref{fig:Snapshots-Energetics} for the time evolutions of the total fluxes). 

Table \ref{tab:Total-particle-flux} gives the values of the total particle flux as the parameters $\kappa$ or $\alpha$ are varied to go from the zonal flow dominated regime to the drift wave turbulence dominated regime, for $\mu=0,10^{-4},10^{-3}$. We did not fill all boxes in the table because the simulations are computationally expensive, and we decided that the data corresponding to the boxes left blank would not give any more insights into the dynamics. The information in the table complements Figure \ref{fig:Total-particle-flux}, from which actual numbers may be hard to extract. We find that:
\begin{itemize}
\item For the relatively large value of the diffusion parameter $\mu$, $\mu=1\times10^{-3}$, the critical values for the onset of the linear instability are$\kappa_{\mathrm{cr}}=0.058$
and $\alpha_{\mathrm{cr}}=1.475\times10^{2}$, while the strong turbulent
transport is triggered at the much later stage at around $\kappa_{\mathrm{turb}}\sim1$
and $\alpha_{\mathrm{turb}}\sim0.5$. 
\item For the smaller value of the collisional diffusion parameter, $\mu=1\times10^{-4}$, the critical values for the onset of the linear instability are $\kappa_{\mathrm{cr}}=0.0185$ and
$\alpha_{\mathrm{cr}}=1.47\times10^{3}$, while strong turbulent
transport is found for $\kappa_{\mathrm{turb}}\sim0.4$ and $\alpha_{\mathrm{turb}}\sim1.5$. 
\item For the diffusionless case $\mu=0$, the system is linearly unstable for all values of $\alpha\neq0,\kappa\neq 0$, but a transition is still observed for the measured level of particle transport, although the transition can be less sharp than for the $\mu\neq 0$ cases depending on the parameter values and on the choice of parameter which is varied.
\end{itemize}
We conclude in this section that the sudden transition in the level of transport characterizing the Dimits shift is successfully captured in the BHW model for a channel domain geometry. As we will show in Section \ref{sec:Dimits-shift-Periodic}, this is also the case for the more commonly considered doubly periodic geometry. This is a key observation of this article, since it is a unique feature of the BHW model among the HW models. Indeed, we will also show in Section
\ref{sec:Dimits-shift-Periodic}, that the MHW model does not produce similarly sharp transitions as one decreases $\alpha$ or increases $\kappa$. This demonstrates that in reduced fluid models, the Dimits shift sensitively depends on the treatment of the electron dynamics parallel to the magnetic field lines, an observation which was also recently made with the Terry-Horton model \cite{stonge2017}. 

\begin{table}
\subfloat{{\footnotesize{}}%
\begin{tabular}{cccccccc}
\toprule 
{\footnotesize{}$\alpha=0.5$, $\kappa$} & {\footnotesize{}$0.1$} & {\footnotesize{}$0.3$} & {\footnotesize{}$0.5$} & {\footnotesize{}$0.8$} & {\footnotesize{}$0.9$} & {\footnotesize{}$1$} & {\footnotesize{}$2$}\tabularnewline
\midrule
\midrule 
{\footnotesize{}$\mu=0$} & {\footnotesize{}0.0031} & {\footnotesize{}0.0155} & {\footnotesize{}1.3688} & {\footnotesize{}5.1319} & {\footnotesize{}-} & {\footnotesize{}6.4705} & {\footnotesize{}-}\tabularnewline
\midrule 
{\footnotesize{}$\mu=1\times10^{-4}$} & {\footnotesize{}0.0066} & {\footnotesize{}0.0316} & {\footnotesize{}1.5238} & {\footnotesize{}3.9184} & {\footnotesize{}-} & {\footnotesize{}6.8305} & {\footnotesize{}53.8872}\tabularnewline
\midrule 
{\footnotesize{}$\mu=1\times10^{-3}$} & {\footnotesize{}0.0064} & {\footnotesize{}-} & {\footnotesize{}0.0448} & {\footnotesize{}0.1143} & {\footnotesize{}0.2320} & {\footnotesize{}3.6692} & {\footnotesize{}54.4245}\tabularnewline
\bottomrule
\end{tabular}}

\subfloat{{\footnotesize{}}%
\begin{tabular}{cccccccc}
\toprule 
{\footnotesize{}$\kappa=1$, $\alpha$} & {\footnotesize{}$0.05$} & {\footnotesize{}$0.1$} & {\footnotesize{}$0.6$} & {\footnotesize{}$1$} & {\footnotesize{}$2$} & {\footnotesize{}$3$} & {\footnotesize{}5}\tabularnewline
\midrule
\midrule 
{\footnotesize{}$\mu=0$} & {\footnotesize{}-} & {\footnotesize{}11.2053} & {\footnotesize{}5.7420} & {\footnotesize{}4.0262} & {\footnotesize{}1.5184} & {\footnotesize{}0.9160} & {\footnotesize{}0.2707}\tabularnewline
\midrule 
{\footnotesize{}$\mu=1\times10^{-4}$} & {\footnotesize{}10.0764} & {\footnotesize{}10.7077} & {\footnotesize{}-} & {\footnotesize{}2.1394} & {\footnotesize{}0.0093} & {\footnotesize{}0.0081} & {\footnotesize{}-}\tabularnewline
\midrule 
{\footnotesize{}$\mu=1\times10^{-3}$} & {\footnotesize{}17.1608} & {\footnotesize{}16.4841} & {\footnotesize{}0.1002} & {\footnotesize{}0.0459} & {\footnotesize{}0.0484} & {\footnotesize{}-} & {\footnotesize{}-}\tabularnewline
\bottomrule
\end{tabular}}

\caption{Total particle flux $\Gamma=\int_{\mathcal{D}}\tilde{u}\tilde{n}dxdy$ with changing
values of $\alpha$ and $\kappa$. Cases corresponding to different values of the collisional diffusion parameter $\mu$, $\mu=1\times10^{-3}$,
$\mu=1\times10^{-4}$, as well as the diffusionless case $\mu=0$ are compared.\label{tab:Total-particle-flux}}
\end{table}

\subsection{Detailed analysis of the transport transition}

\begin{figure}
\subfloat{\includegraphics[scale=0.35]{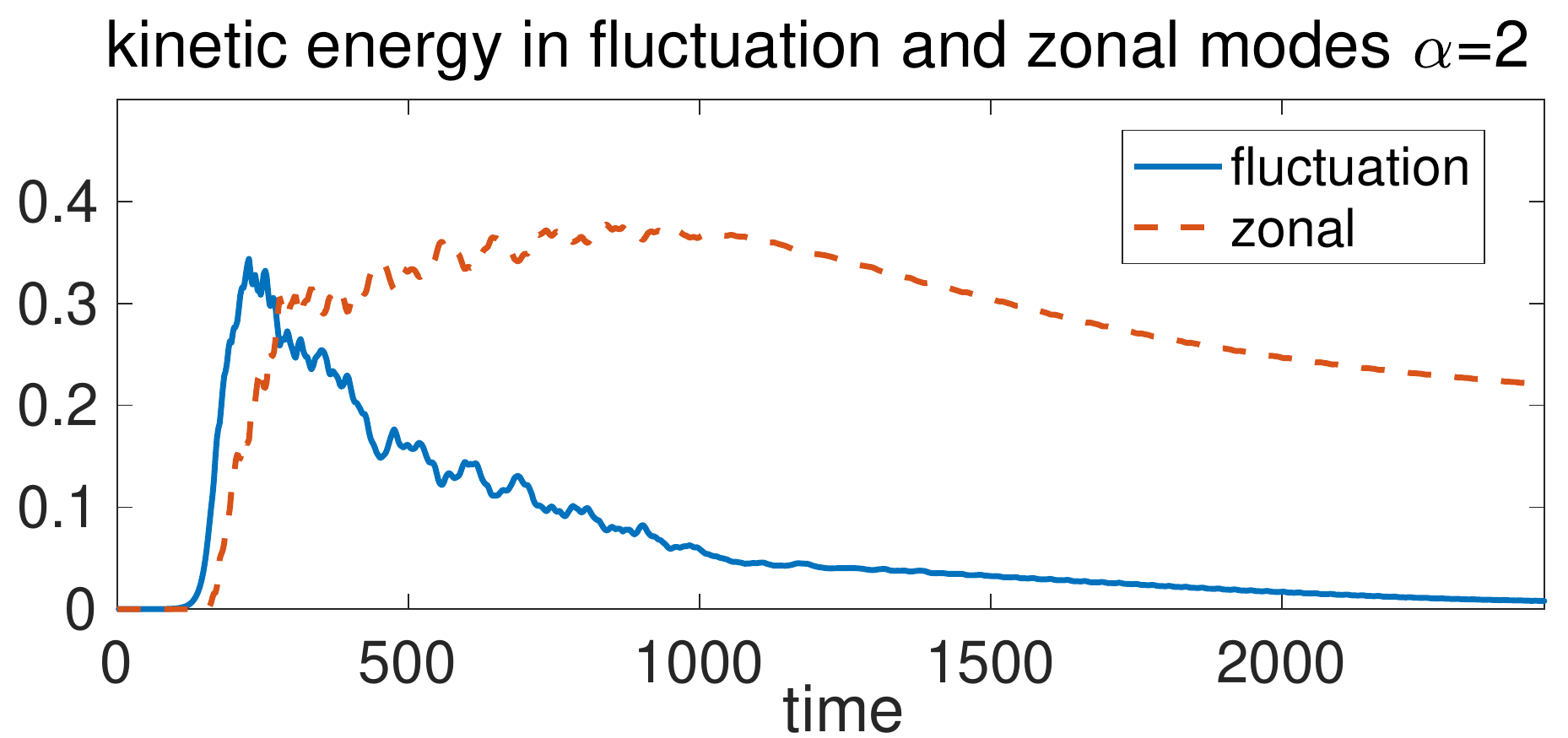}\includegraphics[scale=0.35]{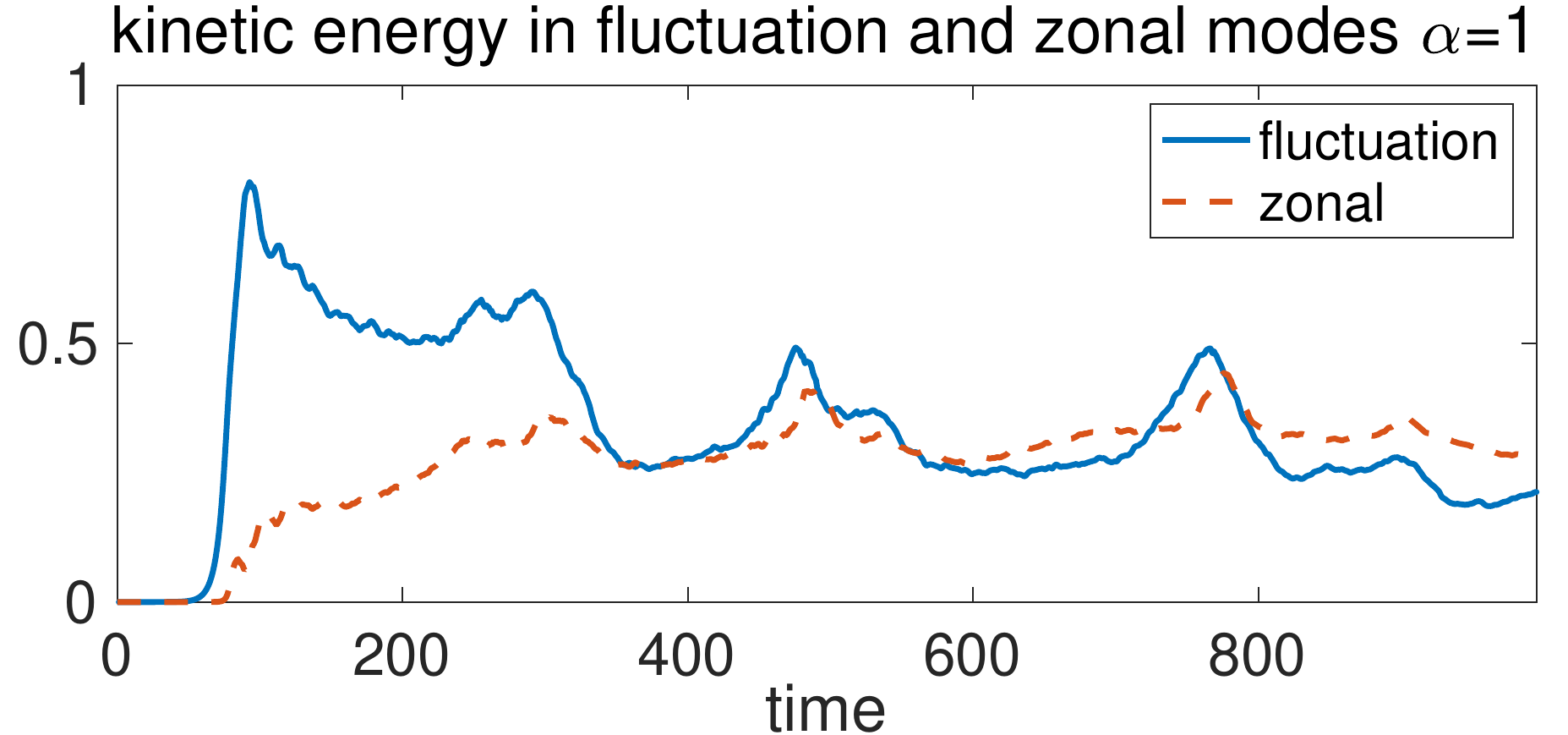}}

\vspace{-1.em}

\subfloat{\includegraphics[scale=0.35]{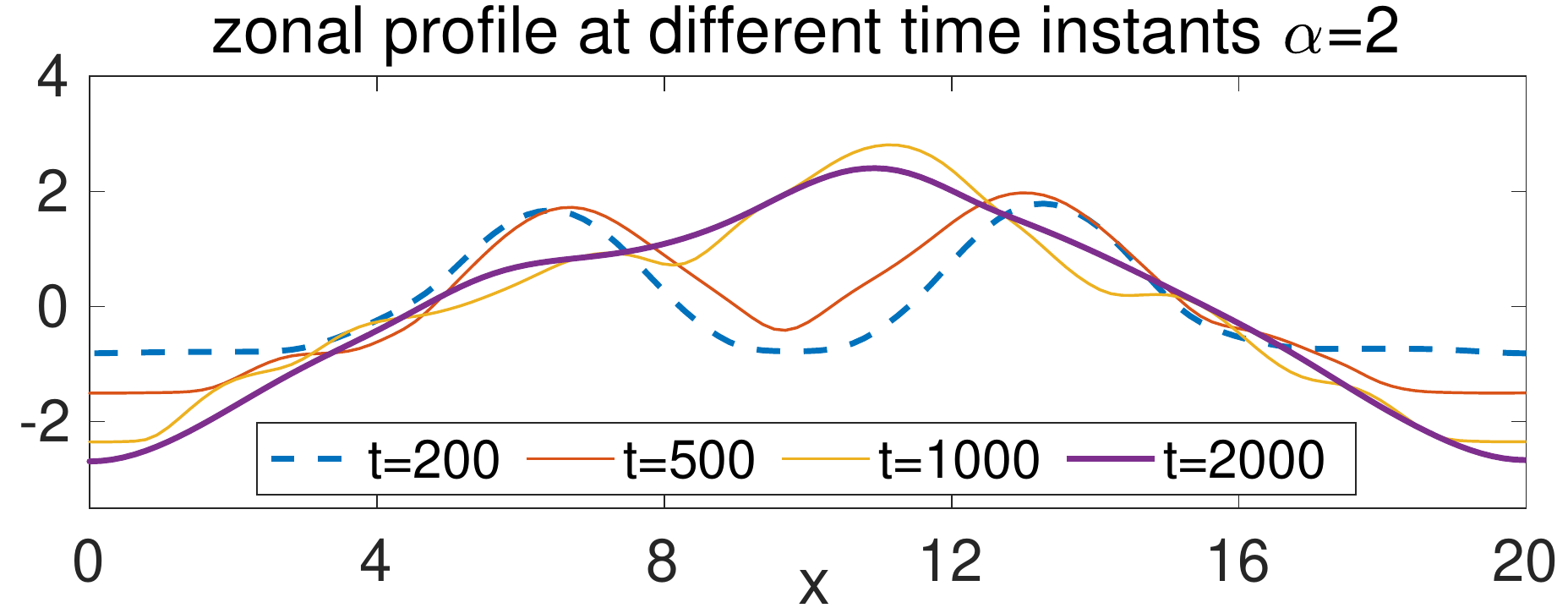}\includegraphics[scale=0.35]{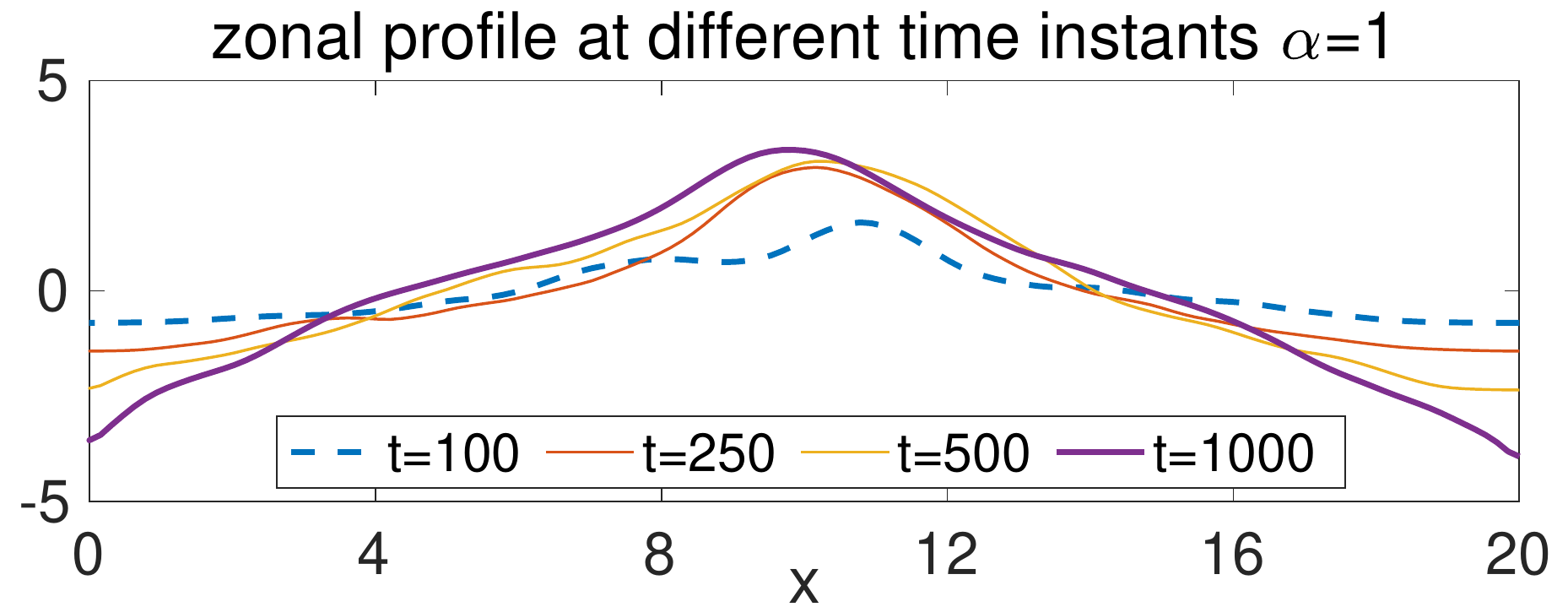}}

\caption{Development of the zonal state from initial fluctuations with small amplitude for the case $\mu=10^{-4}$. Two typical
parameter regimes with dominant zonal flow (left, $\kappa=1,\alpha=2$)
and turbulent flow (right, $\kappa=1,\alpha=1$) are compared. The
upper row compares the time series for the energy in the non-zonal modes $k_{y}\neq0$ and in the zonal modes with $k_{y}=0$. The lower row
shows the development of the zonal mean flow profiles at several times as the zonal flows are generated from the fluctuations.\label{fig:Development-of-zonal}}
\end{figure}

We now take a more detailed look at the dynamical structures in the flow before and after the transition from a low transport regime to a high transport regime. Specifically, we compare the flow solutions and statistics for the different regimes. Figure
\ref{fig:Development-of-zonal} first plots, for $\mu=10^{-4}$, the time-series for the
energy in the non-zonal fluctuation modes with $k_{y}\neq0$ and the total
zonal energy in the modes with $k_{y}=0$ in the zonal flow dominated regime obtained for $\kappa=1,\alpha=2$, which is just before the transition, and in the drift wave turbulence dominated
regime, obtained for $\kappa=1,\alpha=1$, which is shortly after the transition. In both
cases, starting from an initial state with small amplitude fluctuations and with little energy, the energy in the fluctuations first increases dramatically due to the drift instability of non-zonal modes. In the zonal flow regime with $\alpha=2$, the energy in non-zonal modes is then gradually transferred to the zonal
state in the next stage through the secondary energy transfer.
The total energy in the zonal modes reaches approximately the same level
as the maximum excited fluctuation energy in the starting transient
state before the dissipation effect takes over, implying full conversion
of the fluctuation energy to the zonal state. In contrast, in the
drift wave turbulence dominated regime, observed for $\alpha=1$, the non-zonal fluctuations are never completely quenched by the formation of zonal jets. The non-zonal
energy is still transferred to drive the zonal modes through the nonlinear coupling, but the instability intermittently injects energy as bursts into the fluctuation modes. In this way, the turbulent fluctuations are maintained and cannot be completely converted to the zonal state. We note that the zonal energy always increases shortly after the burst in the fluctuation energy, with a short time lag, from which we can infer the direction of the nonlinear energy transfer from non-zonal to zonal modes.

The emergence of the zonal state is also illustrated in the second
row of Figure \ref{fig:Development-of-zonal}. We plot the development
of the zonal velocity profiles $\overline{v}=\partial_{x}\overline{\varphi}$ measured at several differents. For the simulations with $\alpha=2$, corresponding to zonally dominated dynamics, two jets first appear, before the zonal state gradually evolves to a one jet structure. In contrast, a dominant single jet is directly formed in the simulations with $\alpha=1$, which correspond to the turbulent regime, with the development of a stronger positively directed flow in the center of the domain and a stronger negatively directed flow
near the two edges of the computational domain. This observation
isconsistent with the selective decay results in \cite{qi2018selective}
showing a more stable solution with smaller wavenumber, leading to a single dominant large-scale jet.

\begin{figure}
\begin{centering}
\subfloat{\includegraphics[scale=0.32]{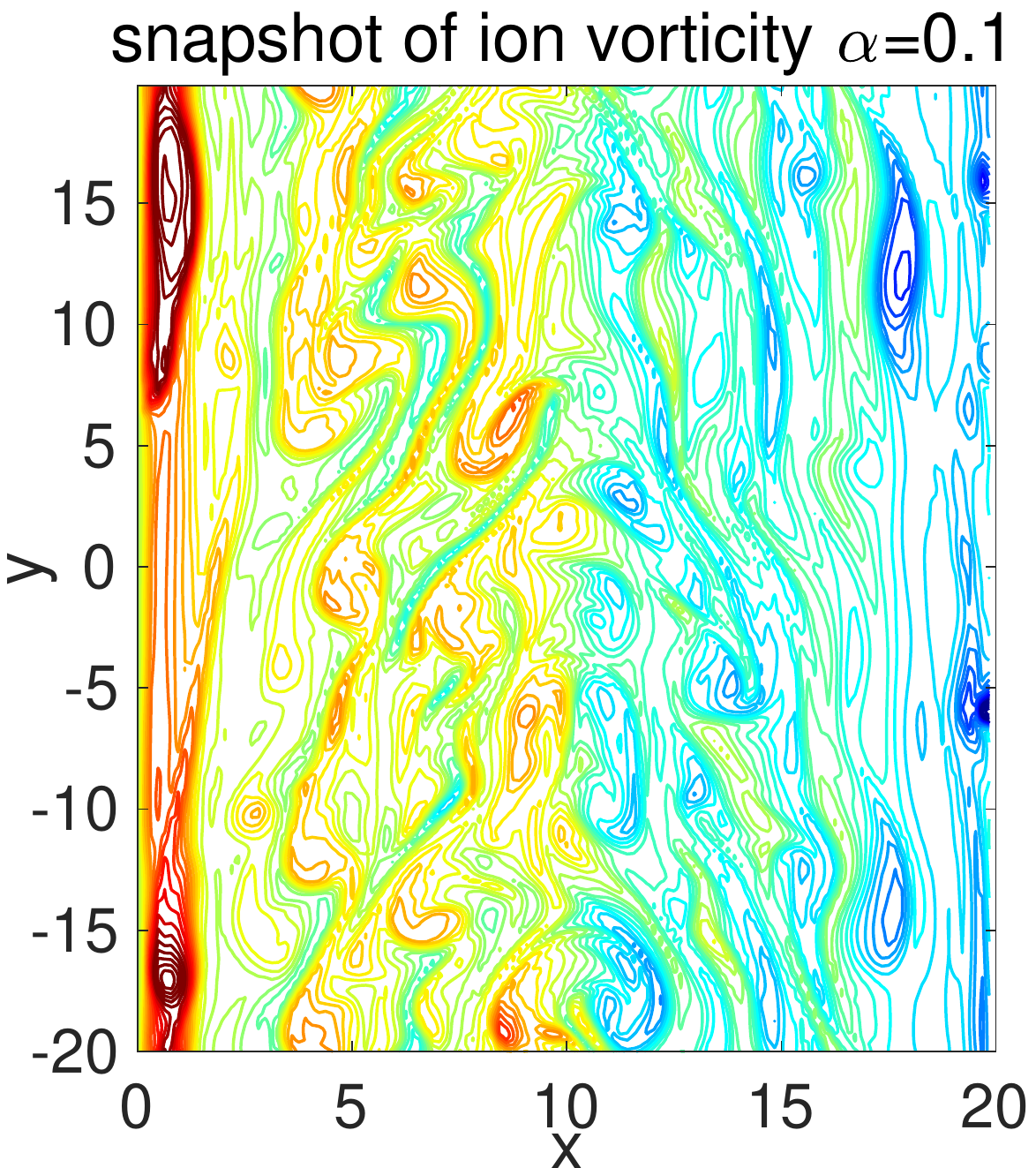}\,\includegraphics[scale=0.32]{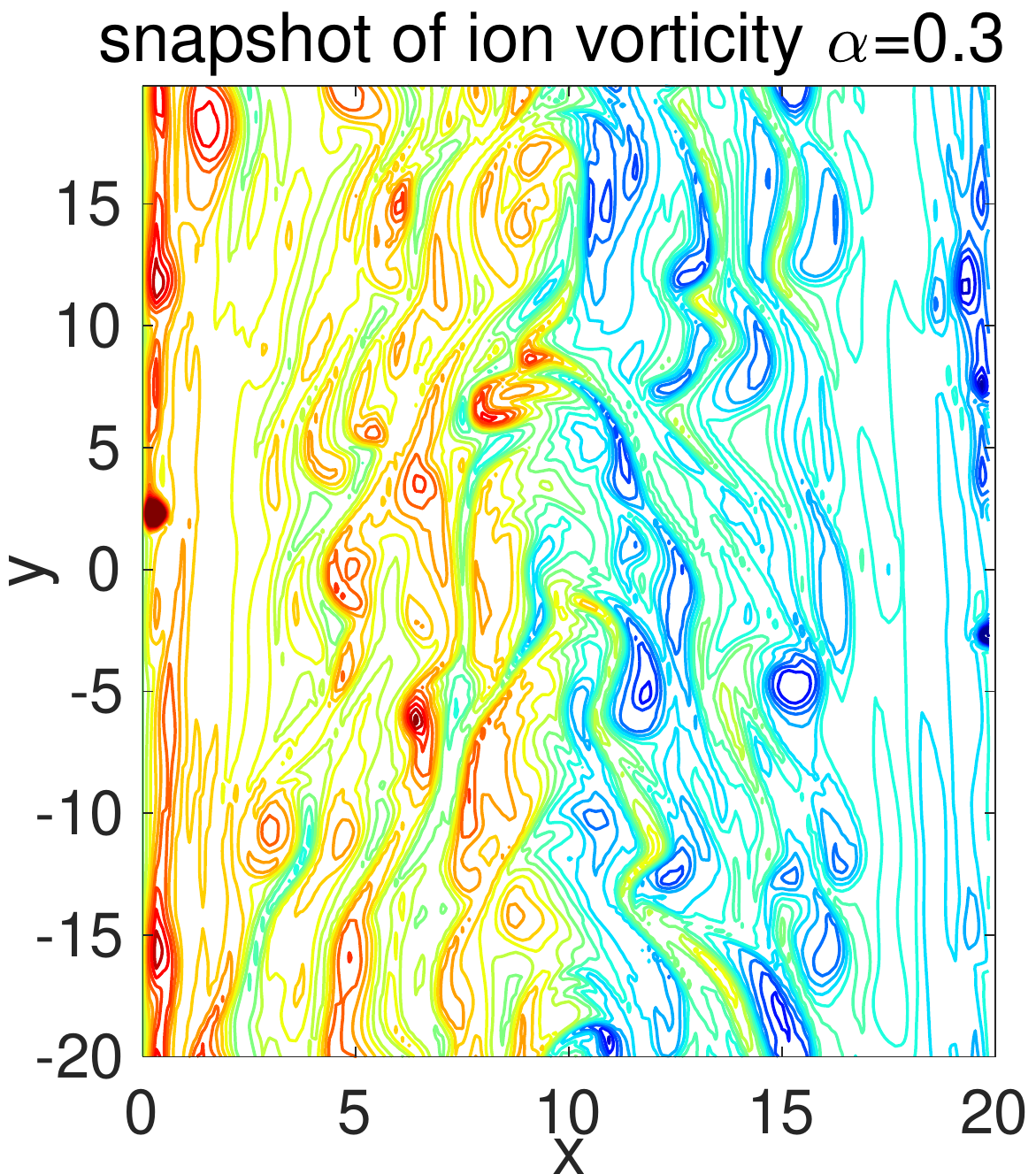}\,\includegraphics[scale=0.32]{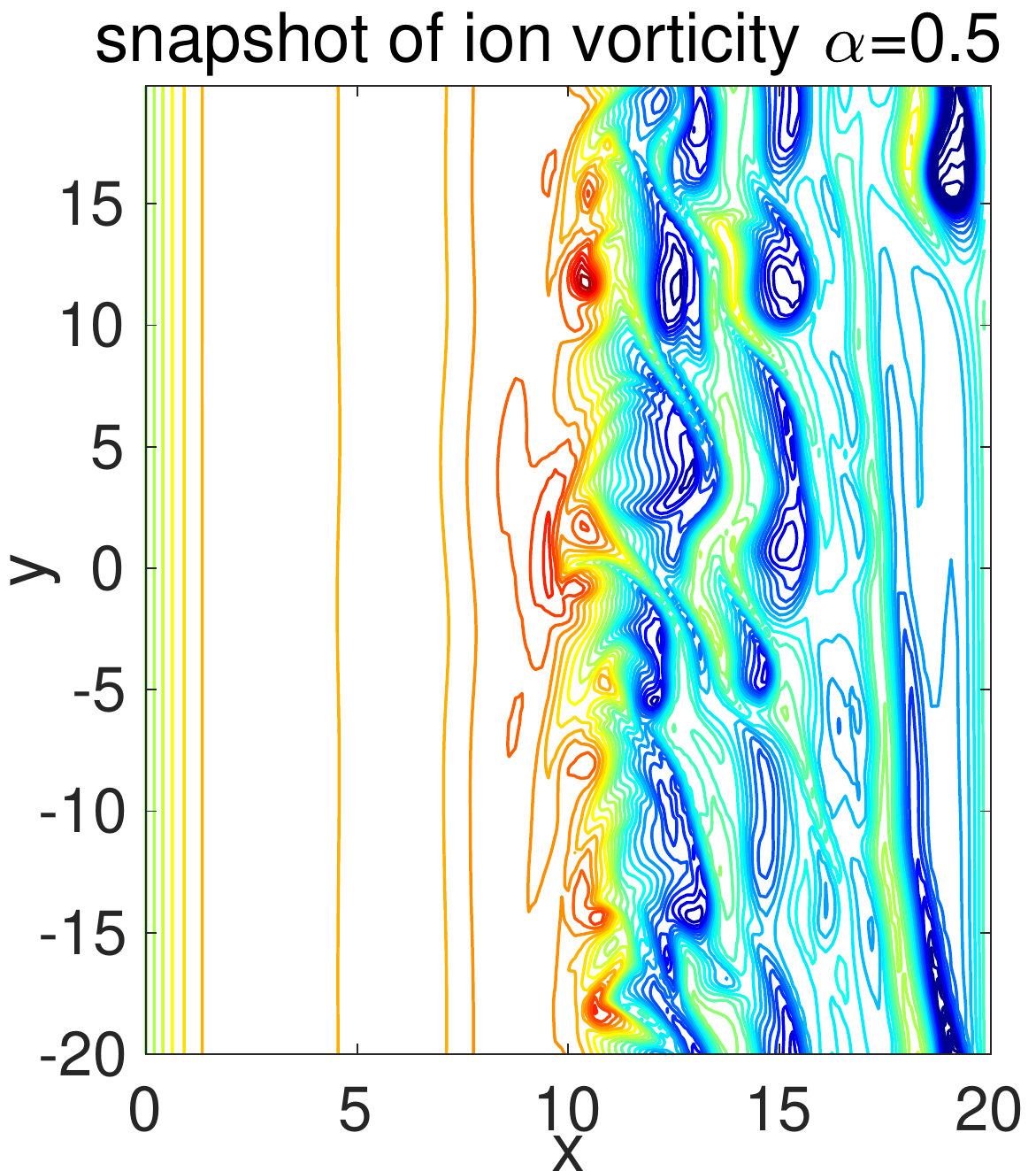}\,\includegraphics[scale=0.32]{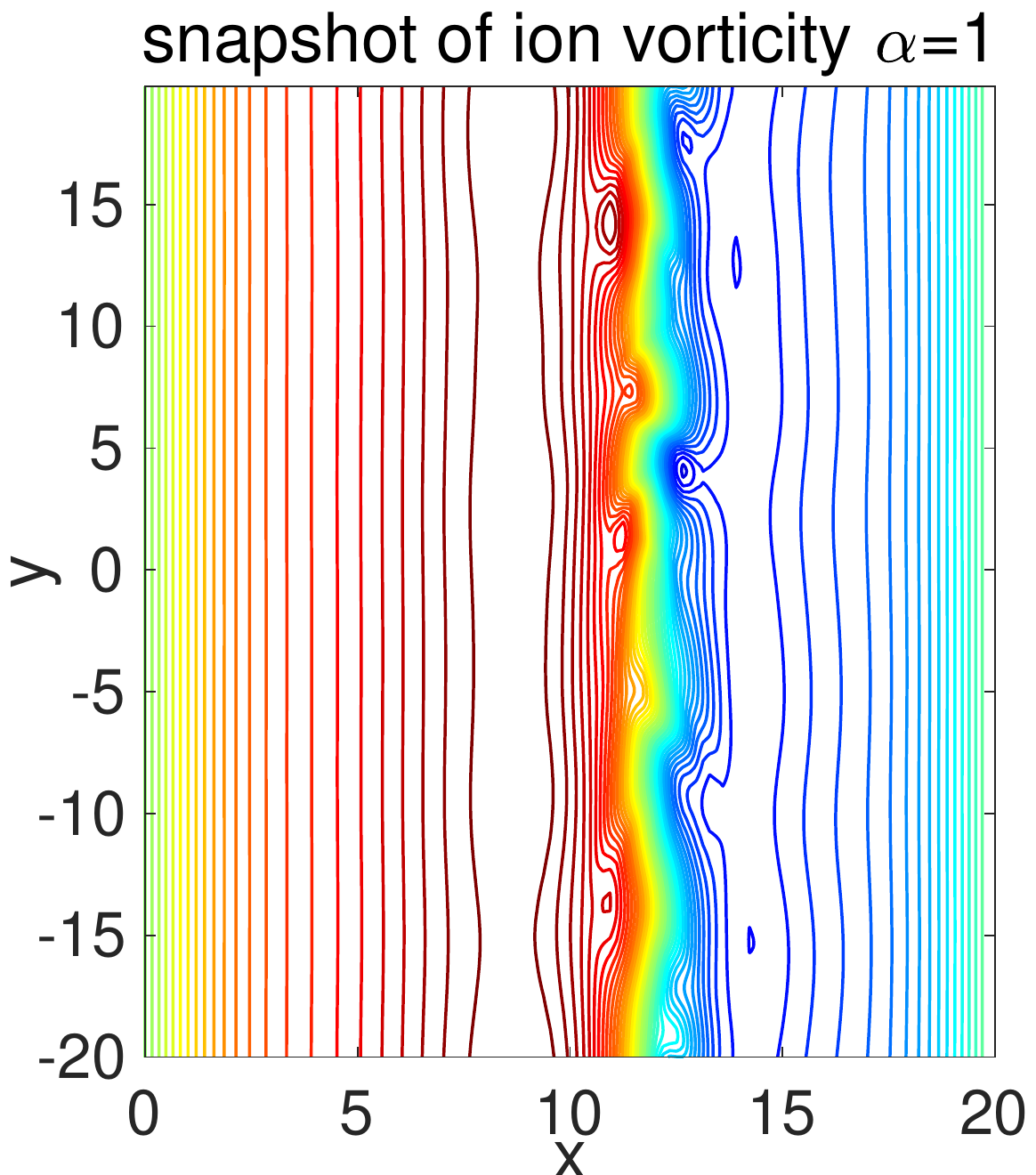}}
\par\end{centering}
\vspace{-1.em}
\begin{centering}
\subfloat{\includegraphics[scale=0.30]{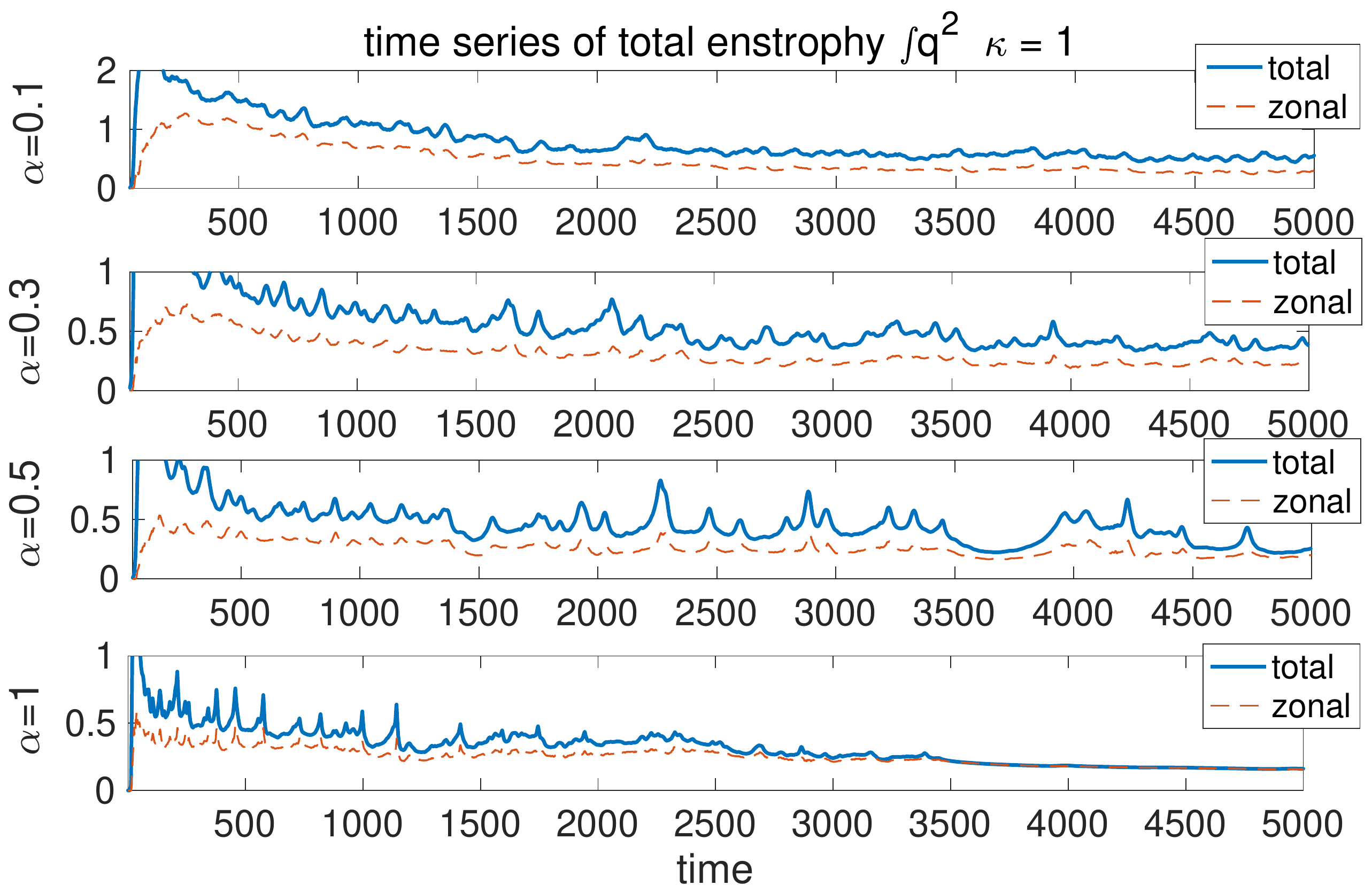}\includegraphics[scale=0.30]{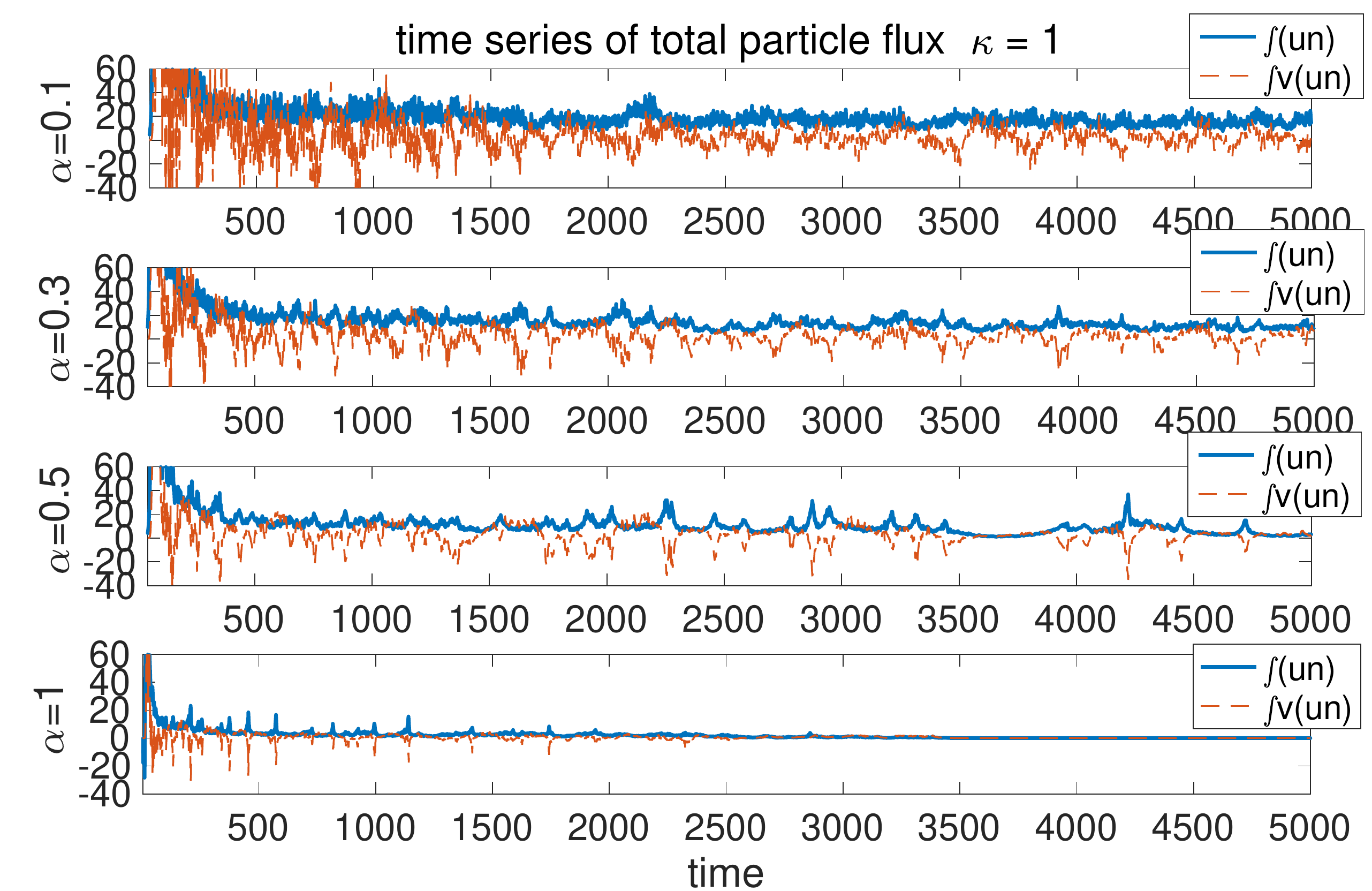}}
\par\end{centering}
\caption{Snapshots of the ion vorticity $\zeta=\nabla^{2}\varphi$ (upper row)
and time series of the 
total enstrophy $W=\int_\mathcal{D} q^{2}dxdy$, as well as the total fluxes $\Gamma=\int_{\mathcal{D}}\tilde{u}\tilde{n}dxdy$
and $\int_{\mathcal{D}}\overline{v}\overline{\tilde{u}\tilde{n}}dxdy$ (lower row) for the
channel domain geometry, for different values of $\alpha$ and fixed $\kappa=1,\mu=1\times10^{-3}$.
More frequent turbulent intermittent bursts are observed as the flow
enters the more turbulent regimes.\label{fig:Snapshots-Energetics}}
\end{figure}

Next, we illustrate the representative flow field snapshots throughout the flow transition, and the corresponding transition for the time series of the enstrophy and the fluxes by comparing the direct simulation results for several different parameter regimes, corresponding to the full range from pure zonal jets to fully turbulent flows. We choose $\kappa=1$ and $\mu=10^{-3}$, which give typical examples of the flow transition, and vary $\alpha$ in order to obtain the desired transition. Snapshots of the ion vorticity and time series of the energetics are shown in Figure \ref{fig:Snapshots-Energetics}. We first confirm with the snapshots that, as the value of $\alpha$ increases, the turbulent flow gradually transitions to regular zonal jets. We can then look at the corresponding effects on the time series for the total enstrophy $W$, the total particle flux
$\Gamma$, and the advected particle flux $\int_{\mathcal{D}}\overline{v}\left(\overline{\tilde{u}\tilde{n}}\right)dxdy$. We recall here that the flux terms act as energy source and sink in the equations for the enstrophy (\ref{eq:eqn_ens}) and for the energy (\ref{eq:eqn_ene}). In the turbulent regime, the enstrophy is excited by the bursty injection from the particle flux. As the value of $\alpha$ decreases, the frequency of intermittent bursts increases, and the solution is kept in a state with high enstrophy for both zonal modes and non-zonal modes. We will return to these features when we discuss avalanche-like bursts in more detail in Section \ref{sec:Avalanche-like-structures}. For large values of $\alpha$, the bursts in the fluxes are much less frequent and have much smaller amplitude. The total enstrophy converges to a constant value with most of the enstrophy in the zonal modes: this is the zonal flow dominated regime.

\begin{figure}
\subfloat{\includegraphics[scale=0.35]{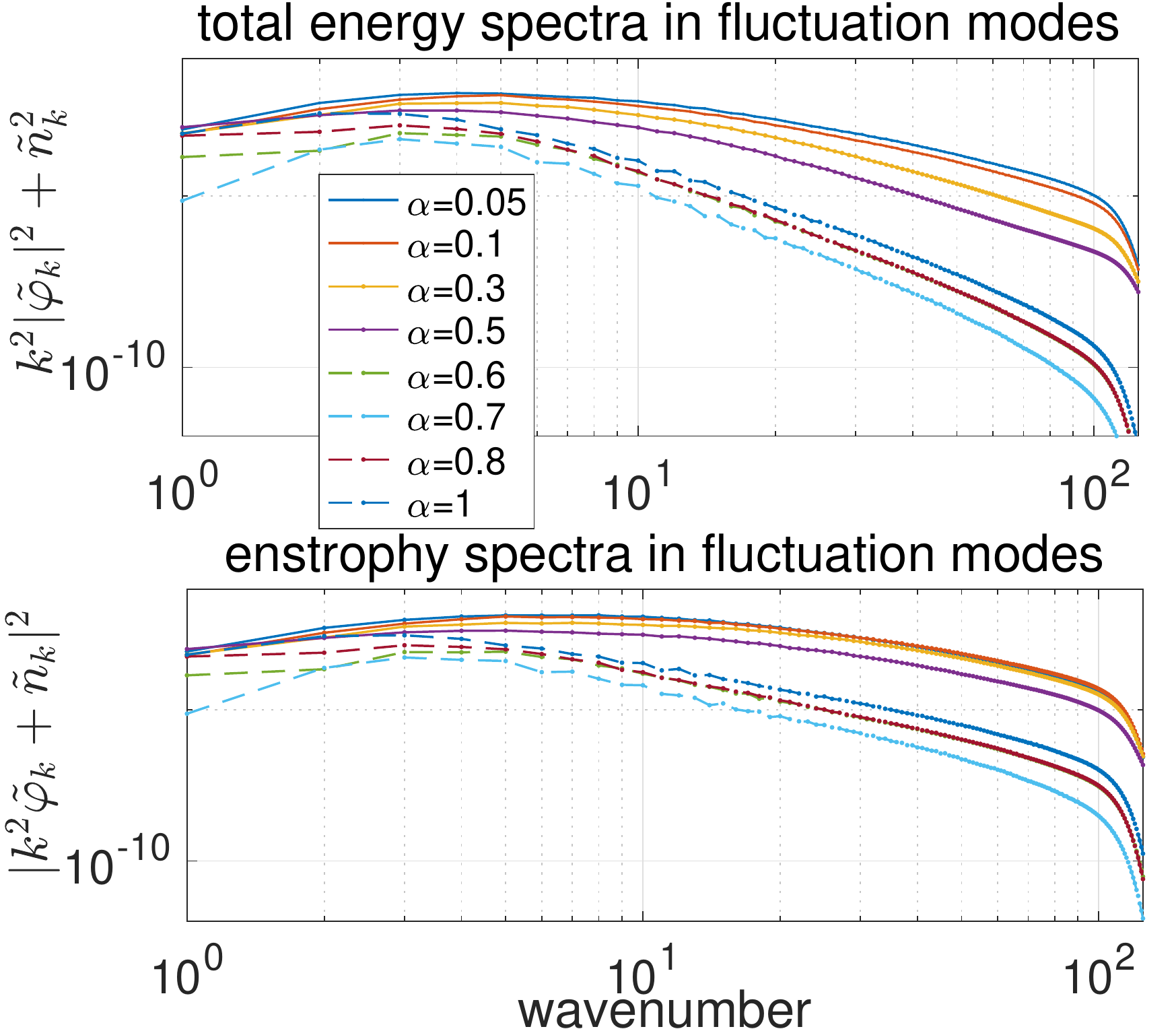}\includegraphics[scale=0.35]{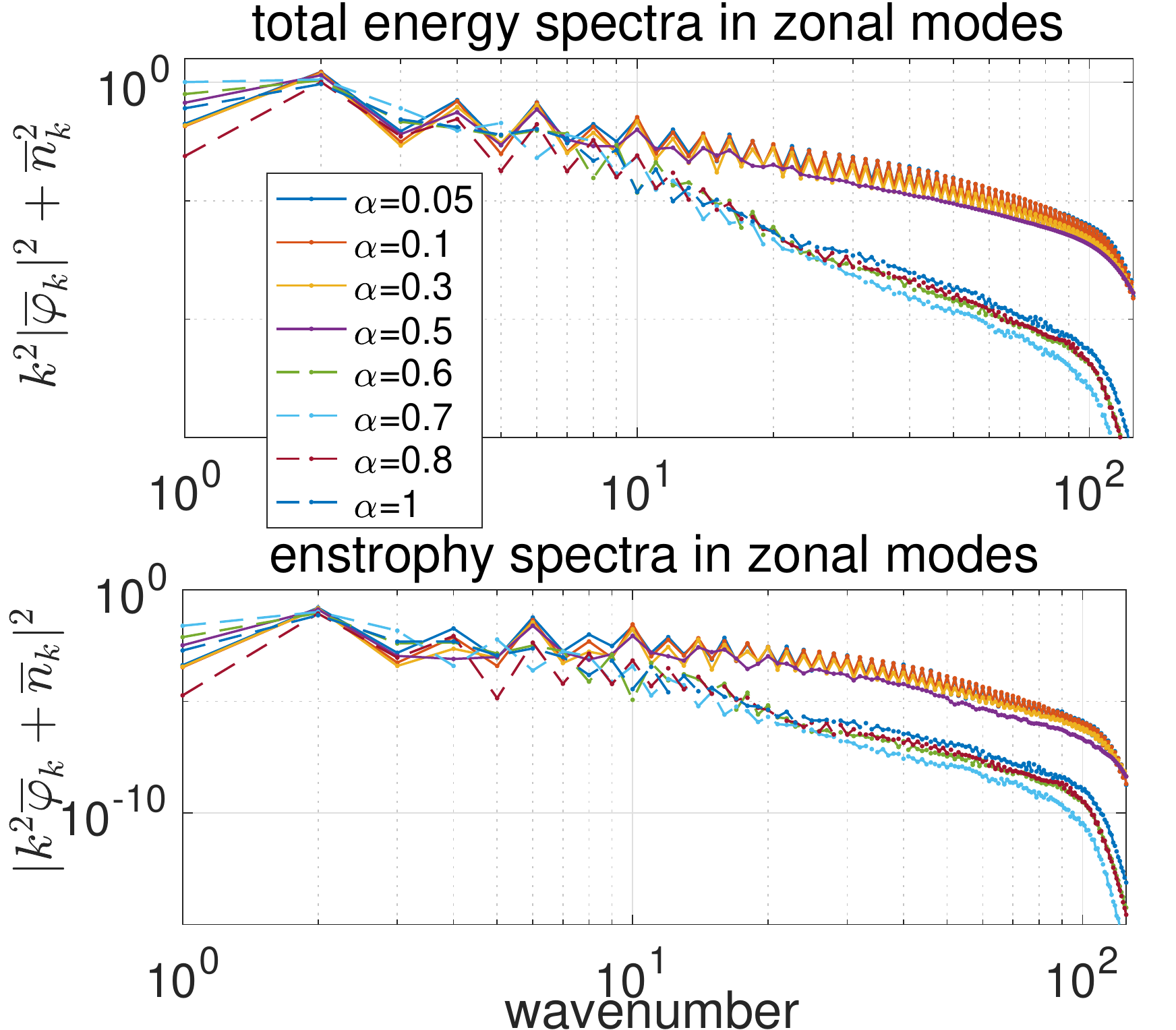}}

\vspace{-1em}

\subfloat{\includegraphics[scale=0.35]{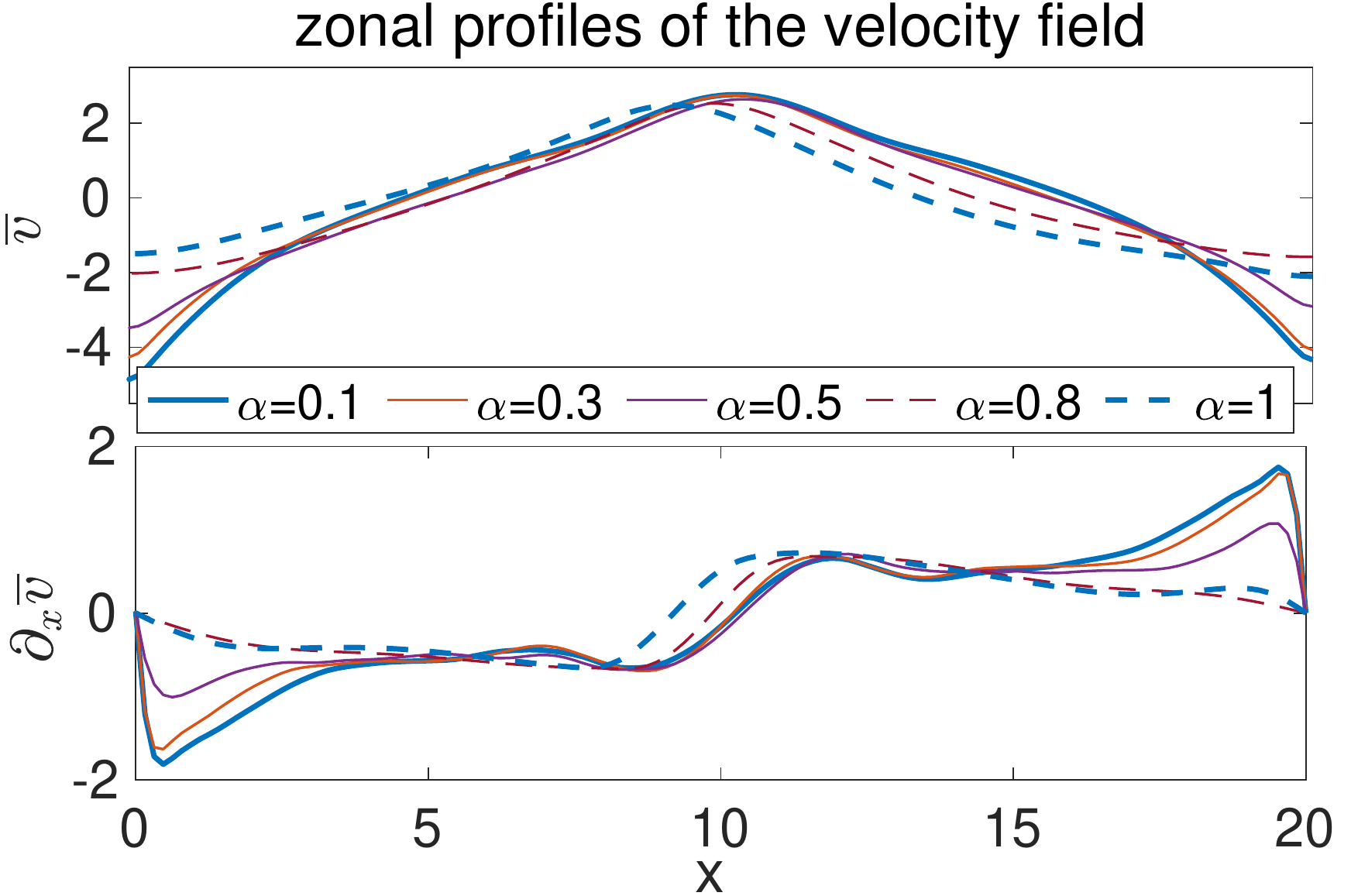}\includegraphics[scale=0.35]{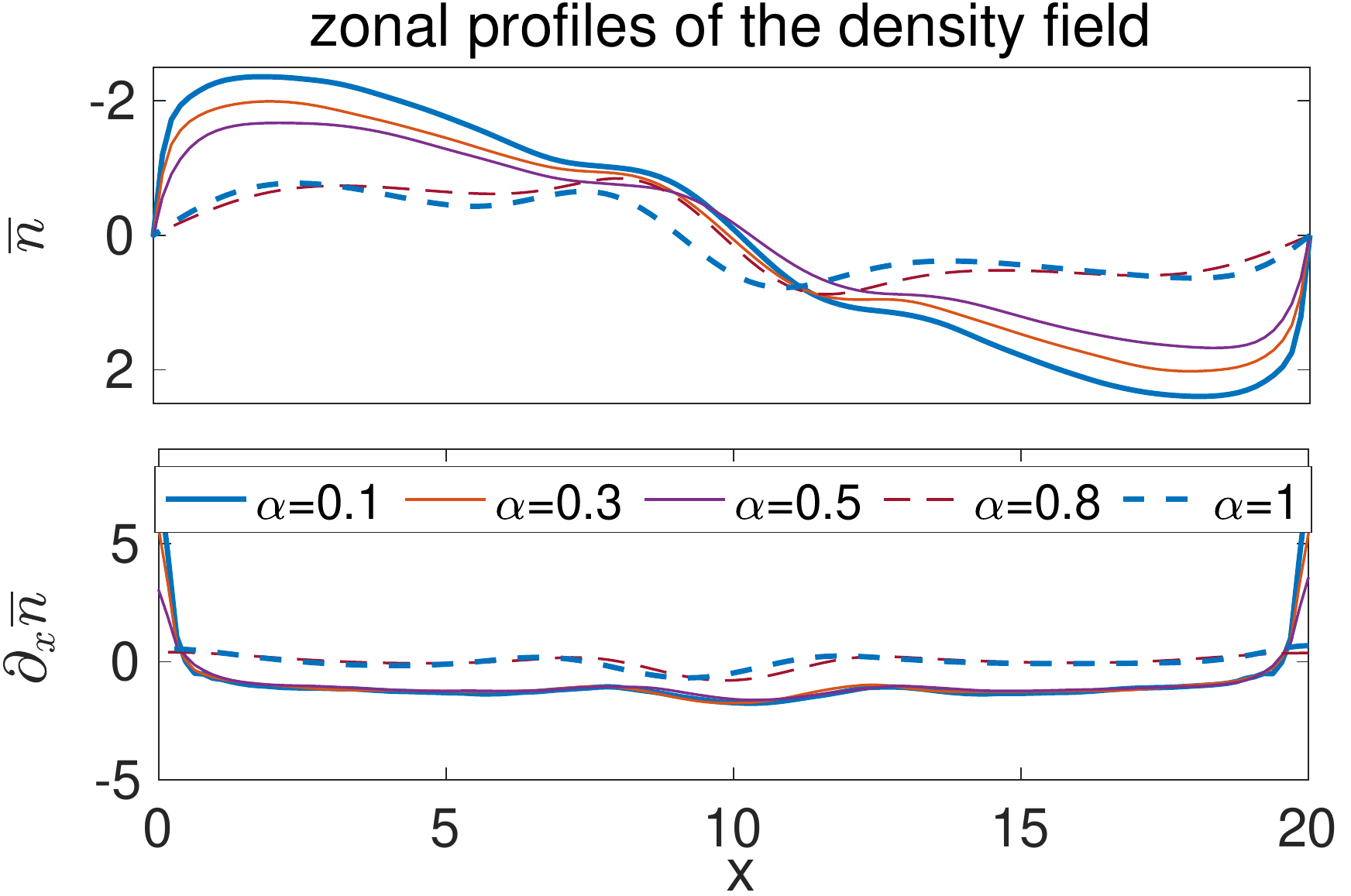}}

\caption{Spectra for the non-zonal fluctuation modes ($k_{y}\protect\neq0$,
left) and the zonal modes ($k_{y}=0$, right) for different
values of $\alpha$, and $\kappa=1$, and $\mu=10^{-3}$. The kinetic energy $k^{2}|\hat{\varphi}_{k}|^{2}+|\hat{n}_{k}|^{2}$,
and the enstrophy $|\hat{q}_{k}|^{2}$ at each scale are compared.
A clear transition can be observed in both the zonal and fluctuation
spectra depending on the parameter values of $\alpha$. The bottom
rows of the figure compare the zonal structures in the velocity and the density
for different values of $\alpha$. The curves corresponding to the turbulence dominated regime are plotted in solid lines, and the curves corresponding to the zonal flow dominanted regime are plotted in dashed lines.\label{fig:Statistical-spectra-chan}}
\end{figure}

Valuable insights can also be obtained by comparing, for values of $\alpha$ corresponding to the different transport regimes, the spectra for the total energy and for the enstrophy, as well as the zonal mean profiles at the end of the simulations, after statistical steady-state has been reached. This is precisely what we display in Figure \ref{fig:Statistical-spectra-chan}, separating non-zonal fluctuation modes $k_{y}\neq0$ and the zonal modes $k_{y}=0$ for the spectra. $\kappa$ is set to 1, $\mu$ is set to $10^{-3}$, and the values of the adiabaticity parameter $\alpha$ are chosen in the interval $\left[0.05,1\right]$ to capture both the turbulent dominated regime and the zonal flow dominated regime. The two regimes can be clearly distinguished from the spectra, with energetic small scales for $\alpha<0.6$, and a sudden drop of both the energy and enstrophy in the small scale modes for $\alpha\geq 0.6$.
Thus, in the zonal flow dominated regime with relatively weak lineary instability, the excited energy and enstrophy are kept in the large-scale zonal modes and the downscale cascade is blocked. When the linear
instability is stronger, turbulent transport is induced by the stronger downscale
cascade to create small-scale fluctuations. This characterizes the statistical transition in the turbulent transport.

The two bottom rows of Figure \ref{fig:Statistical-spectra-chan}
compare the zonal velocity and zonal density profiles measured at the end of the simulation, in statistical steady-state, for $\alpha=0.1,0.3,0.5,0.8,1$. In all cases, a strong flow in the positive $y$ direction and a strong density gradient are seen at the center of the computational domain. The zonal
profiles before and after the transport transition however display different features in the vicinity of the edges of the computational domain. In the turbulent regime with strong transport, stronger zonal flow $\overline{v}$, a larger flow shear and a larger density gradient are observed near the boundaries. These features, which were already found and analyzed in our previous study \cite{qi2019channel} of the BHW model in the channel geometry, are not present for doubly periodic boundary conditions, and are critical to explain the differences between the channel geometry and the doubly periodic geometry we will find for the particle transport in the next sections.

\section{Avalanche-like structures in domains with different aspect ratios\label{sec:Avalanche-like-structures}}

Besides the Dimits shift, other salient aspects of drift wave turbulence driven transport which remain incompletely understood are the importance of non-diffusive, non-local transport mechanisms in the observed level of transport, and whether mesoscale ballistically propagating structures observed for various physical quantities such as the heat flux in ITG simulations \cite{candy2003anomalous,mcmillan2009avalanchelike,difpradalier2010,gorler2011,ivanovArxiv} are the main non-local transport mechanism. An emerging hypothesis, mostly based on the results of numerical simulations, is that these ballistically propagating structures may be sorted into two categories \cite{mcmillan2018}: avalanches \cite{candy2003anomalous,mcmillan2009avalanchelike,difpradalier2010,gorler2011}, whose signatures are large amplitude, rather disordered radially propagating bursts, and more coherent solitary structures \cite{mcmillan2009avalanchelike,vanwyk2016,mcmillan2018,zhou2019solitary,ivanovArxiv}  for which the term ``ferdinons'' was recently coined \cite{ivanovArxiv}. It is not yet clear how avalanches and soliton-like solutions are related, or whether all soliton-like solutions are ``ferdinons'' \cite{ivanovArxiv}. In this article, we will not propose a theoretical explanation for these phenomena. However, we will now show that both disordered avalanches and solitary structures are commonly observed for the particle flux in the BHW model, which is therefore a remarkably simple model for their study. Furthermore, our simulations suggest that avalanches and solitary structures are indeed intimately related, in the sense that the coherent solitary structures frequently emerge from the avalanches in the edge of the computational domain, where the velocity shear and density gradients are large. Finally, by comparing the BHW model and the MHW model, we find that the soliton-like solutions to not emerge from the avalanches in the MHW model, which indicates that electron dynamics parallel to the magnetic field plays a fundamental role in this mechanism.

\subsection{Avalanche-like structures in the zonal particle flux}

\begin{figure}
\subfloat{\includegraphics[scale=0.38]{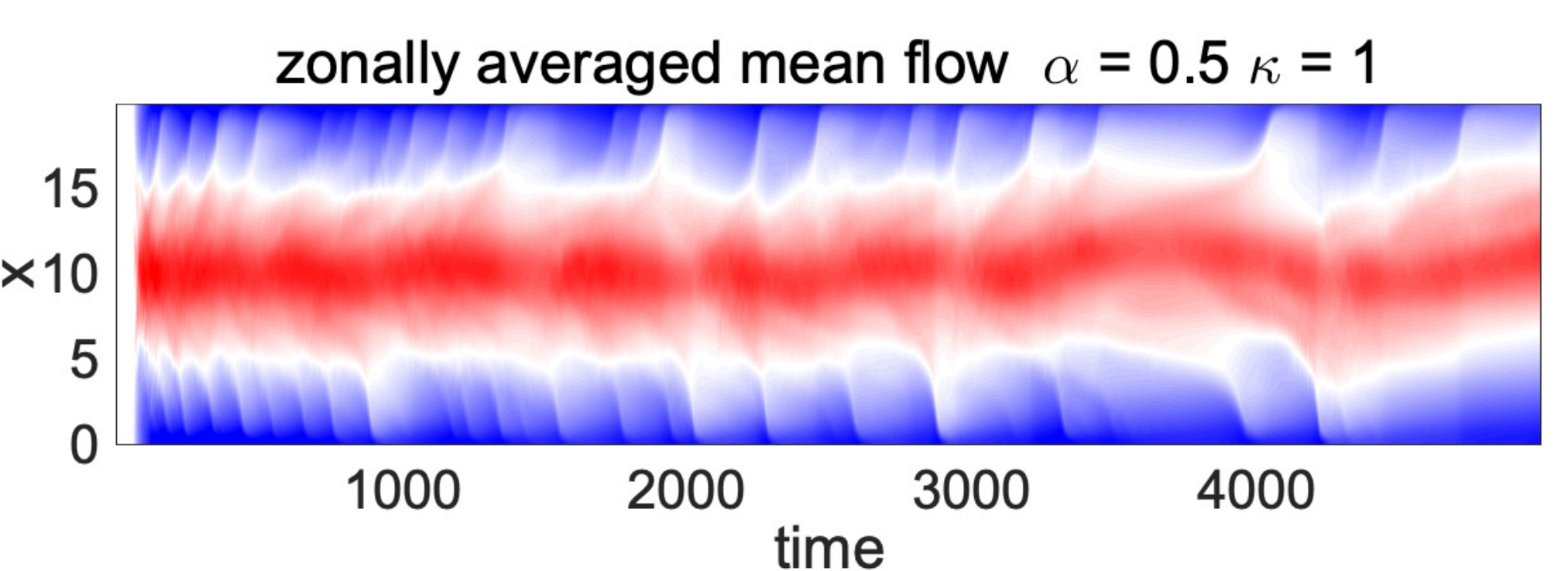}}\subfloat{\includegraphics[scale=0.38]{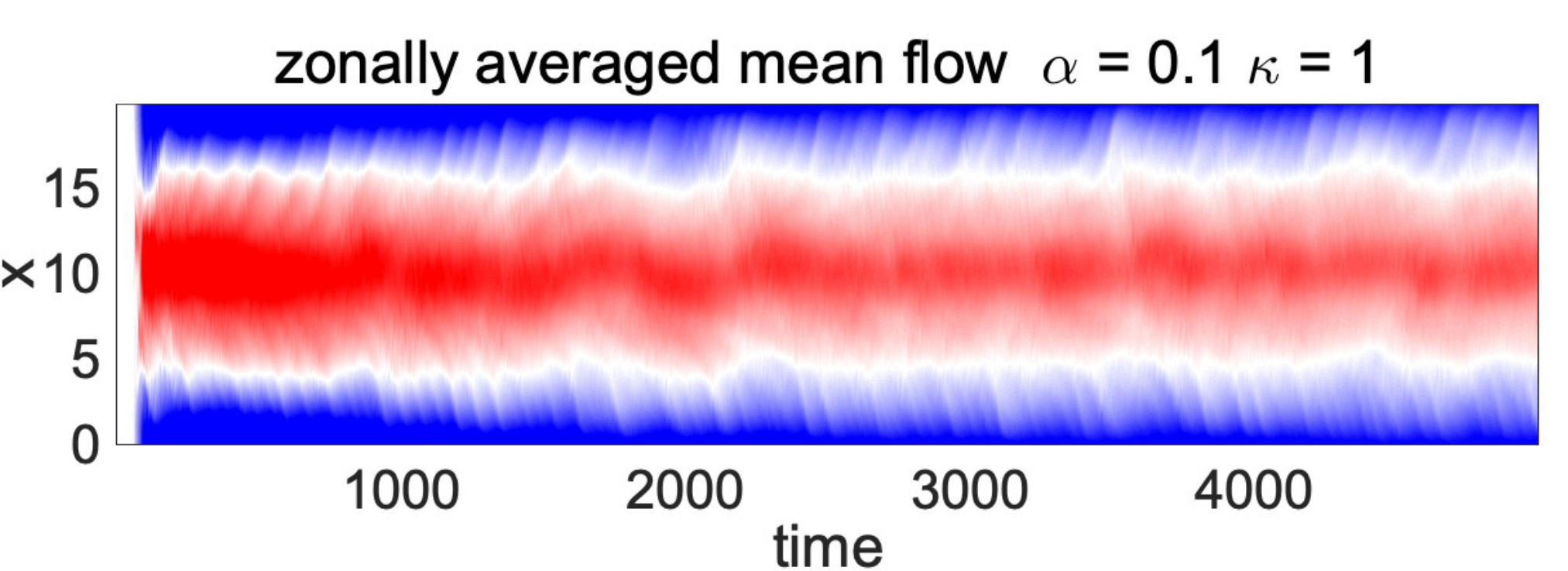}}

\vspace{-1.1em}

\subfloat{\includegraphics[scale=0.38]{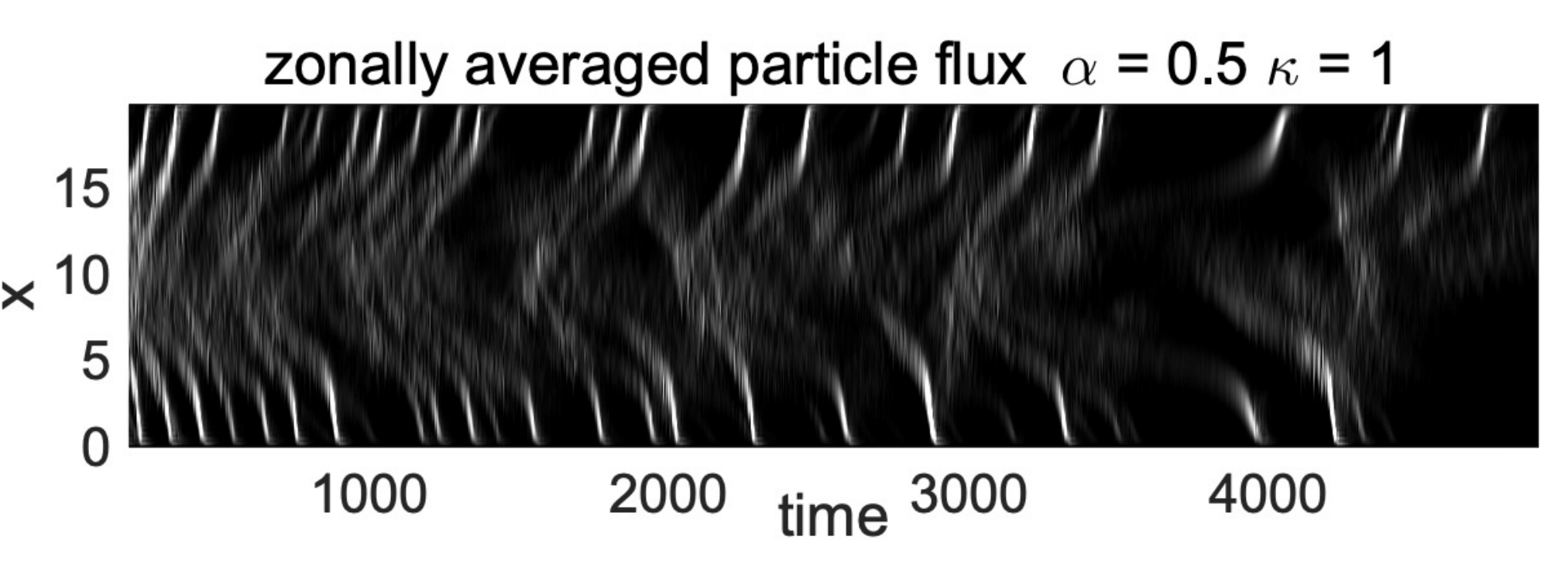}}\subfloat{\includegraphics[scale=0.38]{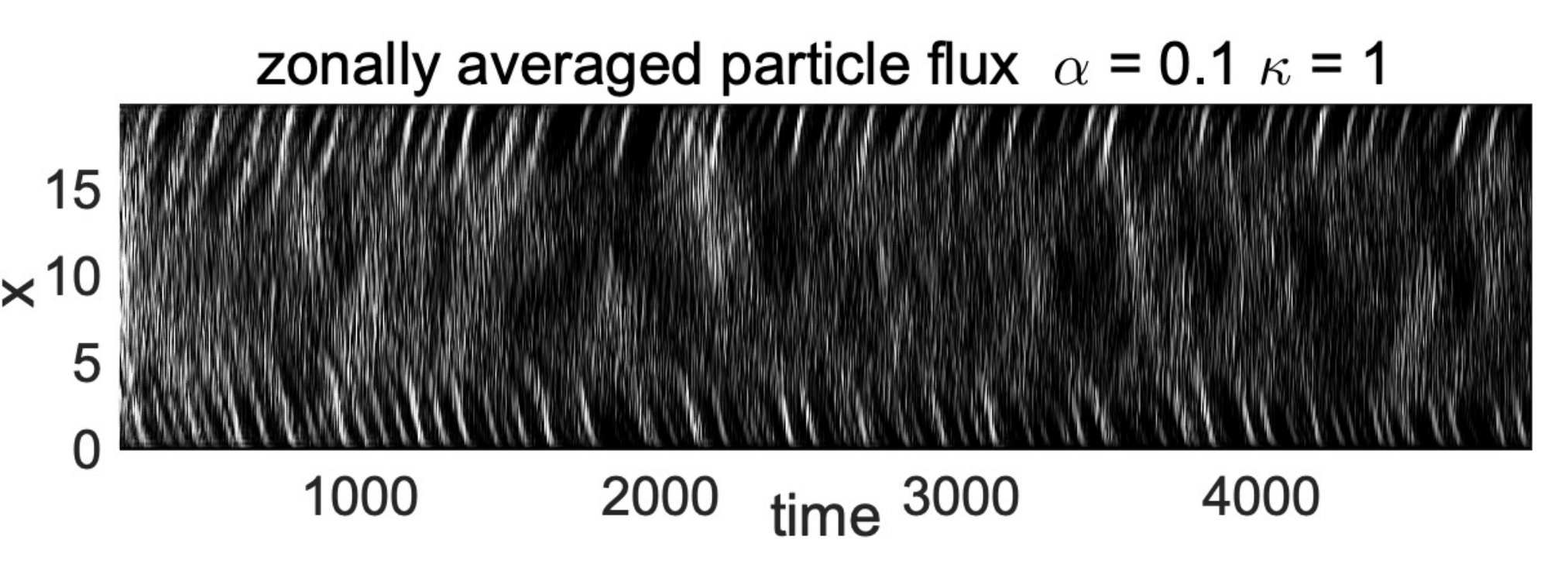}}

\vspace{-1.em}

\subfloat{\includegraphics[scale=0.38]{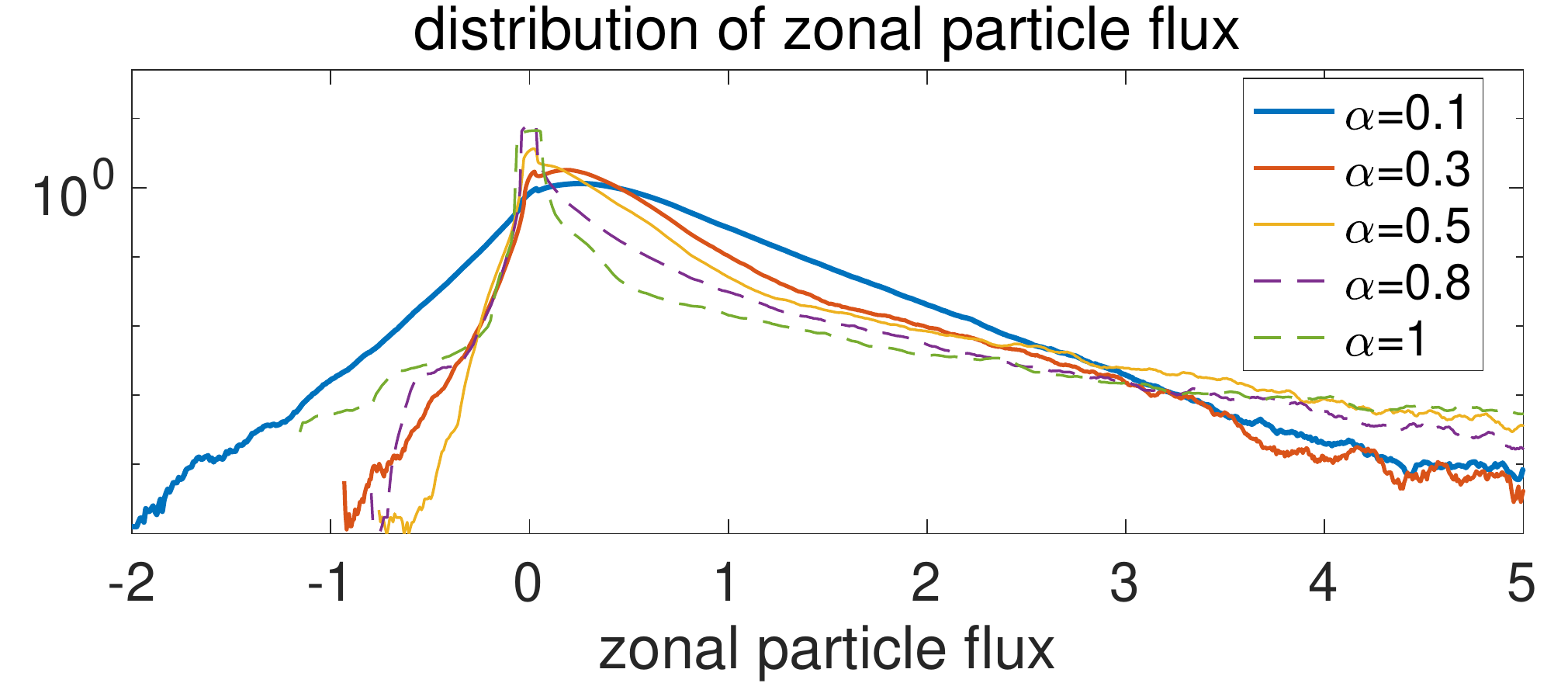}\enskip{}\includegraphics[scale=0.38]{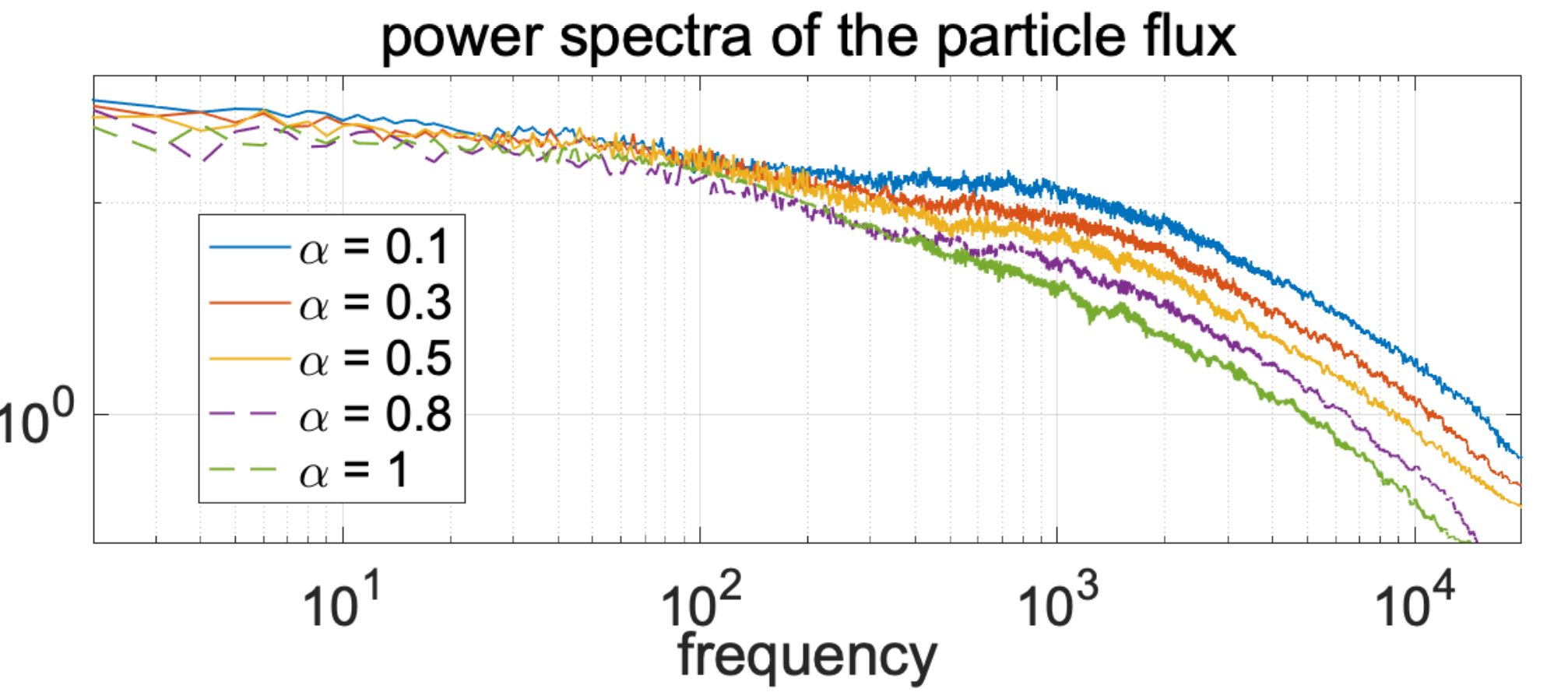}}

\caption{Time evolution of the zonal mean velocity $\overline{v}$ and the
zonal particle flux $\overline{\tilde{u}\tilde{n}}$ from BHW model simulations (Top and middle rows). Two representative
regimes, with relatively sparse bursts of zonal transport ($\alpha=0.5$, left) and more frequent and stronger zonal bursty transport
($\alpha=0.1$, right) are shown. The PDFs of the zonal particle
flux $\overline{\tilde{u}\tilde{n}}$ (left) and power spectra
from the time series of the zonal particle flux (lower right)
for different values of the parameter $\alpha$ are compared in the bottom row. As before, solutions corresponding to
the turbulence dominated regime are plotted with solid lines, while solutions
corresponding to the zonal flow dominated regime are plotted with dashed lines. For all these results, $\kappa=1$ and $\mu =10^{-3}$. \label{fig:Time-evolutions-flux}}
\end{figure}

We focus here on the same channel domain geometry and the same parameter regime as in Section \ref{sec:Dimits-shift}. First, we show the time evolution of the zonally averaged quantities to illustrate the representative avalanche-like bursts emerging in the BHW model in the regime past the Dimits threshold, where significant levels of transport are measured. In the first row of Figure \ref{fig:Time-evolutions-flux}, the time evolution of the zonal flow $\overline{v}$ displays a dominant
large-scale single jet structure frequently shifting in time. The second row of Figure \ref{fig:Time-evolutions-flux} shows the time evolution of the zonal particle flux transport $\overline{\tilde{u}\tilde{n}}$. We observe two major features: 1) a central region, whose extent depends on the value of $\alpha$, in which we see more incoherent avalanche bursts, very similar of the avalanches reported for the heat flux in gyrokinetic simulations \cite{candy2003anomalous,mcmillan2009avalanchelike}; 2) in the vicinity of the edges of the computational domain, coherent radially propagating structures with constant speed. Furthermore, the shifts of the zonal structure and the associated spikes in the zonal flow are strongly correlated with the emergence of the solitary structures from the avalanches. For the smaller $\alpha$ value, $\alpha=0.1$, which corresponds to the more turbulent regime, the heat flux avalanches are stronger in amplitude, have more interacting structures, and the solitary structures are emerging more frequently. For the larger $\alpha$ value, $\alpha=0.5$, the heat flux avalanches are weaker, and solitary structures less frequent. Finally, we note that the solitary structures are most extended for the larger value of $\alpha$, which has smaller velocity and density gradients near the boundary of the domain. 

We next turn to the statistical and spectral signatures of this bursty meso-scale transport. The bottom row of Figure \ref{fig:Time-evolutions-flux} plots the empirical probability density functions (PDF) and the power spectra sampled from time-series of the zonal particle transport $\overline{\tilde{u}\tilde{n}}$ for different values of $\alpha$. The empirical PDFs are constructed by computing the normalized histograms of the time series of the zonal particle flux after time $t=1000$. For most values of $\alpha$, statistical steady-state is established at that point. For the largest values of $\alpha$, some transient events still occur infrequently at the beginning of that time window, which however do not significantly alter the statistics, and are not seen later on. All the PDFs show large positive skewness, implying
a zonal particle transport toward the outer boundary, as expected. Large value
events in the zonal particle flux are much more common than they would be for a normal distribution with the same mean and variance. For the large transport regime, for $\alpha=0.1$, the PDF has a larger variance, and the power spectrum has larger values at high frequencies. As $\alpha$
increases, the PDFs have more of their mass in the vicinity of zero, and the
power spectra decay at a faster rate as the frequency increases, implying weaker zonal transport and fewer occurrences of extreme events. We note that larger values of $\alpha$ have fatter tails, which are due to the existence of more infrequent but larger amplitude, more localized bursts, as seen in the time evolution of the particle flux. Below the Dimits threshold, in the zonal flow dominated regime with near-zero particle flux, these bursts do however not appear any longer, once statistical steady-state is reached.

\begin{figure}
\subfloat{\includegraphics[scale=0.35]{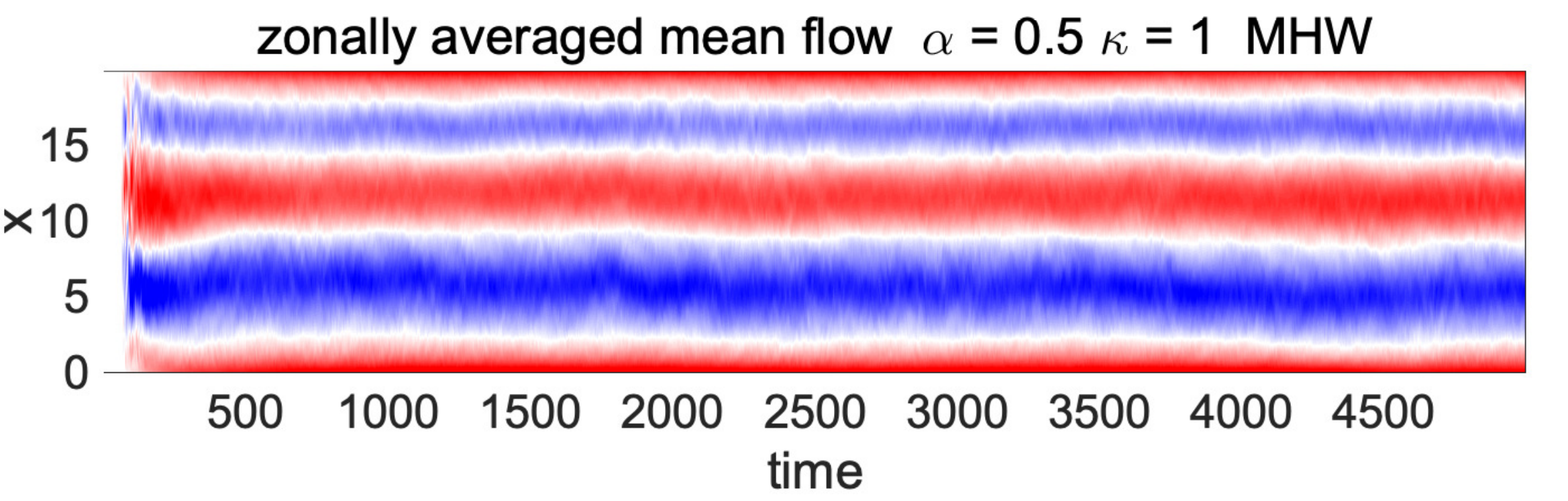}}\subfloat{\includegraphics[scale=0.35]{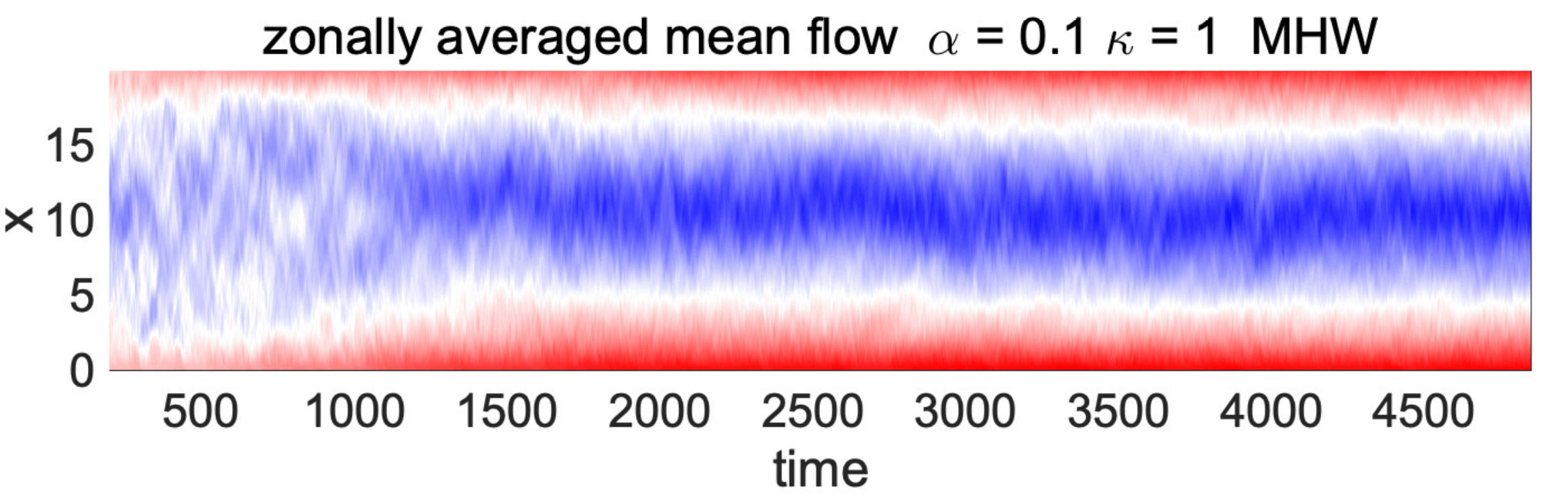}}

\vspace{-1.em}

\subfloat{\includegraphics[scale=0.35]{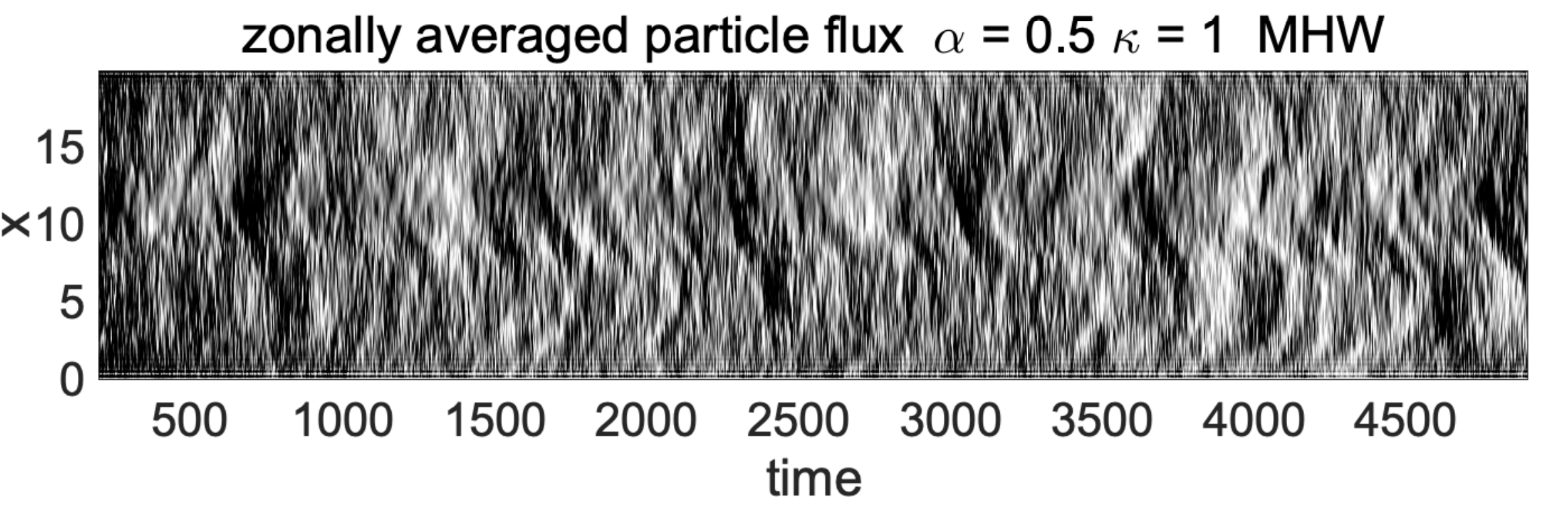}}\subfloat{\includegraphics[scale=0.35]{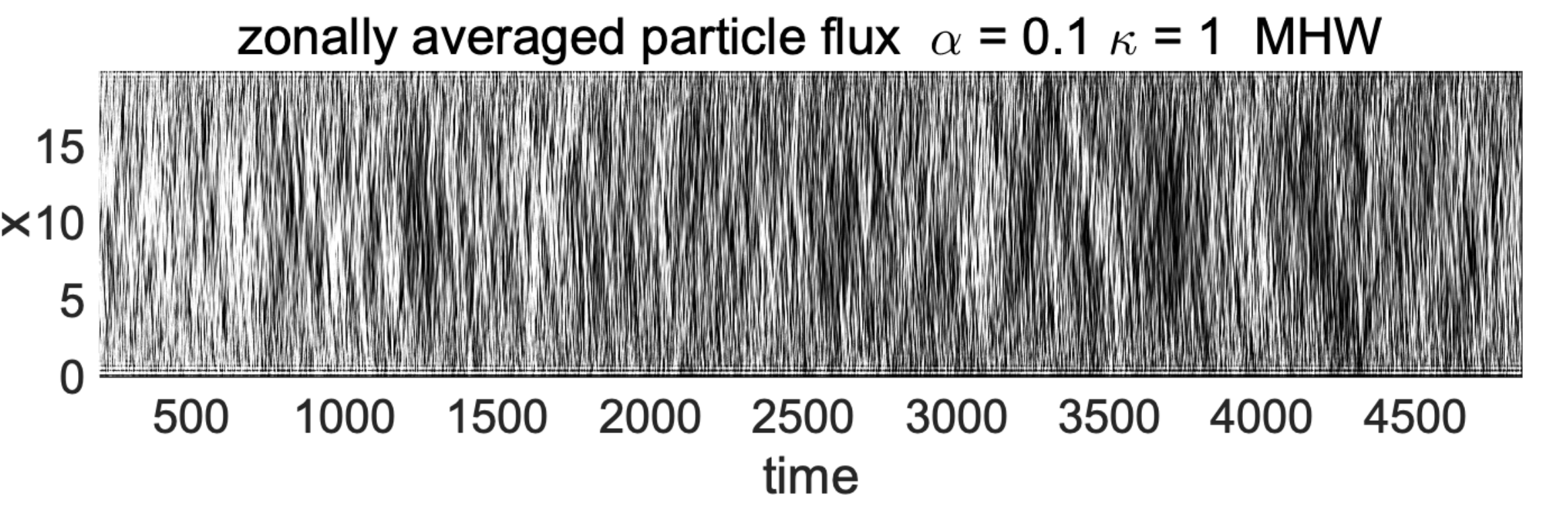}}

\vspace{-.5em}

\subfloat{\includegraphics[scale=0.35]{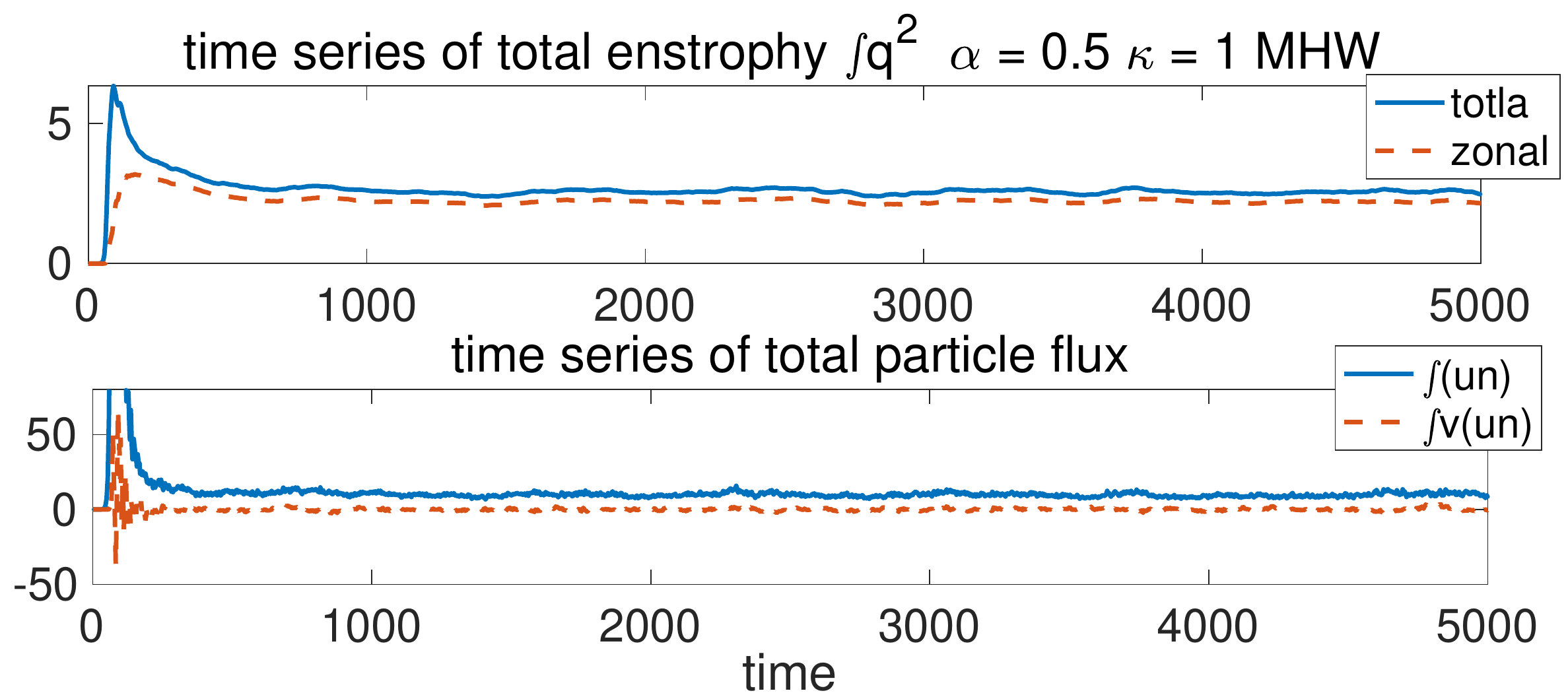}\enskip{}\includegraphics[scale=0.35]{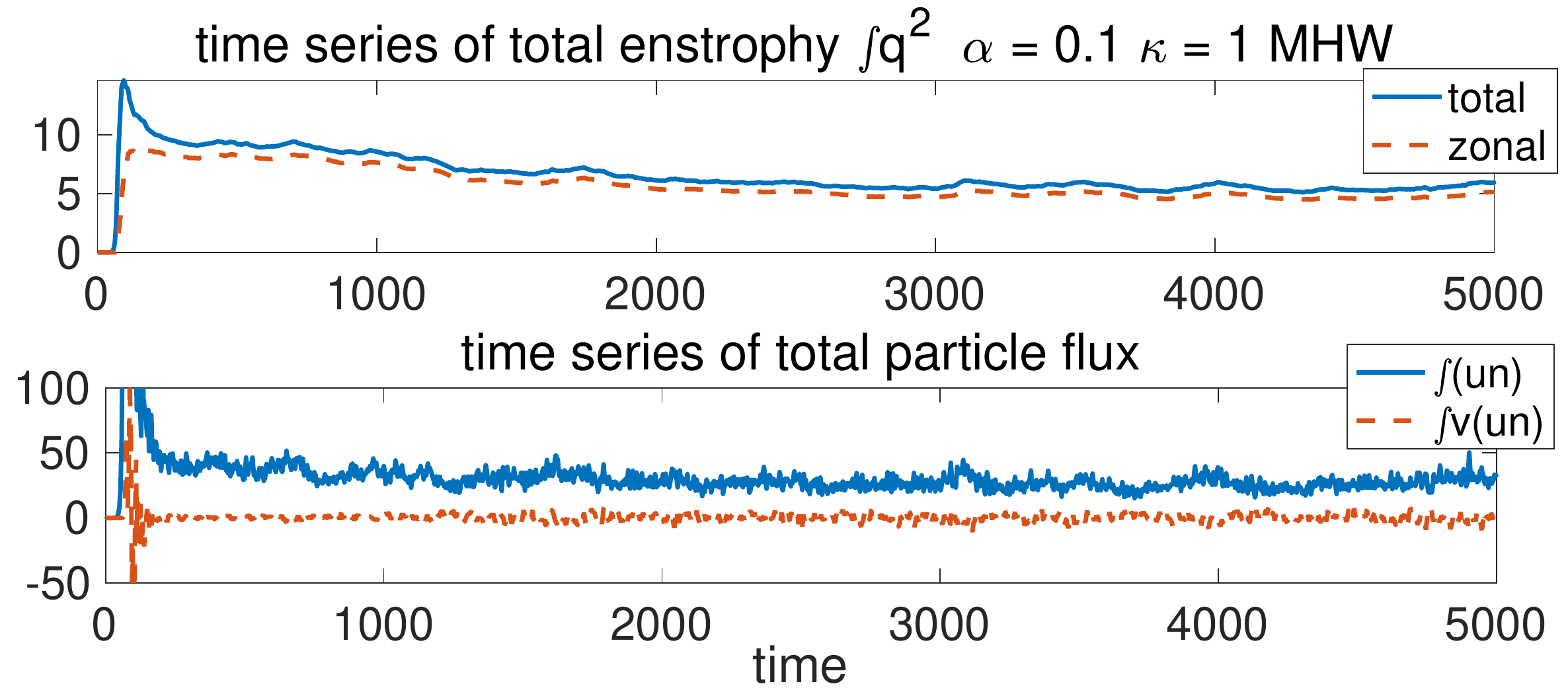}}

\caption{Time evolution of the zonal mean velocity $\overline{v}$ and the
zonal particle flux $\overline{\tilde{u}\tilde{n}}$ from MHW model simulations (Top and middle rows) for the same values of $\alpha$ considered in Figure \ref{fig:Time-evolutions-flux}: $\alpha=0.5$ (left) and $\alpha=0.1$ (right). Time series of the total enstrophy, the enstrophy in the zonal modes (top figures of bottom row) and of the total particle flux $\Gamma$ and of the total advected particle flux $\int_{\mathcal{D}}\overline{v}\left(\overline{\tilde{u}\tilde{n}}\right)dxdy$ (bottom figures of bottom row) also for these values of $\alpha$: $\alpha=0.5$ (left) and $\alpha=0.1$ (right). For all these results, $\kappa=1$ and $\mu=10^{-3}$. \label{fig:Time-evolutions-mhw}}
\end{figure}

Insights regarding the role of the parallel electron dynamics in the observed avalanches and solitary structures in the BHW model can be gained by plotting the corresponding simulation results for the MHW model in Figure \ref{fig:Time-evolutions-mhw}, for the same set of values $\alpha$, $\kappa$, and $\mu$. Zonal jets are also developed in the MHW model case. However, after the zonal mean flow $\bar{v}$ reaches a statistical steady state, the flow has much weaker time variability than in the BHW model. Correspondingly, the zonal particle flux in the MHW model has the relatively incoherent avalanche structures present in the BHW in the center of the computational domain, but does not have the solitary structures emerging from the avalanches in the vicinity of the boundaries.

In the bottom half of Figure \ref{fig:Time-evolutions-mhw}, we also show the time series for the total enstrophy $W$ as well as the fluxes $\Gamma$ and $\int_{\mathcal{D}}\overline{v}\overline{\tilde{u}\tilde{n}}dxdy$ from the MHW model, which can be compared with the corresponding BHW results in Figure \ref{fig:Snapshots-Energetics}. We see that in the MHW model, the zonal contribution dominates largely for the enstrophy. Furthermore, for both values of $\alpha$, the fluxes lack the intermittent bursts we had observed for the time series in the BHW model. In the turbulent regime, the total particle flux thus has a larger mean value in the MHW model, but has less variability and less intermittency. This result supports the following conjecture: the different treatment of the parallel electron dynamics in the BHW and MHW models is responsible for the presence or not of the radially propagating coherent solitary structures near the boundaries of the computational domain, and these solitary structures are responsible for the largest intermittent bursts measured for the fluxes in the BHW model, while the avalanches are not. In this article, we will not present a theory or additional arguments to demonstrate the validity of this hypothesis, but it is the subject of ongoing work, to be reported in a future article.

\subsection{The effect of different domain aspect ratios}

In previous sections, our computational domain was always such that $L_{x}=L_{y}=L$. As a first investigation of the role of boundary conditions and of Fourier resolution on our results, we consider the effect of a different size of the computational domain on our observations. Specifically, we extend the channel width in the $x$ direction to $\left[0,aL\right]$, where we call $a=L_{x}/L_{y}>1$ the aspect ratio. Our past results suggest that by doing so, we may observe the generation of more zonal jets, which may interact with each other, with multi-scale dynamics \cite{qi2019flux}.

\subsubsection{Multi-scale jet dynamics and large scale avalanche evolution}

We first consider a computational domain with aspect ratio
$a=5$, for the sets of parameters $\kappa=1,\alpha=0.5$, corresponding to the strong turbulence regime, and $\kappa=0.5,\alpha=0.5$, corresponding to a weaker turbulence regime. The collisional diffusion coefficient is $\mu=10^{-3}$ for both test cases. The snapshots of the ion vorticity
and the time series of the zonal mean flow are compared for these two regimes in Figure
\ref{fig:Model-statistics-ratio}. Differences with our results for the computational domain with aspect ratio $a=1$, shown in Figure \ref{fig:Snapshots-Energetics},
appear clearly. For both values of $\alpha$, larger numbers of zonal jets are created in the starting transient stage instead of a dominant single jet structure as before. The multiple jets then interact and
merge into more dominant zonal structures. At the same time, the vorticity plots indicate that the flow is more turbulent, with more small-scale vortical structures. The multi-scale jet dynamics is correlated with large-scale dynamics for the particle flux avalanches, also shown in Figure
\ref{fig:Model-statistics-ratio}. The avalanches are particularly strong and widespread in the starting transient phase, before gradually shrinking to a narrower region in the center of the computational domain as the zonal jets merge to form stronger and more localized zonal structures. Furthermore, we observe that as before, simple travelling structures are seen emerging from the avalanches in the zonal particle flux, although they are not as coherent and localized as for the computational domain with aspect ratio $a=1$.

We conclude this section with a comparison with simulations from the MHW model for the same values of the parameters $\alpha$ and $\kappa$, and the same computational domain. The results for the zonal mean velocity $\overline{v}$ and the zonal particle flux $\overline{\tilde{u}\tilde{n}}$ are shown in Figure \ref{fig:Model-statistics-ratio-mhw}. In the MHW simulations, we also see initial mergings of zonal jets, but these happen during a short transition period. After this transient, the zonal jets have much less variability than in the BHW model, they do not merge any more, and we do not observe any evidence of strong large scale dynamics and multi-jet interactions. This difference in behavior between the BHW model and the MHW model was already highlighted for doubly-periodic boundary conditions in \cite{qi2019flux}. As a result of the weakness of the multi-scale interactions, there is little large scale dynamics in the particle flux avalanches as compared to the BHW model, and the avalanches remain spread across most of the computational domain. Finally, as for the simulations with aspect ratio $a=1$, in the MHW model we do not see solitary structures emerging from the particle flux avalanches. 

\subsubsection{Statistical invariance with different aspect ratios}

\begin{figure}
\subfloat[snapshots of ion vorticity]{\includegraphics[scale=1.]{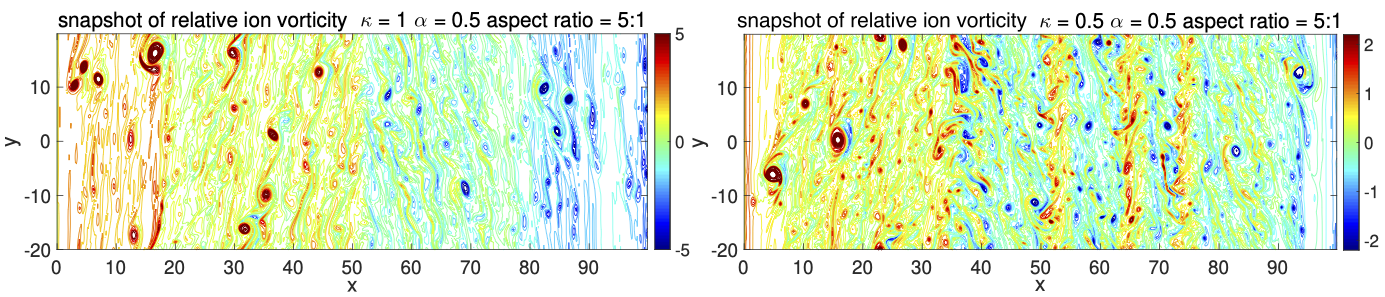}

}

\subfloat[zonally averaged time series $\alpha=0.5,\kappa=1$]{\includegraphics[scale=0.27]{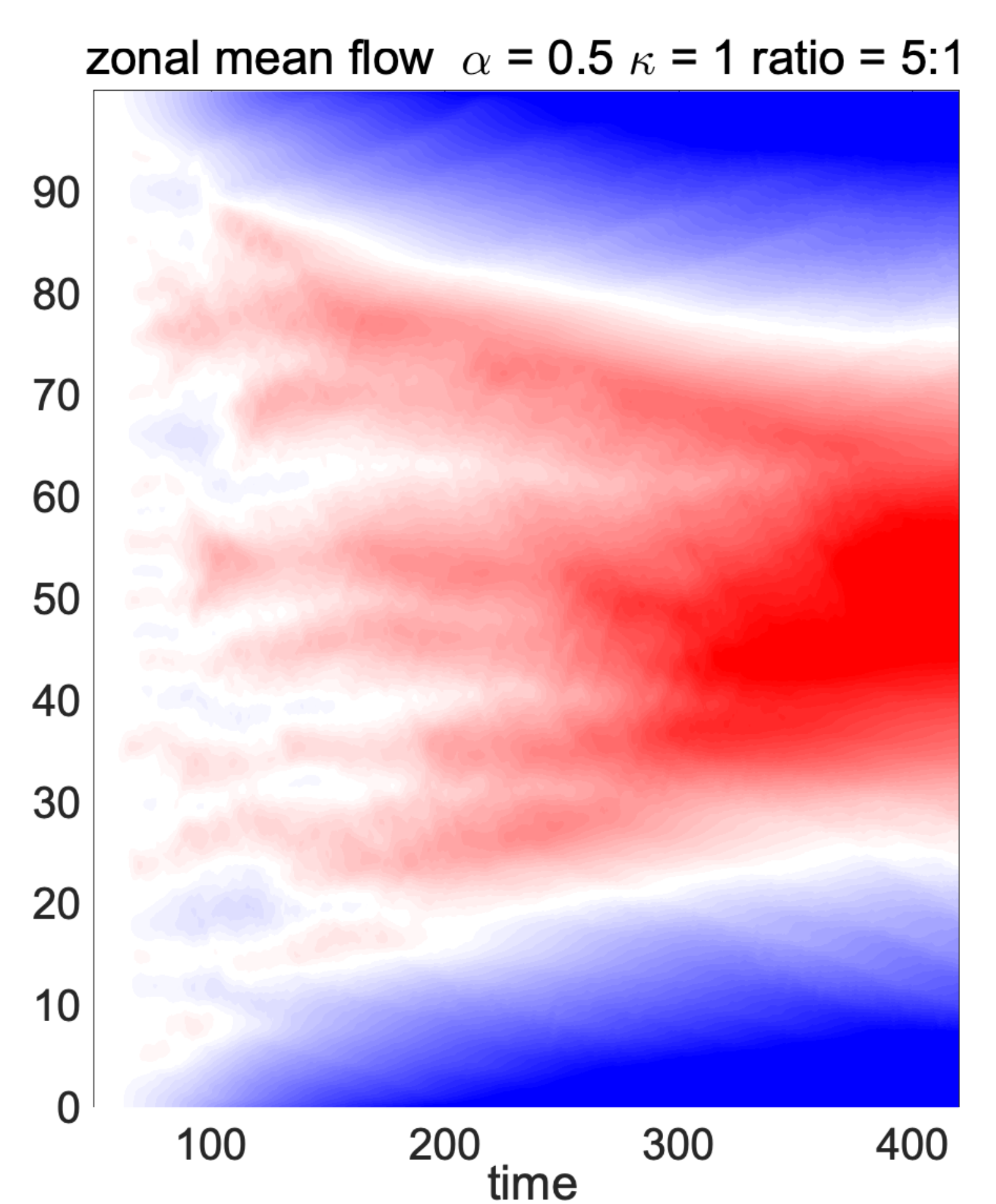}\includegraphics[scale=0.27]{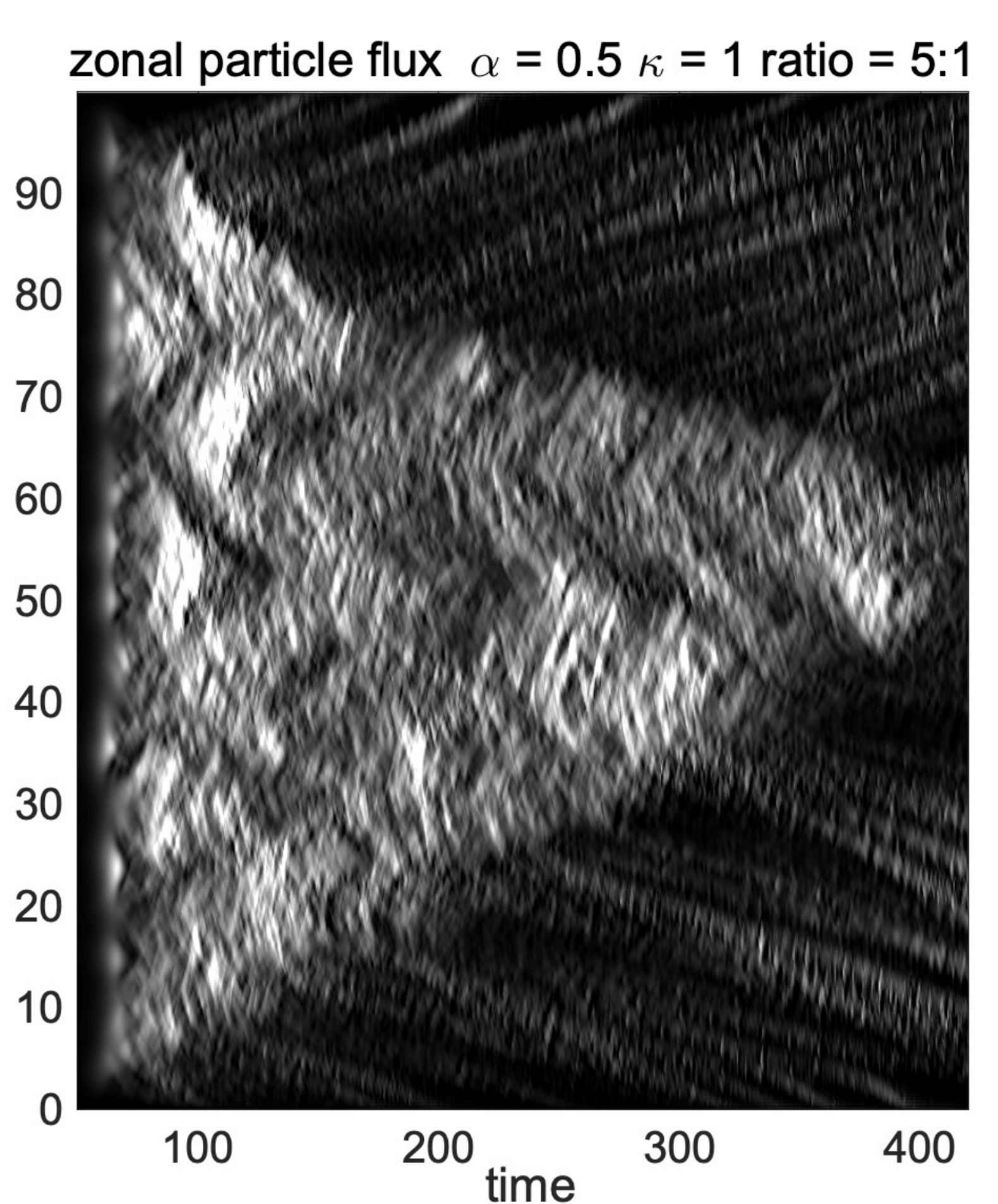}

}\subfloat[zonally averaged time series $\alpha=0.5,\kappa=0.5$]{\includegraphics[scale=0.27]{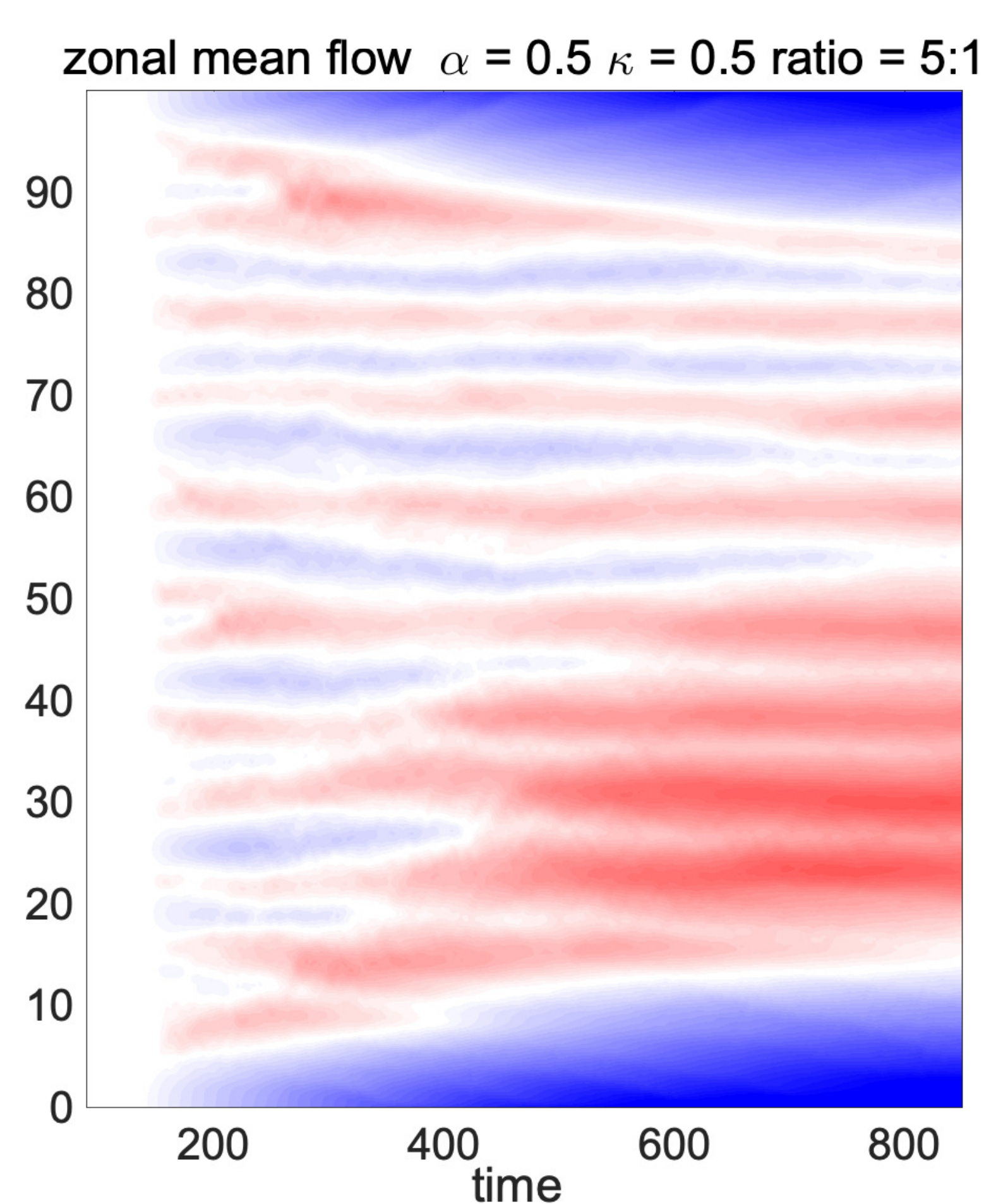}\includegraphics[scale=0.27]{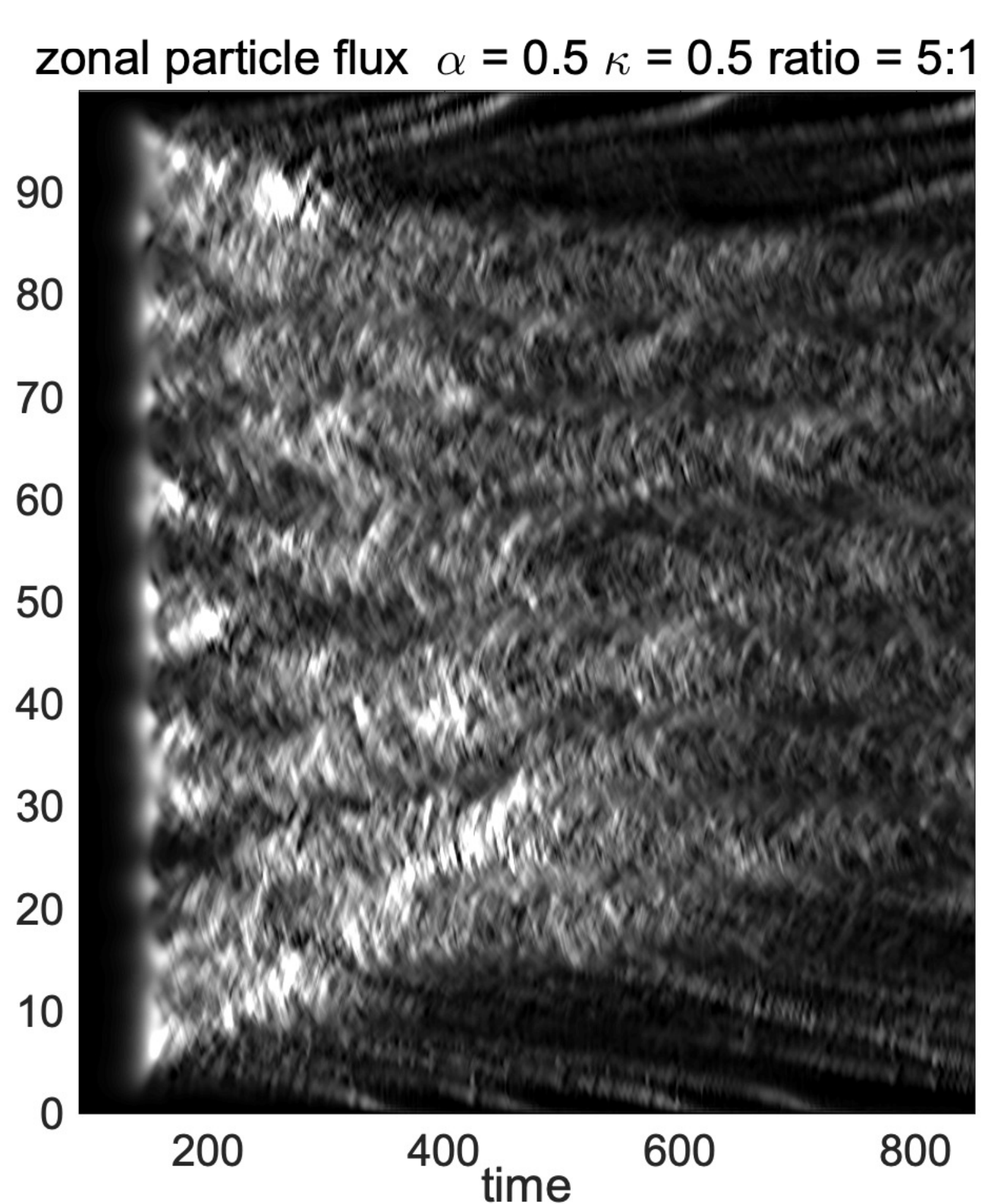}

}

\subfloat[equilibrium statistical spectra with different aspect ratios]{\includegraphics[scale=0.35]{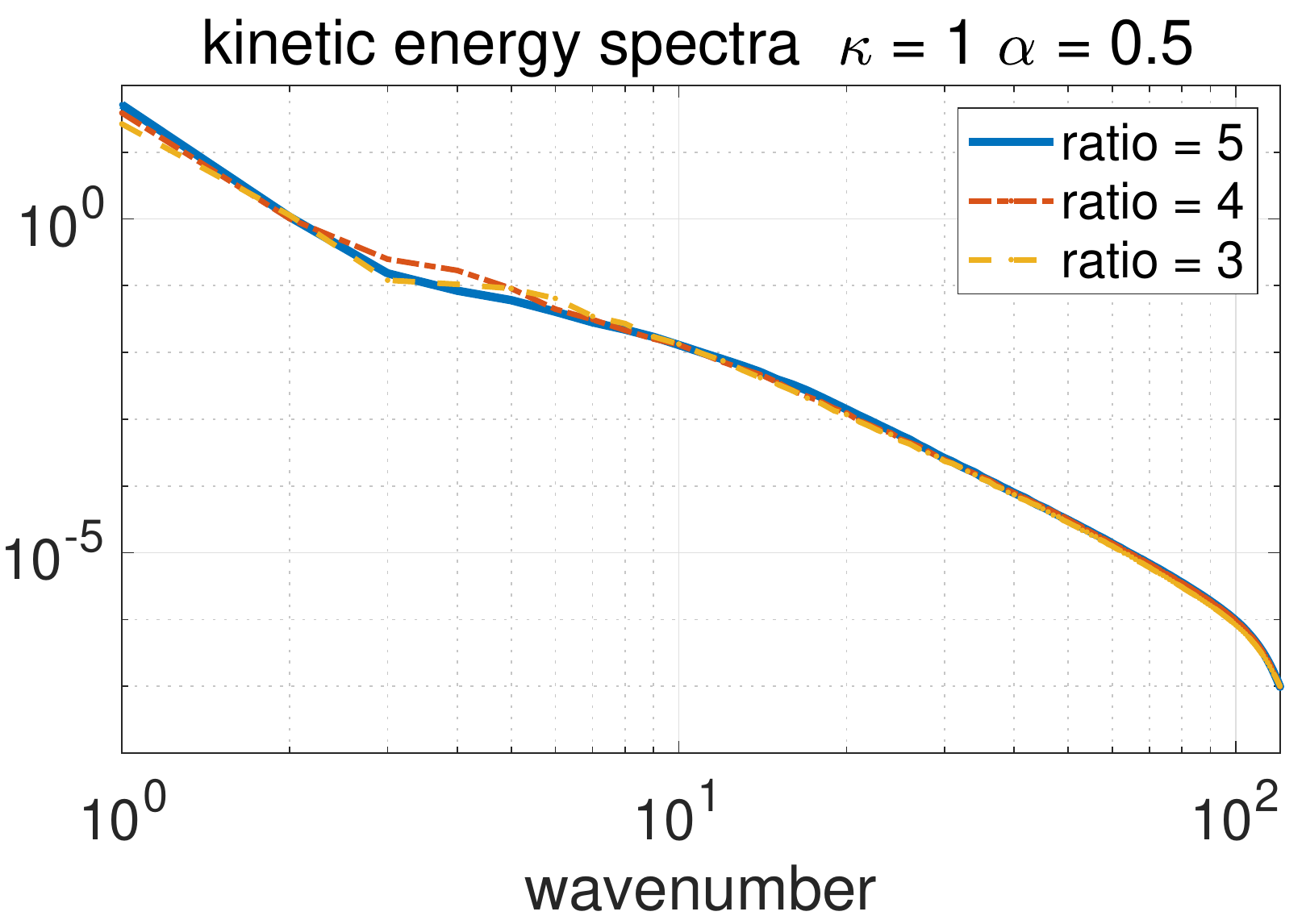}\includegraphics[scale=0.35]{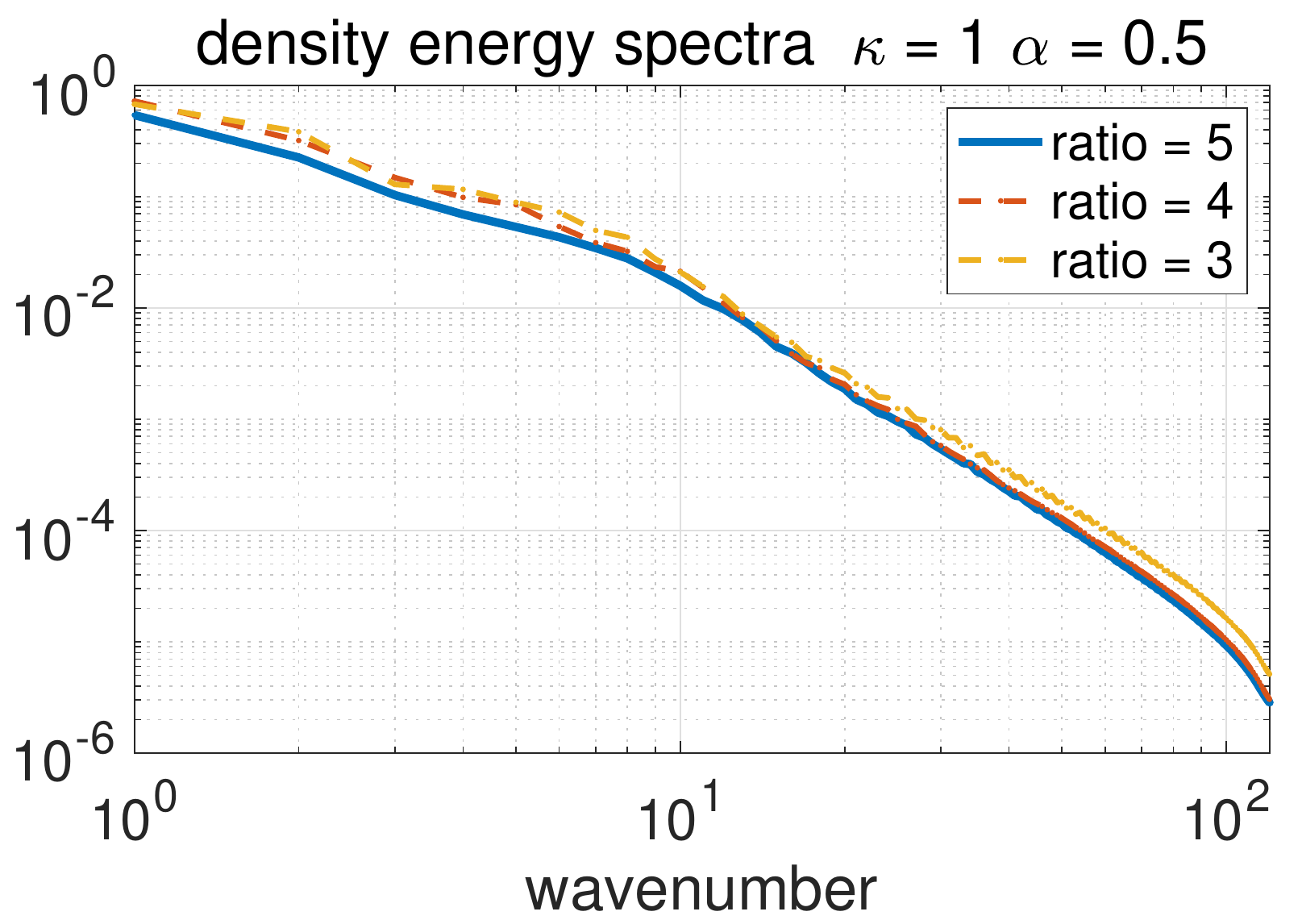}\includegraphics[scale=0.35]{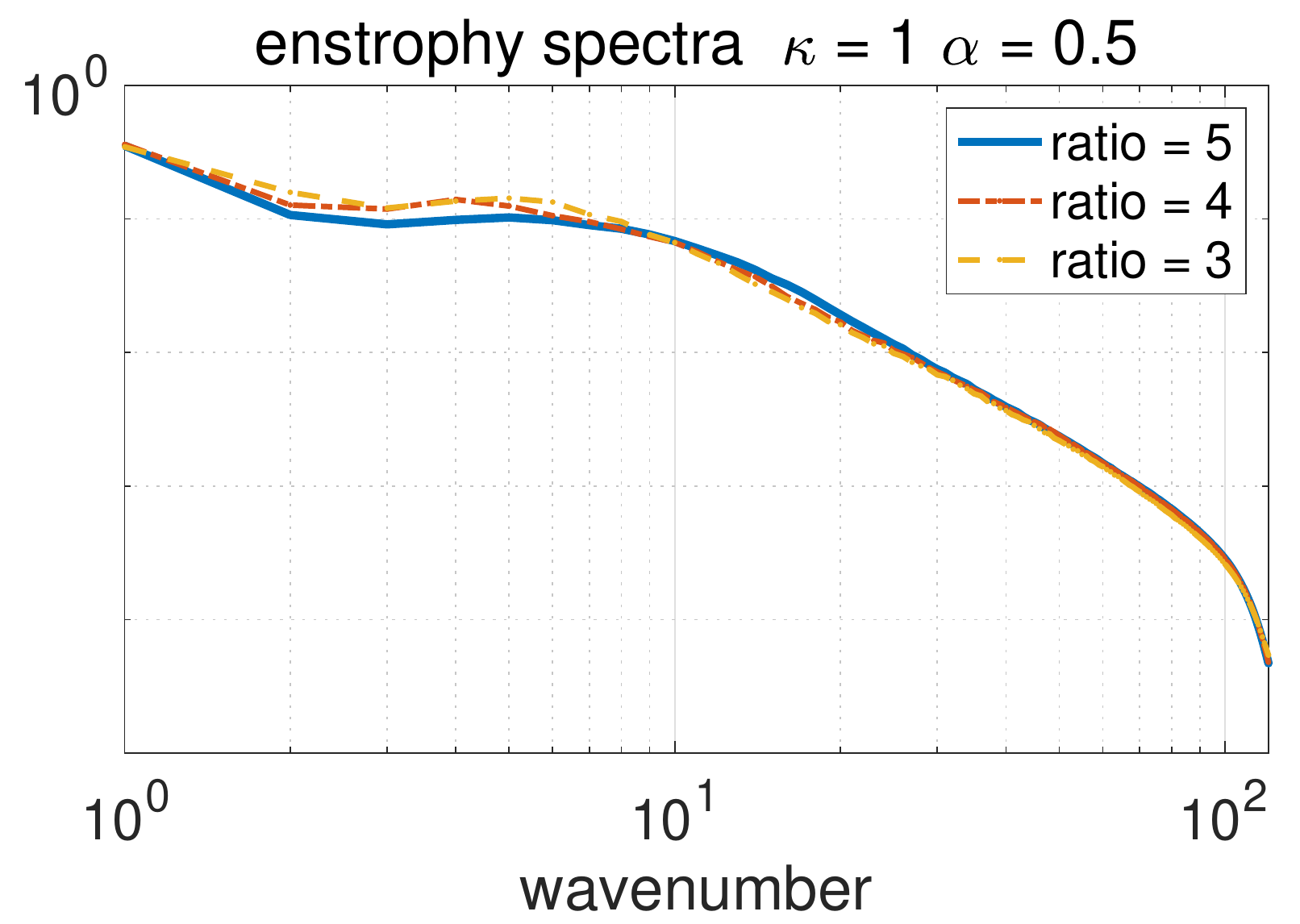}

}

\caption{Snapshots of the ion vorticity (upper row), and time-series of the
zonal mean velocity $\overline{v}$ and the zonal particle flux $\overline{\tilde{u}\tilde{n}}$
(middle row) from the BHW model for a computational domain with aspect ratio $a=5$. The figures on the left-hand side were obtained for $\kappa=1,\alpha=0.5$, and the figures on the right-hand side for $\kappa=0.5,\alpha=0.5$, with $\mu=10^{-3}$ in both cases. Kinetic energy spectra (left), density energy $\frac{1}{2}\int_{\mathcal{D}}n^2dxdy$ spectra (middle), and enstrophy spectra (right) for different domain aspect ratios $a=5,4,3$, $\kappa=1,\alpha=0.5$, once statistical steady-state was reached (lower row).\label{fig:Model-statistics-ratio}}
\end{figure}

\begin{figure}

\subfloat[zonally averaged time series $\alpha=0.5,\kappa=1$]{\includegraphics[scale=0.27]{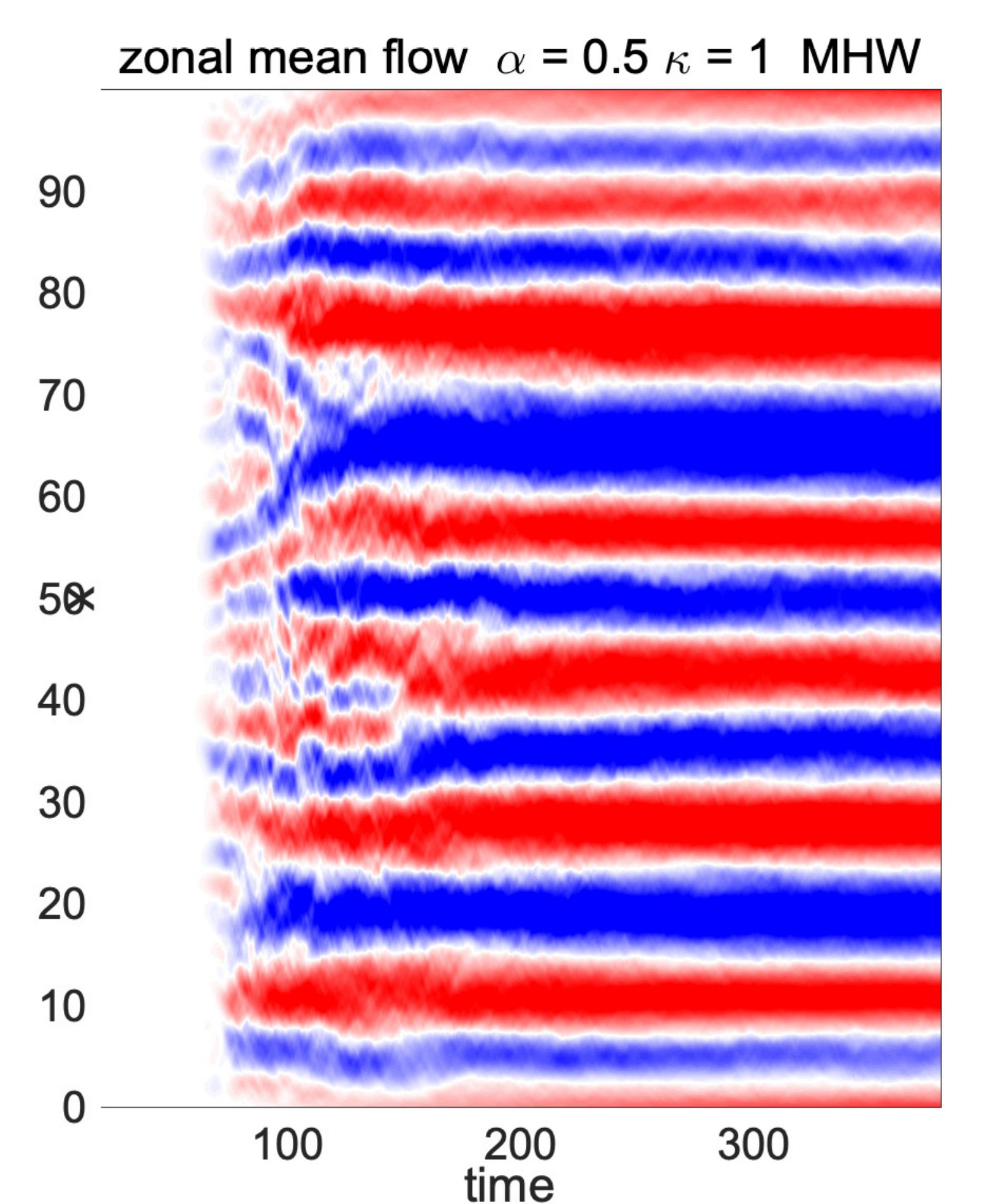}\includegraphics[scale=0.27]{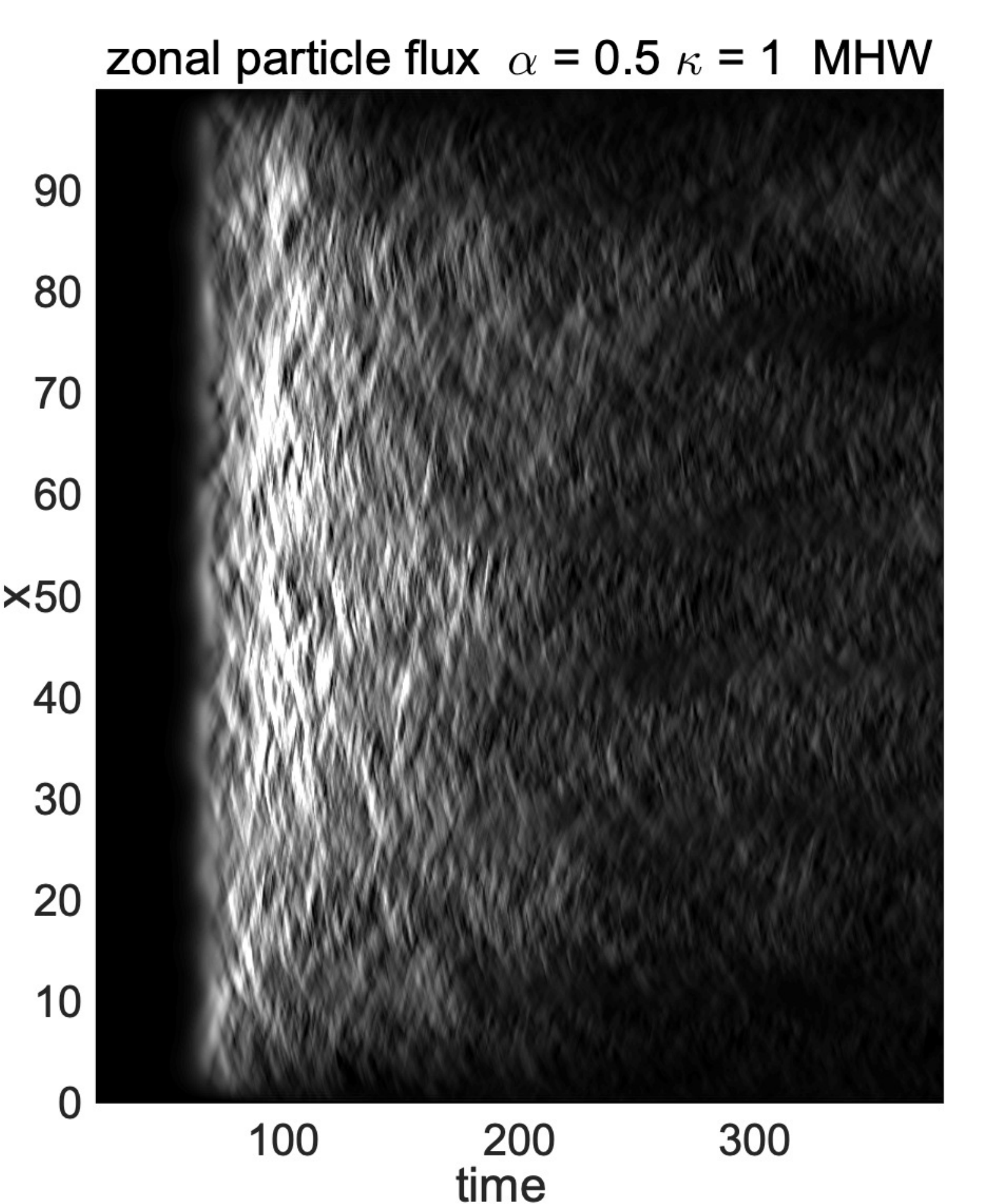}

}\subfloat[zonally averaged time series $\alpha=0.5,\kappa=0.5$]{\includegraphics[scale=0.27]{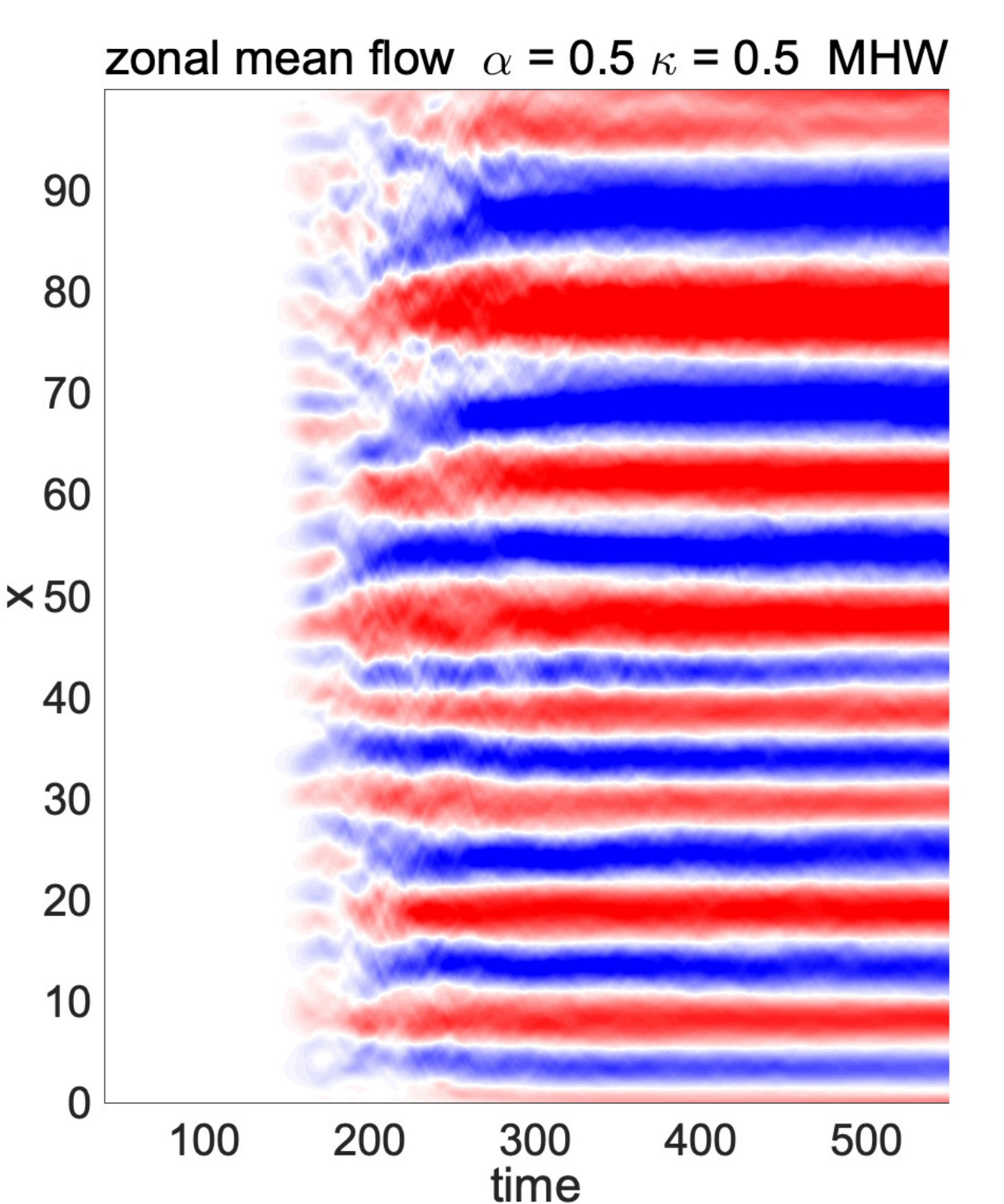}\includegraphics[scale=0.27]{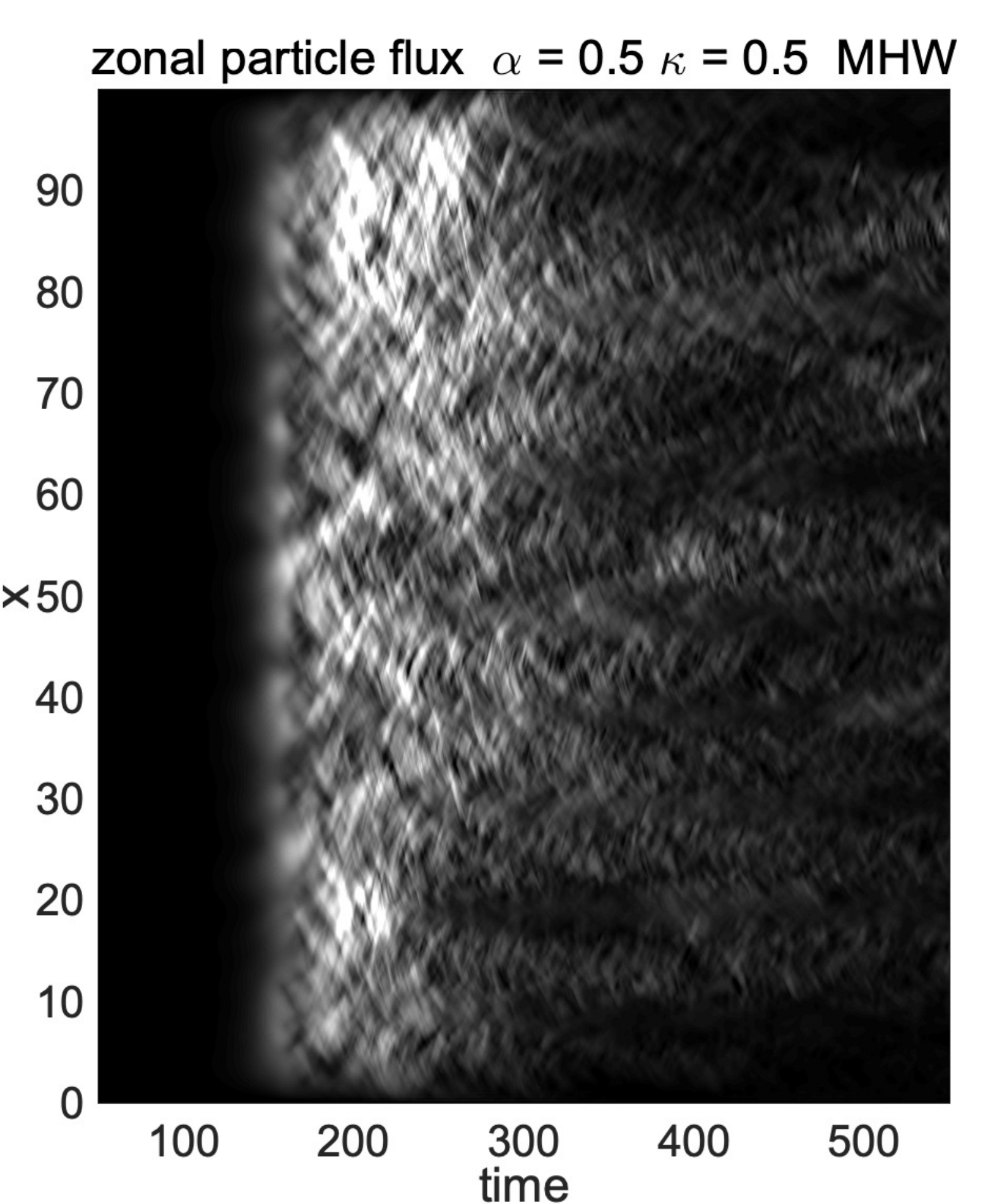}

}

\caption{Time series of the zonal mean velocity $\overline{v}$ and the zonal particle flux $\overline{\tilde{u}\tilde{n}}$ from the MHW model for a computational domain with aspect ratio $a=5$. The figures on the left and middle left were obtained for $\kappa=1, \alpha=0.5$, and the figures on the middle right and right were obtained for $\kappa=0.5, \alpha=0.5$.\label{fig:Model-statistics-ratio-mhw}}
\end{figure}

In this final subsection, we show the statistical invariance of the simulations for different aspect ratios $a$ of the computational domain once the transient multi-scale evolution we just discussed has taken place, and statistical steady-state has been reached. The bottom row of Figure
\ref{fig:Model-statistics-ratio} displays the spectra of the kinetic energy $\frac{1}{2}\int_{\mathcal{D}}|\nabla\varphi|^2dxdy$, the density energy $\frac{1}{2}\int_{\mathcal{D}}n^2dxdy$ and of the enstrophy once statistical equilibrium was reached, for the following aspect ratios of the computational domain: $a=3,4,5$. A longer $x$ domain length contains larger energy and also stronger particle flux in the transient state. Thus it generates stronger turbulent
dynamics. However in the final statistical equilibrium state, it is
observed that the radially averaged steady-state energy and enstrophy spectra are approximately independent of the  aspect ratio. This invariance
confirms the existence of a consistent statistical structure in the channel domain flows, even though much richer dynamics are created
with a longer $x$ domain length in the transient phase.

\section{Dimits shift and avalanche structures in a doubly periodic geometry\label{sec:Dimits-shift-Periodic}}

Many studies of turbulent transport in magnetic confinement devices, either with reduced fluid models or the gyrokinetic equations, impose doubly periodic boundary conditions for their simulations. We now investigate how imposing doubly periodic boundary conditions instead of the channel geometry boundary conditions modifies our results. We have already studied the BHW model with doubly periodic boundary conditions extensively in \cite{majda2018flux,qi2019flux}. In this section, we therefore focus on the Dimits shift and signatures of non-local, non-diffusive transport, which are the central subjects of this article. For all the simulations, the computational domain is $\left[-L_{x},L_{x}\right]\times\left[-L_{y},L_{y}\right]=\left[-20,20\right]\times\left[-20,20\right]$

\subsection{Dimits shift and comparison between the BHW and MHW models}

\begin{figure}
\subfloat[total particle flux as a function of $\alpha$ or $\kappa$]{\includegraphics[scale=0.34]{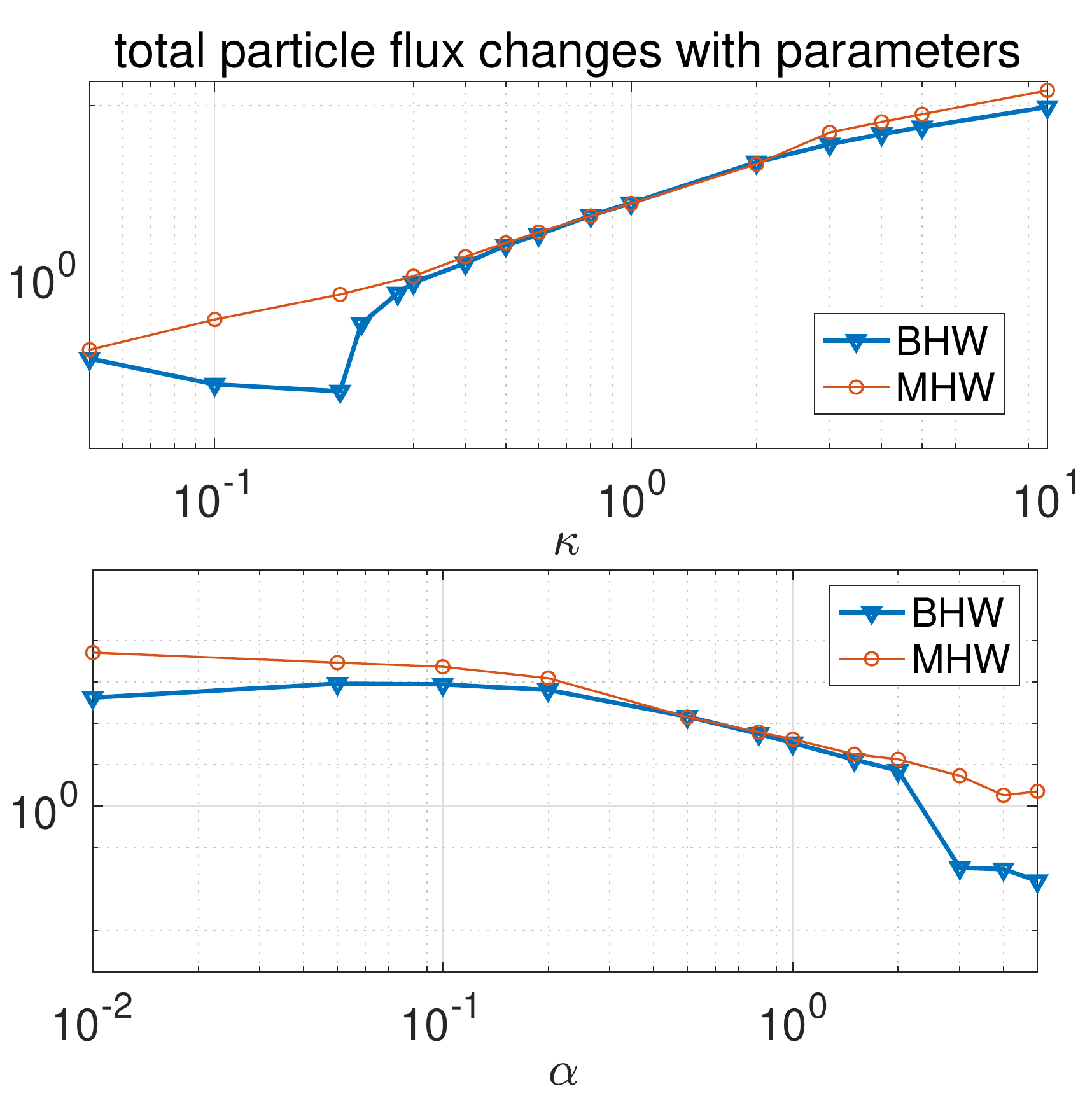}

}\subfloat[solutions with $\alpha=0.5$ and $\kappa=1$]{\includegraphics[scale=0.34]{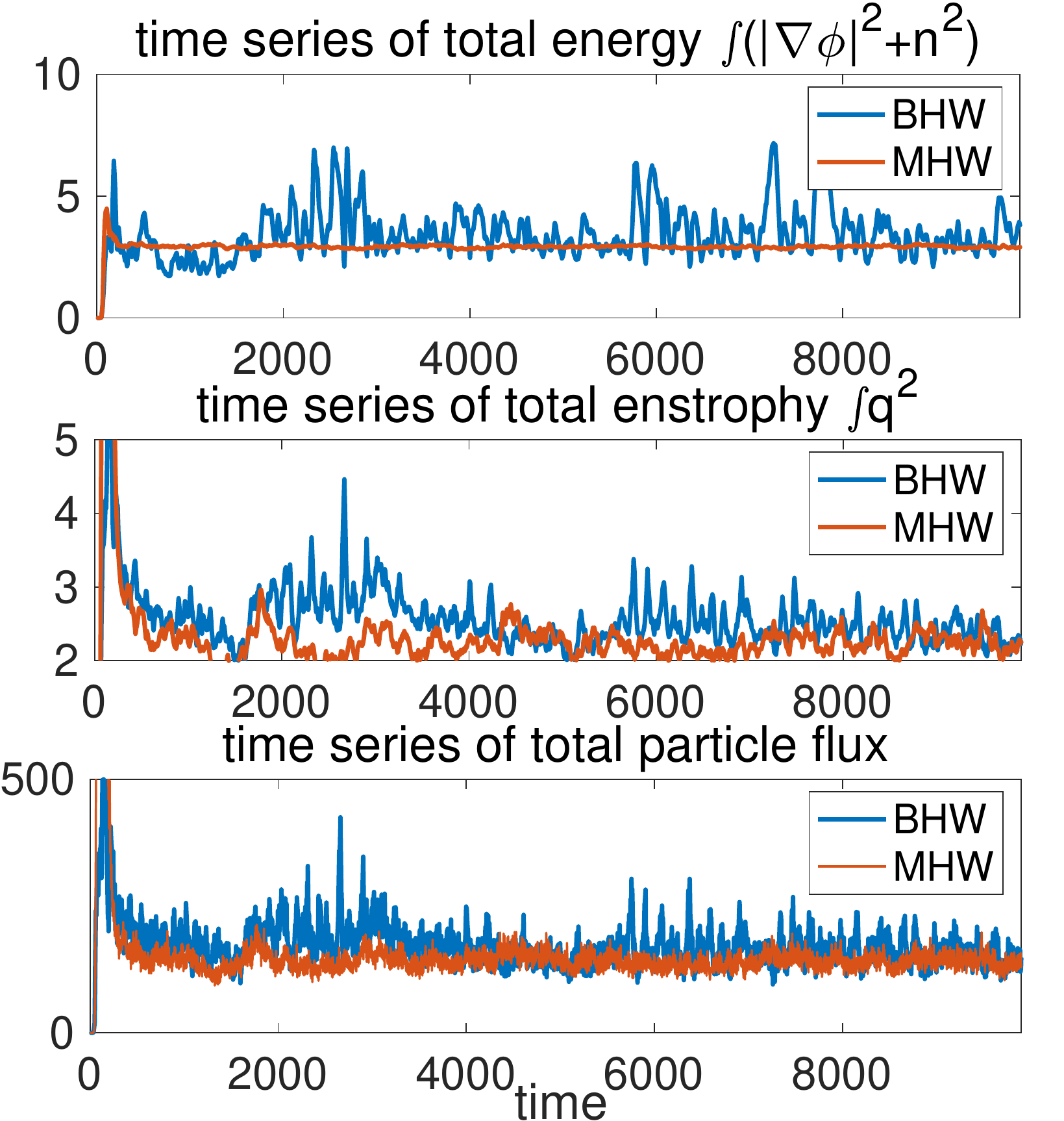}\negthickspace{}\includegraphics[scale=0.34]{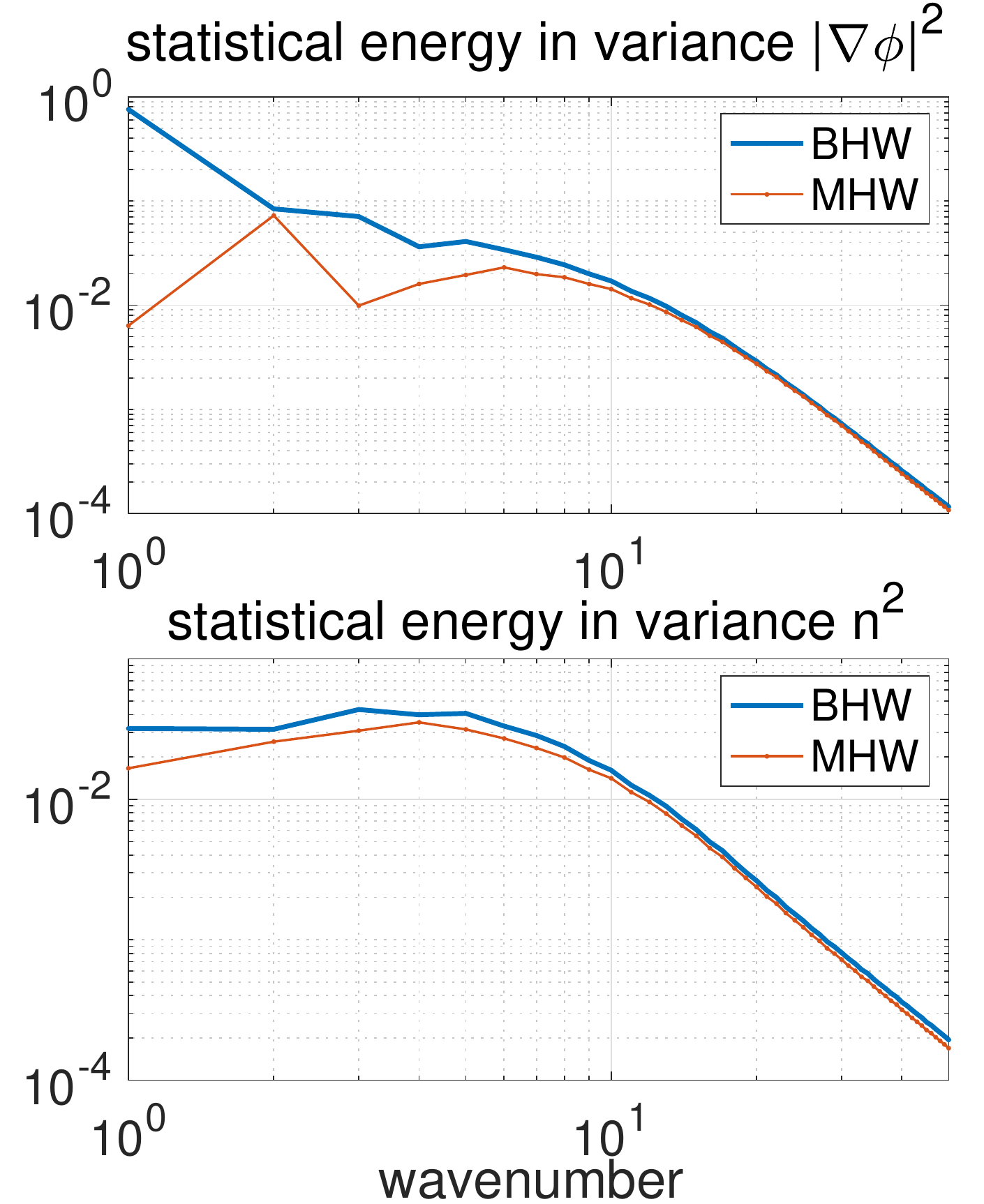}

}

\caption{Comparison of the total particle flux from the BHW and MHW model simulations
(left column), and the time-series of the total energy and particle
flux and the energy spectra for variance from the BHW and MHW model
simulations with $\alpha=0.5$ and $\kappa=1$ (right column). For all these simulations, $\mu=10^{-3}$. \label{fig:Time-series}}
\end{figure}

We first compare the transition between the low transport regime and the high transport regime in the BHW model and the MHW model. A comprehensive summary of our results is given in Figure \ref{fig:Time-series}, which shows, for each model, the total particle flux $\Gamma$ as a function of $\kappa$ for the fixed values $\alpha=0.5$ and $\mu=10^{-3}$, and as a function of $\alpha$ for the fixed values $\kappa=1$ and $\mu=10^{-3}$, as well as the times series for the total energy and enstrophy and for the particle flux for $\alpha=0.5$ and $\kappa=1$, and the spectra for the energy in variance $\langle\vert f-\langle f\rangle\vert^{2}\rangle$, where we compute the second-order moment of the velocity and density fields $f=\nabla\varphi, n$ by taking the long time average $\langle f \rangle = \frac{1}{T}\int_0^{T} f\left(t\right) dt $ of the state variables for each scale with the same wavenumber $k$. 

In both models, a transition between a zonal flow dominated regime with low particle transport and a turbulent regime with strong particle transport is observed as we vary the physical parameters $\alpha$ and $\kappa$. However, the transition is much less sharp in the MHW model: for a fixed value of $\alpha$, the particle flux increases with $\kappa$ as a power law, and for a fixed value of $\kappa$, the particle flux increases with decreasing values of $\alpha$ as a power law as well. In other words, the MHW model does not have what one would characterize as a true Dimits shift, marked by a sudden transition in the particle flux \cite{dimits2000simulation,dimits2000comparisons,stonge2017}, as obtained in the BHW model, which has near-zero particle flux until the Dimits threshold is reached. This result can be understood in the light of our previously published results \cite{majda2018flux,qi2019flux}, and in particular Figure 4 in \cite{majda2018flux}. In the MHW model, purely zonal states are never generated, even when the linear instability is weak. Small scales vortices are always present, which drive some amount of turbulent transport. They are responsible for the absence of a true nonlinear upshift for the threshold for strong particle transport in the MHW model. In contrast, when the linear instability is weak, pure zonal states are generated in the BHW model, which fully impede particle transport.

In addition, we identify further differences between the two models in the turbulence dominated regime, as illustrated by the time-series of the total energy and particle flux in the center column of Figure \ref{fig:Time-series}
as well as the energy spectra are compared in detail in the right
column of Figure \ref{fig:Time-series}. The time series for the BHW model are 
characteristic of strong but intermittent turbulent transport, while the time series for the MHW model rapidly converge to a statistical steady state with much weaker fluctuations. In the turbulence dominated regime, transport in the MHW model is strong in averaged value but less bursty and variable than in the BHW model, as we had already found previously for the channel geometry. This is further confirmed by observing the variance energy spectra of the two models shown in the right column of Figure \ref{fig:Time-series}. The variances in the leading modes (that is, the first two largest scales dominated by the variance in zonal modes $k_x=1, 2$) have much larger values in the BHW model than in the MHW model case. This is consistent with the much stronger variability in the zonal modes already reported in previous studies of the BHW model \cite{majda2018flux,qi2019flux,qi2019channel}.

\subsection{Transport structures for doubly periodic boundary conditions}

\begin{figure}
\begin{centering}
\subfloat{\includegraphics[scale=0.3]{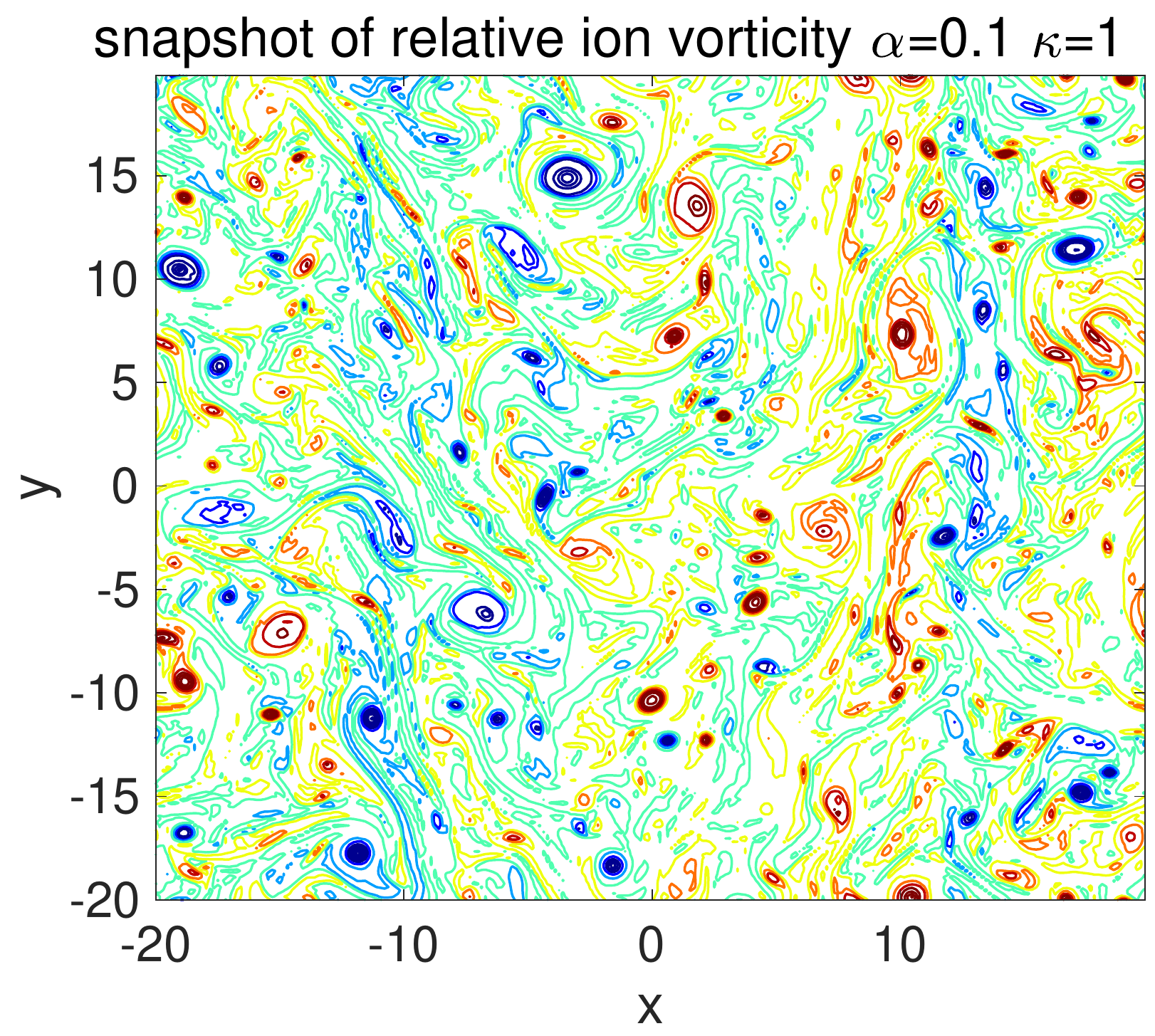}\qquad{}\includegraphics[scale=0.3]{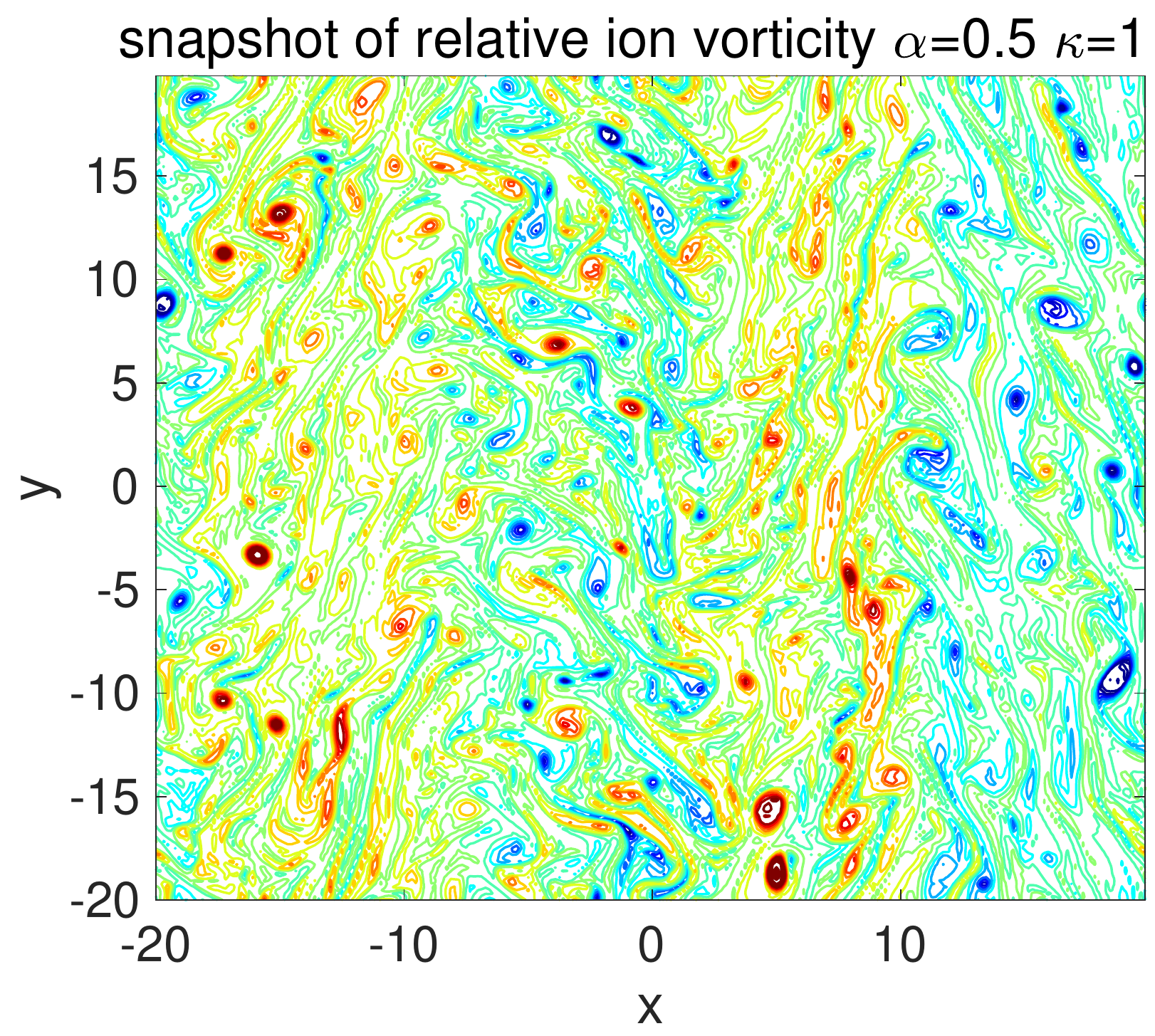}\qquad{}\includegraphics[scale=0.3]{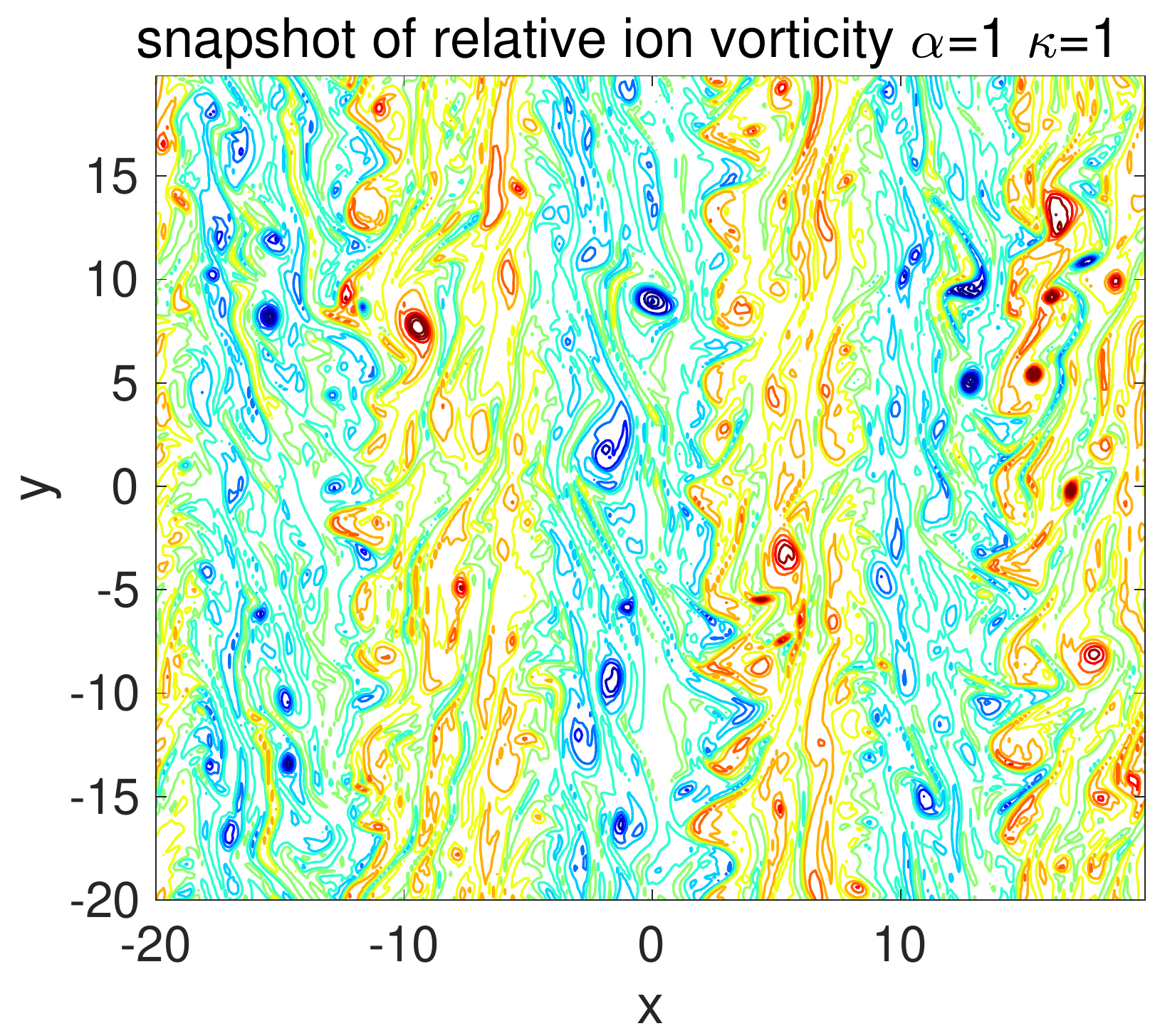}}
\par\end{centering}
\vspace{-1.em}
\begin{centering}
\subfloat{\includegraphics[scale=0.32]{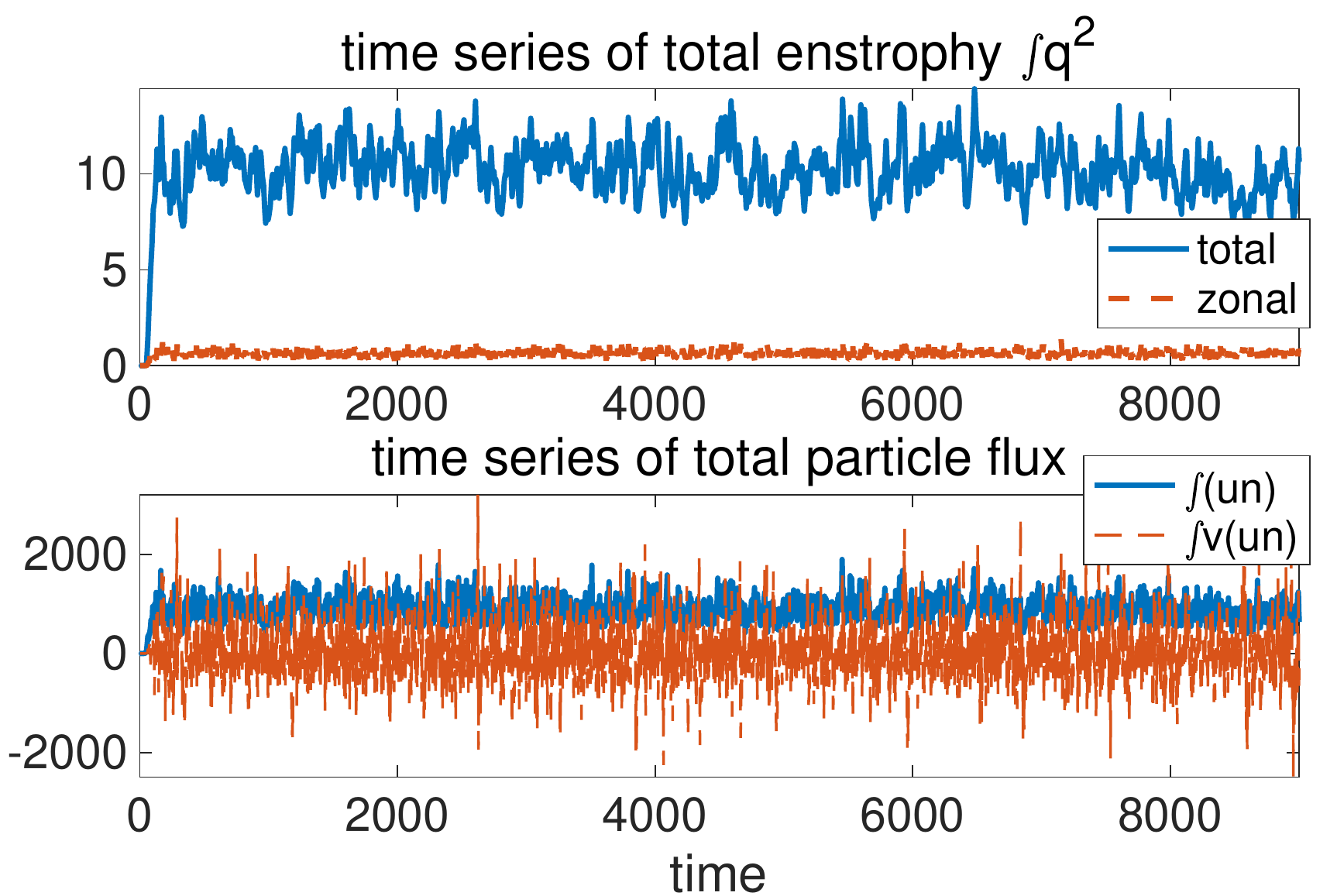}\includegraphics[scale=0.32]{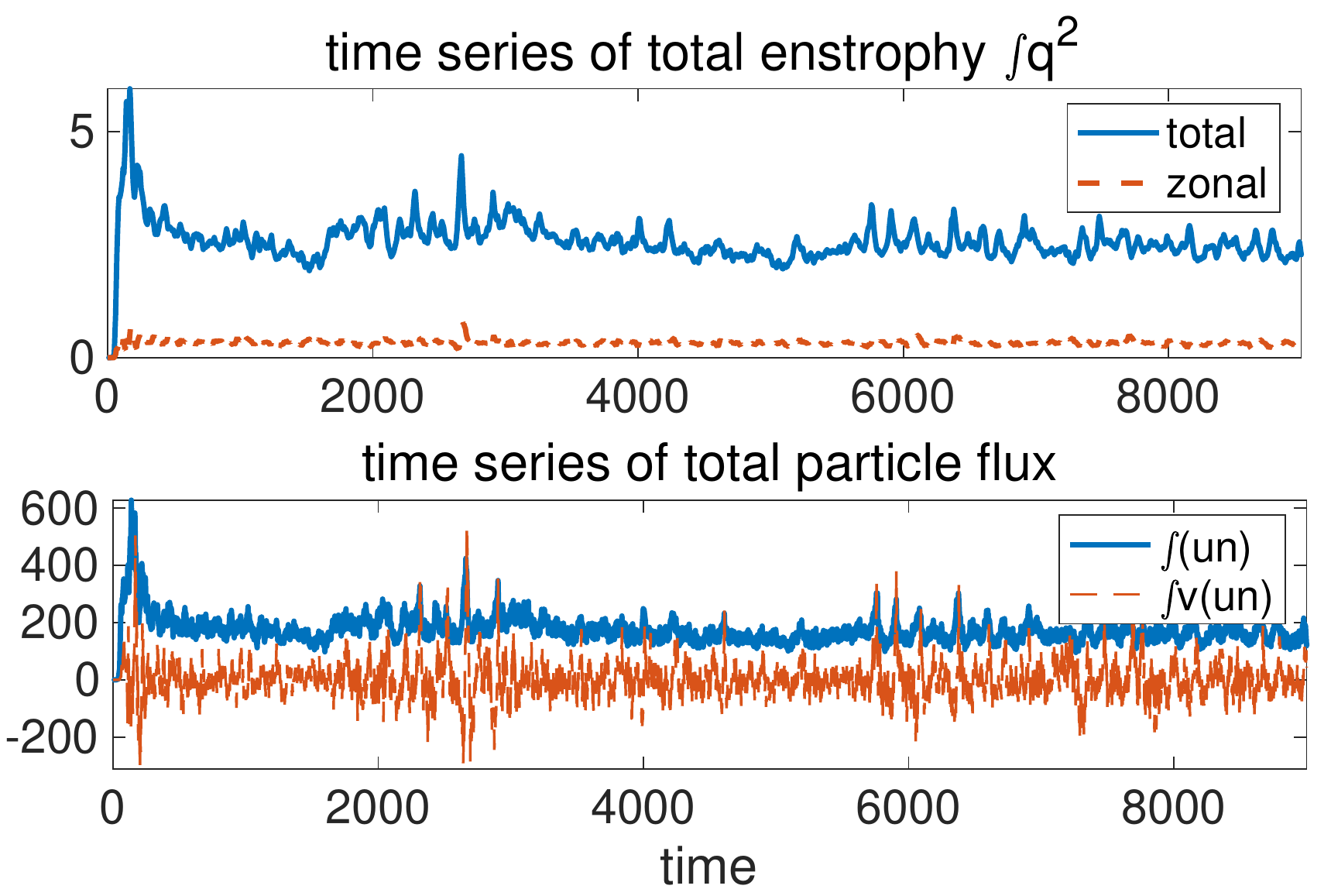}\includegraphics[scale=0.32]{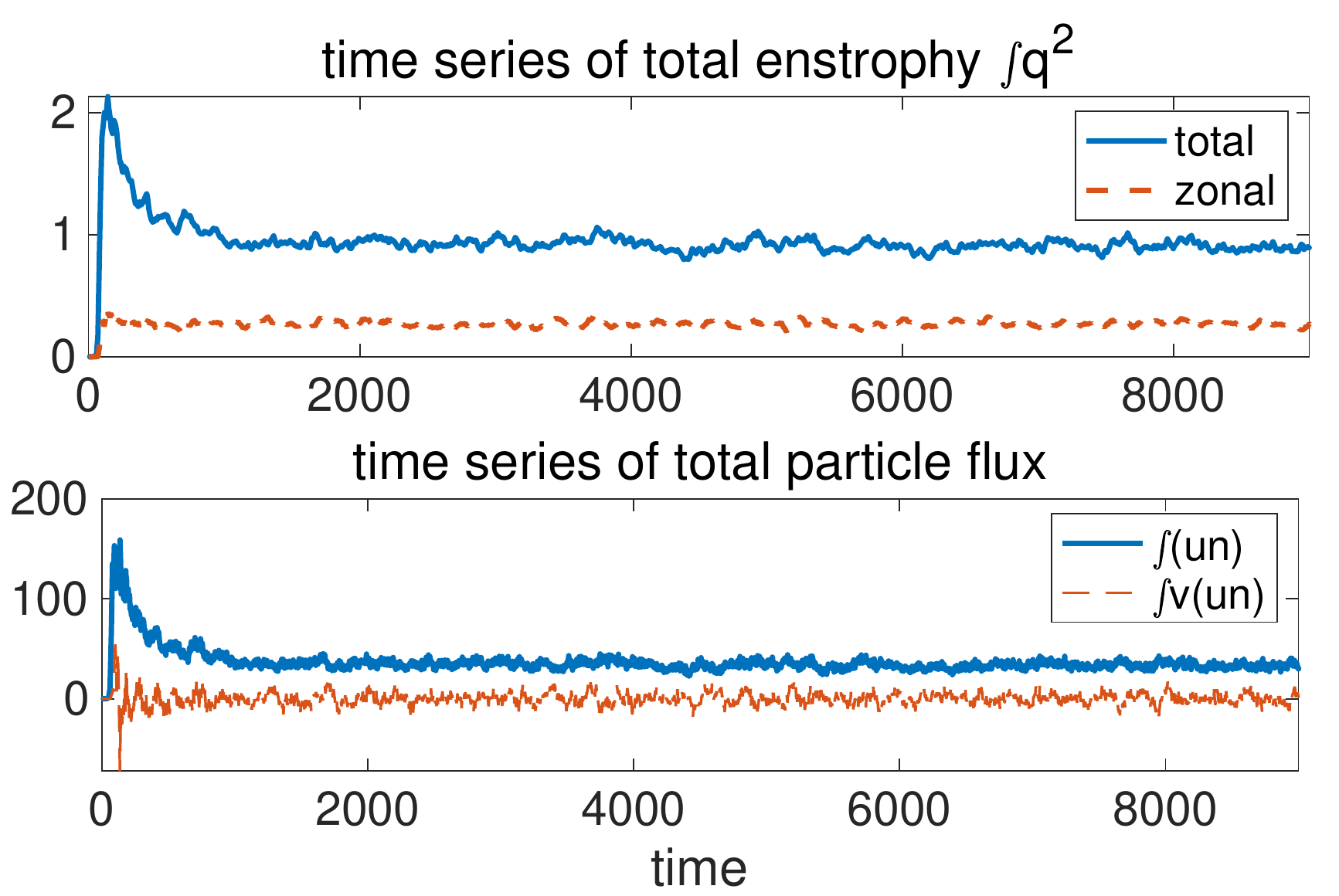}

}
\par\end{centering}
\caption{Snapshots of the ion vorticity field $\zeta=\nabla^{2}\varphi$ (upper
row) and time series of the total enstrophy $W=\int_\mathcal{D} q^{2}dxdy$, as well as the total fluxes $\Gamma=\int_{\mathcal{D}}\tilde{u}\tilde{n}dxdy$
and $\int_{\mathcal{D}}\overline{v}\overline{\tilde{u}\tilde{n}}dxdy$ (lower row) from the BHW model with doubly periodic boundary condition, for fixed $\kappa=1,\mu=1\times10^{-3}$ and different values of $\alpha$: $\alpha=0.1$ (left), $\alpha=0.5$ (center), and $\alpha=1$ (right).\label{fig:Snapshots-peri}}
\end{figure}

We now look at the flow and transport features in the BHW model in more detail, to compare them with the equivalent results we obtained in the channel geometry. In Figure \ref{fig:Snapshots-peri}, we present the flow snapshots and time series of the total enstrophy and total particle flux for $\kappa=1, \mu=10^{-3}$ and three values of $\alpha$, $\alpha=0.1$, $\alpha=0.5$, and $\alpha=1$, which can be directly compared with Figure \ref{fig:Snapshots-Energetics}. Just like in the channel domain case, and in agreement with our observations from the previous section, as the value of $\alpha$ increases, the flow goes through a transition from a fully turbulent regime with strong non-zonal vorticity and strong and bursty particle transport to a zonal flow dominated regime with nearly zero particle flux. We nevertheless note that for the same values of $\alpha$, $\kappa$, and $\mu$, the flow computed for doubly periodic boundary conditions is more turbulent, with more small scale vortices. This explains why, for fixed $\kappa$ and $\mu$, the Dimits threshold is found for a larger value of $\alpha$ for the doubly periodic boundary conditions than for the channel geometry.

\begin{figure}
\subfloat{\includegraphics[scale=0.36]{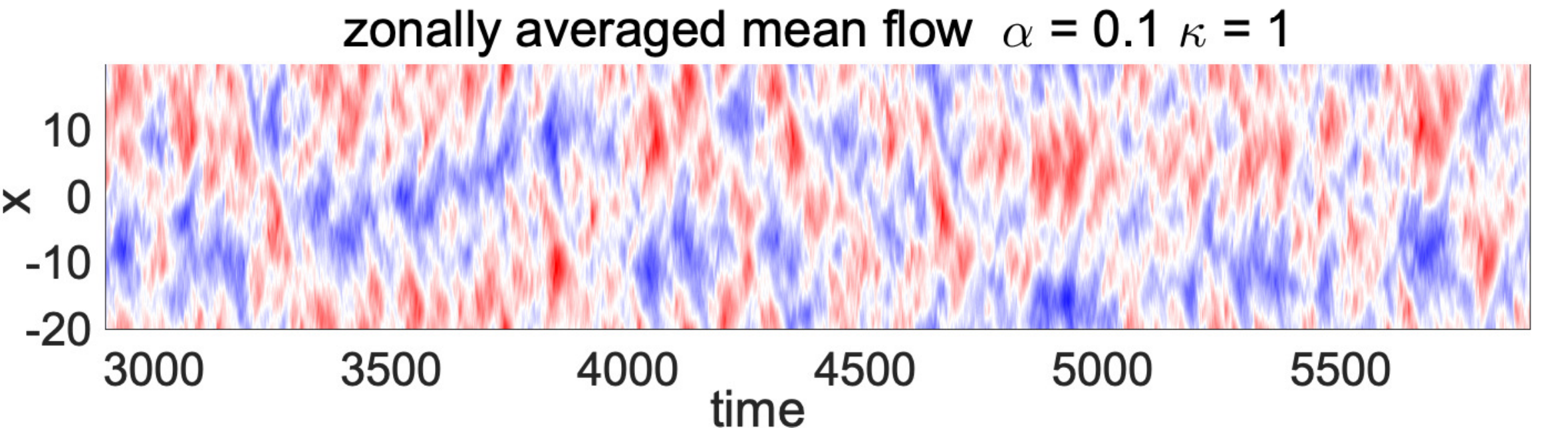}\includegraphics[scale=0.36]{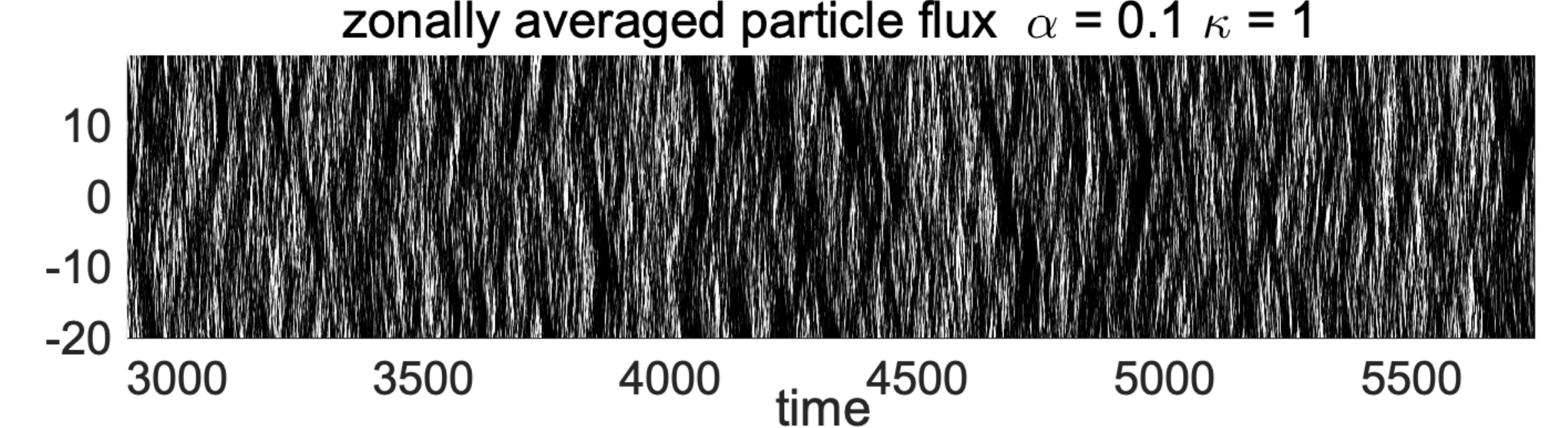}

}

\vspace{-1em}

\subfloat{\includegraphics[scale=0.36]{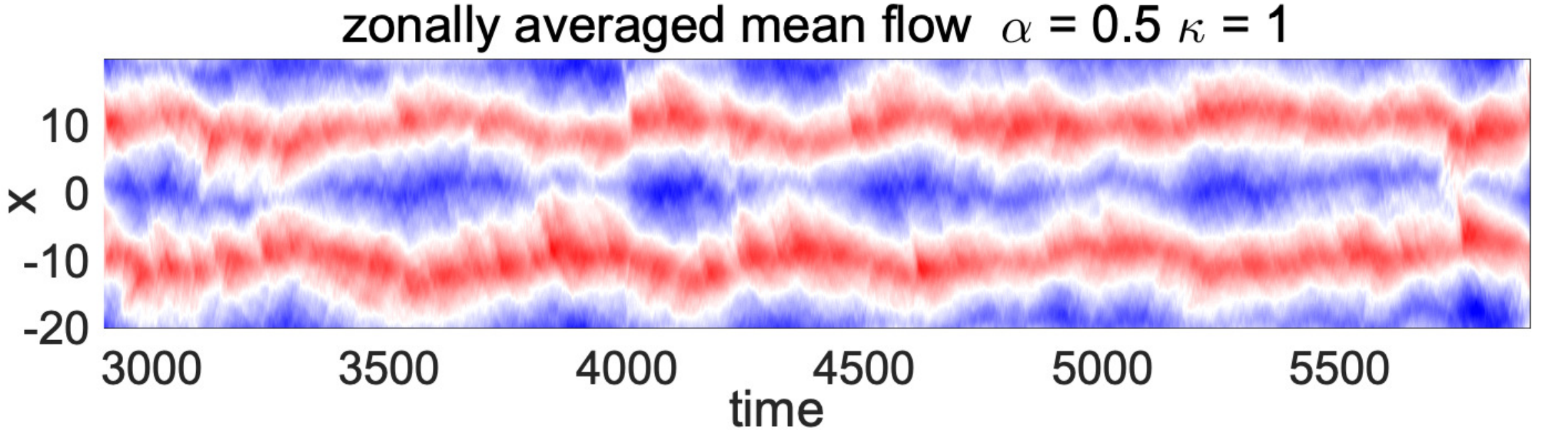}\includegraphics[scale=0.36]{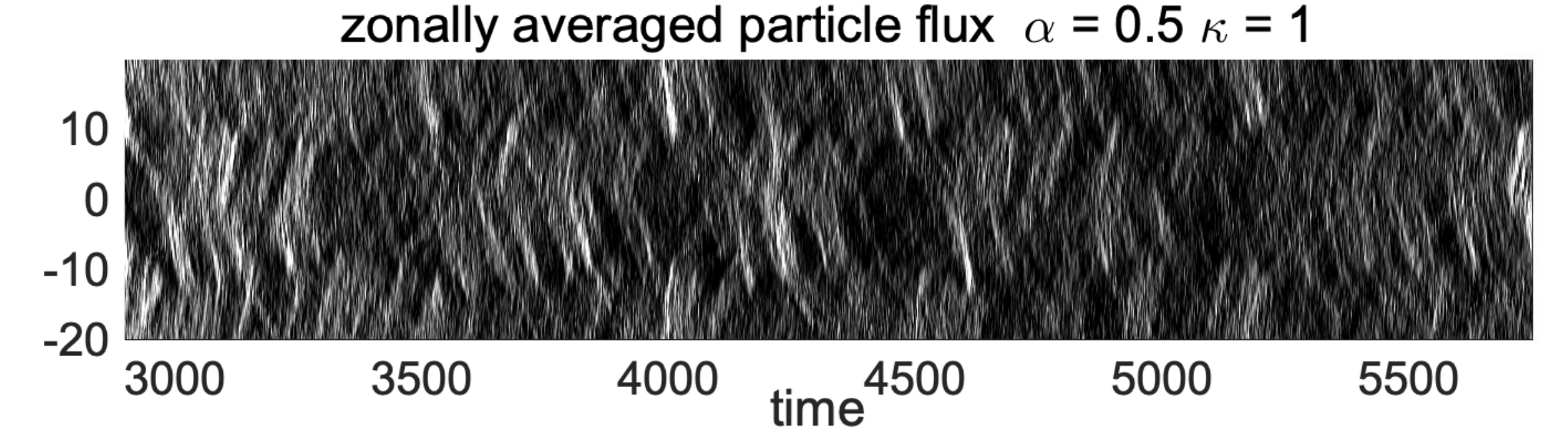}}

\vspace{-1em}

\subfloat{\includegraphics[scale=0.36]{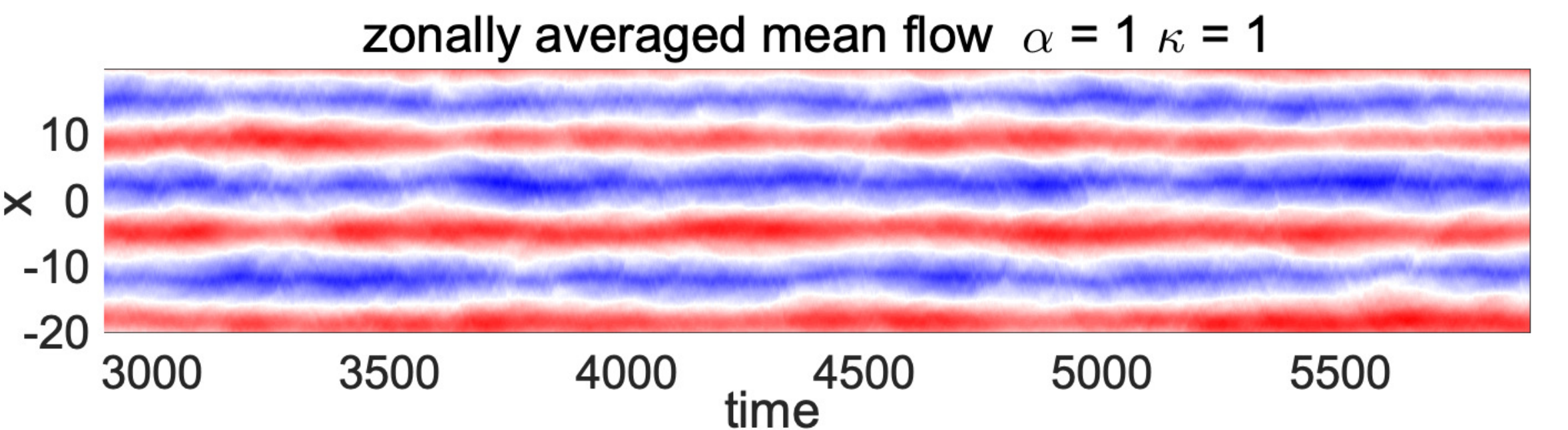}\includegraphics[scale=0.36]{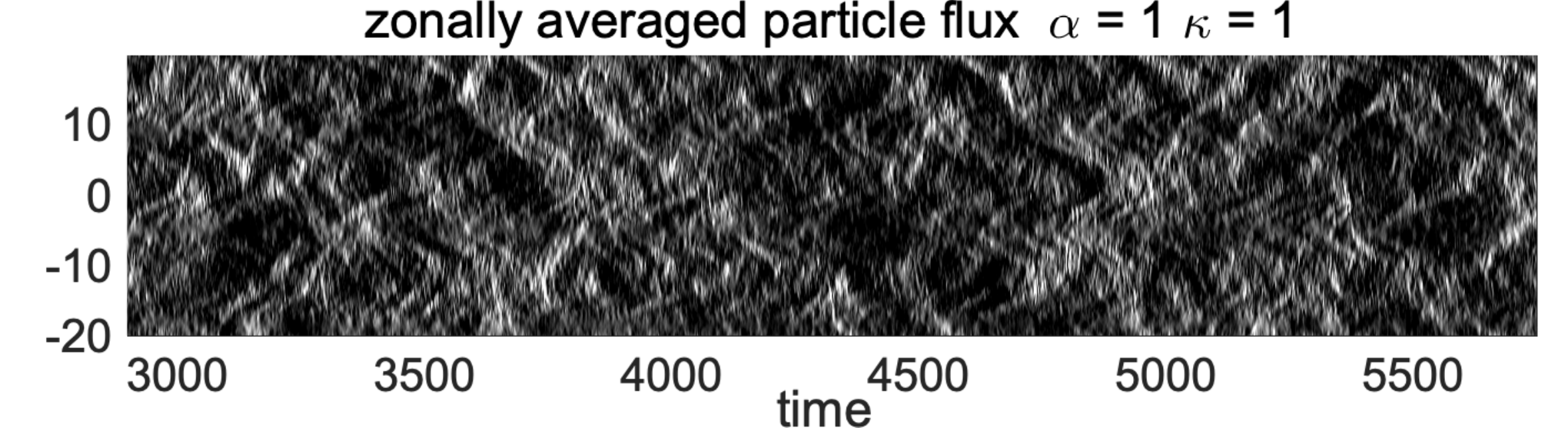}}

\caption{Time evolution of the zonal velocity $\overline{v}=\partial_{x}\overline{\varphi}$
(left column) and the zonal particle flux $\overline{\tilde{u}\tilde{n}}$
(right column) from the BHW model with doubly periodic boundary conditions, for $\kappa=1$, $\mu=1\times10^{-3}$, and different values of the adiabaticity parameter $\alpha$: $\alpha=0.1$ (top), $\alpha=0.5$ (middle), and $\alpha=1$ (bottom). The maximum values of zonal particle flux in the three cases are: 1895.2 in $\alpha=0.1$; 303.63 in $\alpha=0.5$; and 44.07 in $\alpha=1$. \label{fig:Time-evolution-peri}}

\end{figure}

The time evolution of the zonally averaged velocity
$\overline{v}$ and of the zonal particle flux $\overline{\tilde{u}\tilde{n}}$
for the doubly periodic domain case are shown in Figure \ref{fig:Time-evolution-peri}, for the same sets of parameters as in the previous figure: $\kappa=1$ and $\alpha=0.1$, $\alpha=0.5$, and $\alpha=1$, $\mu=10^{-3}$. Figure \ref{fig:Time-evolution-peri} may be analyzed in the light of the equivalent plots for the channel
geometry presented in Figure \ref{fig:Time-evolutions-flux}. With periodic boundary conditions in the $x$-direction as well, the zonal jets and the zonal particle
flux $\overline{\tilde{u}\tilde{n}}$ can cross the boundary, and appear on the other side of the computational domain. The zonal particle flux is thus observed to travel from both right to left and left to right across the edges of the domain, leading to avalanches throughout the computational domain. The zonal jets have the same variability as in the channel geometry, but we do not see the emergence of the soliton-like solutions which were localized in the regions of strong velocity shear and large density gradient. The strong turbulence regime with $\alpha=0.1$ displays strong, coherent and frequent radially propagating fronts, with a very strong maximum particle flux $\max_{t,x} \overline{\tilde{u}\tilde{n}} = 1895.2$; in the weaker turbulence regime with $\alpha=0.5$, the radially propagating fronts are still relatively strong, but appear less frequent and less coherent, and the maximum particle flux is reduced to 303.63; finally in the zonal flow dominant regime with $\alpha=1$, the particle flux is much weaker in amplitude, with maximum value 44.07, and much less coherent.

\section{Summary\label{sec:Summary}}

We have conducted direct numerical simulations with the flux-balanced Hasegawa-Wakatani (BHW) model \cite{majda2018flux} to study turbulent transport driven by a resistive drift-wave instability and mediated by the emergence of zonal flows. We have considered two different geometries, a channel domain geometry \cite{qi2019channel}, and a doubly periodic geometry, and different aspect ratios of the computational domain. For all geometries, we have found the existence of a Dimits shift for the particle flux, i.e. the existence of a region in parameter space in which linear instability is predicted yet near-zero particle flux is measured, and of a sudden increase in the particle flux as the density gradient or the parallel electron resistivity exceed the values corresponding to the Dimits threshold for our problem. Below the Dimits threshold, the flow is quasi-laminar, in an almost pure zonal state. The energy in the non-zonal modes, which are driven by the resistive-drift instability, is almost fully transferred to the zonal modes by the secondary instability. 

For all geometries, we also observe signatures of non-local transport mechanisms once the Dimits threshold is crossed, characterized by avalanches reminiscent of the heat flux avalanches reported in gyrokinetic simulations of the ITG instability \cite{candy2003anomalous,mcmillan2009avalanchelike,difpradalier2010,gorler2011}. These are large amplitude, large and meso-scale disordered radially propagating bursts, which are responsible for the strong intermittency measured in the time series of the particle flux. For domains which are extended in the radial direction,   we emphasized the rich multi-scale dynamics we identified in our simulations, where the large scale structure of the avalanches evolves in time in the transient phase during which zonal jets merge and interact with each other. In the channel domain geometry, where strong velocity shear and large density gradients are naturally created by the turbulent dynamics in the vicinity of the radial edges of the computational domain, coherent solitary structures with constant speed of propagation are seen to frequently emerge from the particle flux avalanches. Our results suggest that avalanches and soliton-like solutions have a common origin. The coherent solitary structures are responsible for larger amplitude bursts measured in the total particle transport.

We have made direct comparisons with the modified Hasegawa-Wakatani model (MHW) for both geometries. Particle flux avalanches are also seen in the MHW model, but without significant large-scale dynamics, which is consistent with the absence of multi-jet interactions and jet mergings in that model, except during a short initial transient phase. In the MHW model, zonal flows also reduce the level of particle transport, and can fully impede transport for certain choices of the physical parameters, but the transition from low transport to high transport is gradual as one increases the density gradient or parallel resistivity. In that sense, we can say that the MHW model does not have a true Dimits shift, unlike the BHW model. Another major difference with the BHW model is the absence of soliton-like solutions accompanying the avalanches in the particle flux. These differences highlight the key role of the treatment of the electron parallel dynamics in reduced fluid models to capture the key mechanisms of drift turbulence driven transport in magnetically confined devices.

Our results demonstrate that the BHW model is one of the simplest model supporting a drift instability and capturing all at once the existence of a Dimits shift, avalanches, and coherent solitary structures. It is therefore a convenient model to complete our theoretical understanding of the Dimits shift, for which much progress was recently made with other reduced fluid models \cite{stonge2017,ivanovArxiv,zhu2019theory}. When the BHW model is considered in the channel domain geometry, it is also a useful model for the study of soliton-like solutions, and their link with avalanches, which are topics which are even less well studied and understood than the Dimits shift. Specifically, one may wonder if all soliton-like solutions  \cite{mcmillan2009avalanchelike,vanwyk2016,mcmillan2018,zhou2019solitary,ivanovArxiv} observed in models of drift wave turbulence are ``ferdinons" \cite{vanwyk2016,ivanovArxiv} and if avalanches are made of multiple ferdinons which interact with each other, or if the radially propagating structures in avalanches are inherently different from the soliton-like solutions. Finally, with the rich multi-scale dynamics and abrupt phase transitions supported by the BHW model, it is an ideal testbed to investigate the capability of novel model reduction techniques to construct inexpensive models able to accurately predict the level of turbulent transport in magnetically confined plasmas. We are currently working on all of these questions, and expect to present results in the near future.

\section*{Acknowledgement}

This research of A. J. M. is partially supported by the Office of
Naval Research N00014-19-1-2286. D. Q. is supported as a postdoctoral
fellow on the grant. A.J.C. was partially
supported by the U. S. Department of Energy, Office of Science,
Fusion Energy Sciences under Award Nos. DE-FG02-86ER53223
and DE-SC0012398 and the Simons Foundation/SFARI(560651, AB).

\bibliographystyle{plain}
\nocite{*}
\bibliography{refs_dimits}

\end{document}